\documentclass{article}
\usepackage{graphicx}
\usepackage{amsfonts}
\def\ligne#1{\hbox to\hsize{#1}}
\def\leurre{\noindent\leftskip0pt\small\baselineskip 10pt}
\newtheorem{thm}{\textbf{Theorem}}
\newtheorem{fig}{\textbf{Figure}}
\newtheorem{tab}{\textbf{Table}}
\def\boxempty{\hbox{\vbox{\hsize=7pt\offinterlineskip
\ligne{
\vrule height 7pt depth 0pt width 0.6pt
\vbox to 7pt{\hsize=5.8pt
\hrule height 0pt depth 0.6pt width 5.8pt
\vfill
\hrule height 0.6pt depth 0pt width 5.8pt
}\hskip-0.5pt
\vrule height 7pt depth 0pt width 0.6pt
}}
}}

\author{Maurice {\sc Margenstern}}
\title{A weakly universal cellular automaton on the pentagrid with two states}
\begin{document}
\maketitle

\begin{abstract}
In this paper, we prove that there is a weakly universal cellular automaton on the pentagrid 
with two states. This paper improves in some sense a previous result with three states.
Both results make use of \textit{\`a la Moore} neighbourhood. However, the result with
three states is rotation invariant while that with two states is not.
In both cases, at each step of the computation, the set of non quiescent states has always
infinitely many cycles. 
\end{abstract}

\section{Introduction}

Self-contentedness is a convenience for the reader: he/she is exempted from searching definitions, 
statements and sometimes proofs from previous papers, many times from the same author(s).
Now, if I comply to this constraint, anti-plagiarism software will warn the reader that there is
a definite percent of overlap of this paper with other papers of the author. The struggle
against plagiarism is definitely necessary. I just mention that it may have some inconvenience.
Now, I am convinced that the reader will understand these inconveniences, so that he/she will
play the game of looking into the papers indicated in the references instead of throwing the
paper because some context is not at hand.
   Accordingly, I do not repeat most of the introduction of~\cite{mmpenta3st}. 
 As mentioned in the abstract,
the improvement with respect to~\cite{mmpenta3st} is obtained by relaxing the constraint
of rotation invariance. Again, we repeat that the new result is a true planar cellular
automaton,which means that at each step of the computation, the set of non quiescent states 
has always infinitely many cycles, as already mentioned in the abstract. Also, the
reason for not repeating the introduction is that we make use of the same model of
railway computation. Also for this reason, we do not mention the reminder sections 
of~\cite{mmpenta3st} about the railway model and its implementation in the hyperbolic
tiling we use in this paper. We also not repeat the introduction to the approach to
hyperbolic tilings used in this paper, again, the reader is referred 
to~\cite{mmbook1,mmbook2,mmbook3} and to several papers as~\cite{mmchapter} where other
references are available.

In Section~\ref{scenar}, I indicate the new features for implementing the model with the
constraint of using two states only.
In Section~\ref{rules}, we give the
rules of the automaton, insisting in the way we defined these rules in a context where
rotation invariance is no more required, which gives a definite advantage as indicated by
the result itself:

\begin{thm}\label{letheo}
There is a weakly universal cellular automaton on the pentagrid with Moore neighbourhood
which is 
truly planar and which has two states.
\end{thm}

Presently, we turn to the proof of this result.

\section{The scenario of the simulation}
\label{scenar}

    As in previous papers, sensors and control devices are no more immediate
neighbours of the cells where tracks arrive at the point which define a switch.
This reinforce the careful study of the tracks themselves as their role for conveying 
key information is more and more important.

Here too, tracks are blank cells marked
by appropriate milestones. We carefully study this point in Sub-section~\ref{tracks}.
Later, in Sub-sections~\ref{roundabout} and~\ref{forcontrol}, we look at the changes
introduced with respect to~\cite{mmpenta3st}.

\subsection{The tracks}\label{tracks}

In this implementation, the tracks are represented in the same way as in~\cite{mmpenta3st}.
For the convenience of the reader, we repeat the illustration of this implementation given
by Figure~\ref{elemtrack}.
 
\vtop{
\ligne{\hfill 
\includegraphics[scale=1.3]{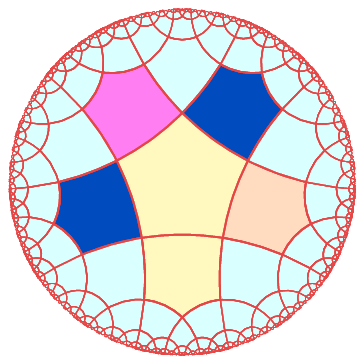}
\hfill}
\vspace{-20pt}
\begin{fig}\label{elemtrack}
\leurre
Element of the tracks.
\end{fig}
}

    However, some care has to be given to this structure. With two states, we have
to thoroughly check that all constraints we have to define horizontal and vertical tracks can
be satisfied. We refer the reader to Sub-section~\ref{trackrules} were we give the rules
and in which appropriate figures allow us to check the correctness of the implementation
given in this paper.
We just remember that for this implementation, the locomotive is implemented as
a single black cell: it has the same colour as the milestones of the tracks. Only the position
of the locomotive with respect to the milestones allow us to distinguish it from the milestones.
As clear from the next sub-section, we know that besides this
\textbf{simple locomotive}, the locomotive also occurs as a \textbf{double one} in some
portions of the circuit. Moreover, the circuit also makes use of signals which are implemented
in the form of a simple locomotive. So that at some point, it may happen that we have three
simple locomotives travelling on the circuit: the locomotive and two auxiliary signals involved
in the working of some switch. For aesthetic reasons, the black colour which is opposed
to the blank is dark blue in the figures.

\subsection{The round-about}
\label{roundabout}

    The round-about replaces the crossing, a railway structure, by a structure inspired by
road traffic. At a round-bout where two roads are crossing, if you want to keep the 
direction arriving at the round-about, you need to leave the round-about at the second 
road. Figure~\ref{roundaboutfig} illustrates this features, a figure from~\cite{mmpenta3st}
which is repeated here for convenience.

\vtop{
\ligne{\hskip 75pt
\includegraphics[scale=0.6]{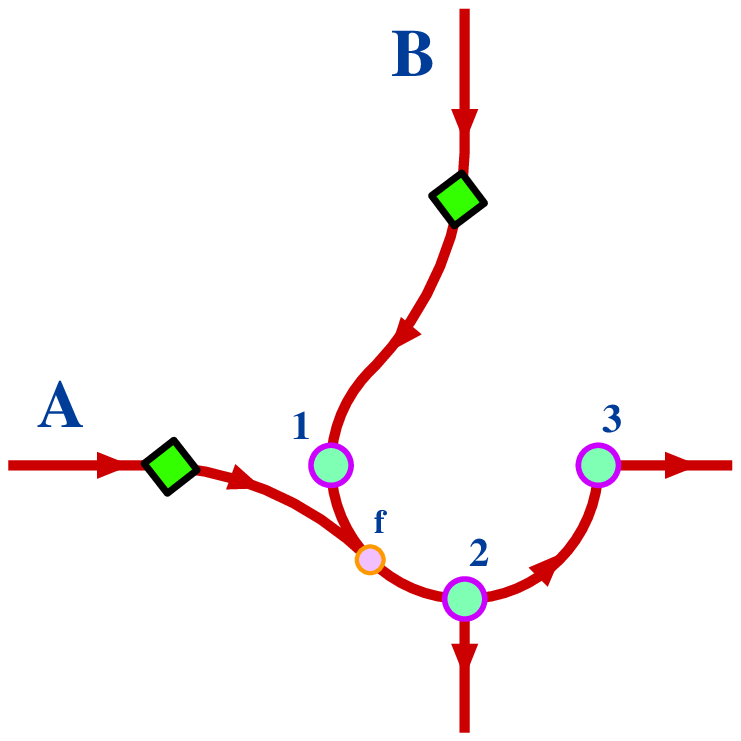}
\hfill}
\vspace{-5pt}
\begin{fig}\label{roundaboutfig}
\leurre
Implementation scheme for the round-about.
\end{fig}
}

Now, this strategy requires the cellular automaton knows how to count up to two. 
As in~\cite{mmpenta3st}, we use the three auxiliary structures which are represented in 
Figure~\ref{roundaboutfig} by a rhombus, a small circle and a bit larger ones.

   These structures are~: the fixed switch,
see Sub-subsection~\ref{fix}, the doubler and the selector, see Sub-subsection~\ref{doublesel}. 
When the locomotive arrives close to the round-about, it first meets the doubler: it 
transforms the simple locomotive into a double one which consists of two contiguous black cells
occupying blank cells of the tracks. Then, the locomotive may arrive at the fixed switch, 
depending on whether it arrived from~\textbf{A} or from\textbf{B}. Then,
the double locomotive arrives at the first selector: the structure recognizes a double 
locomotive. It kills one of the cells and the surviving simple locomotive is sent further
on the round-about. When it meets the second selector, the structure recognizes a simple
locomotive. Accordingly, it sends it on the track which leaves the round-about at that point.

   In order to use the illustrations of this section and of the next one, me make use of a 
numbering of the tiles based on what is indicated by Figure~\ref{pentanum}. The central tile
is numbered~0. Its side are increasingly and counter-clockwise numbered from~1 to~5. Side~1
is defined as follows. Let~$V$ be the upper vertex of tile~0 which is on the vertical 
diameter~$\delta$ of the right-hand side figure. The side~1 of tile~0 is the tile which shares~$V$ 
with tile~0 and which is on the left-hand side of~$\delta$.

\vskip 10pt
\vtop{
\ligne{\hskip 25pt
\includegraphics[scale=0.65]{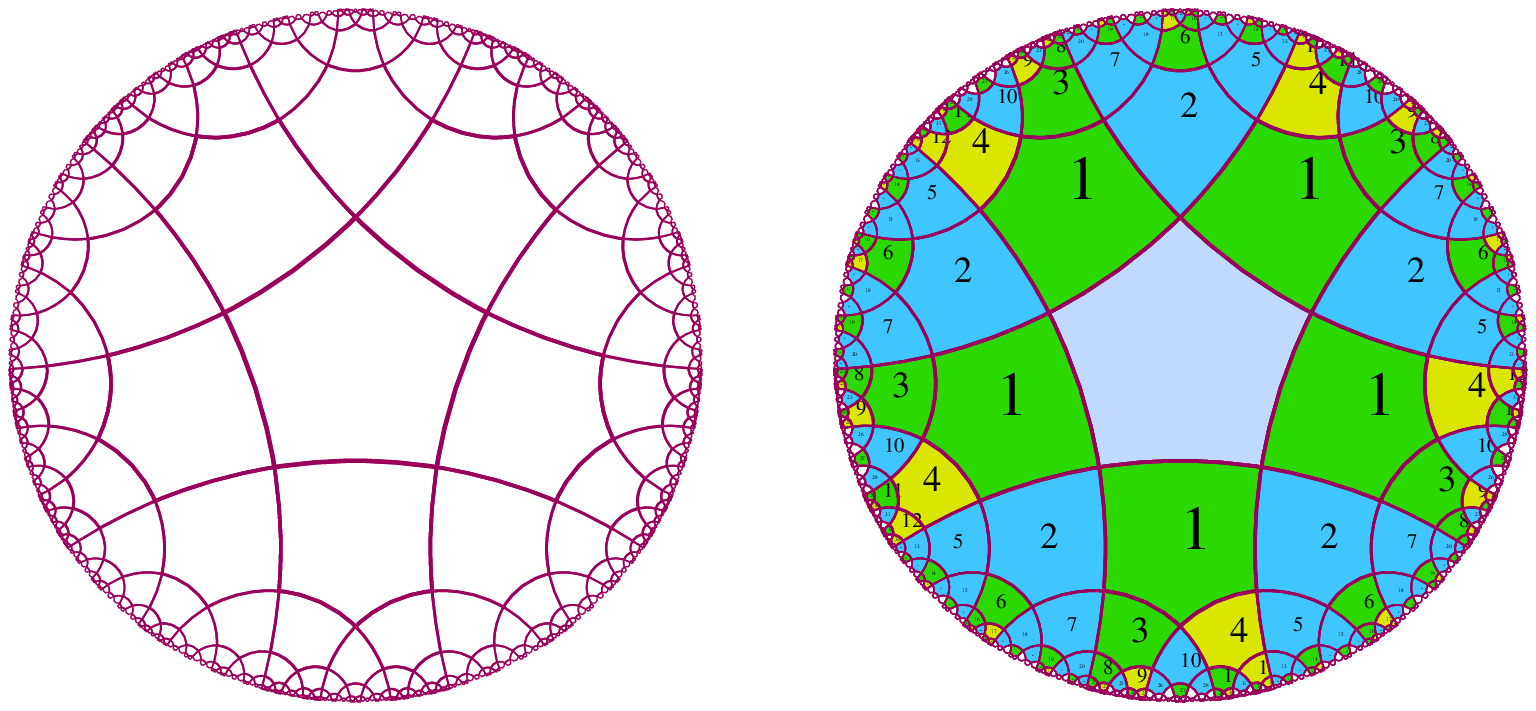}
\hfill}
\vspace{-5pt}
\begin{fig}\label{pentanum}
To left, the pentagrid. To right, the numbering of the tiles of the pentagrid.
\end{fig}
}

Each side of tile~0 delimits a sector: sector~1 is headed by the tile 1(1) obtained from tile~0
by a reflection on its side~1. The sector is delimited by the rays starting from the other
end of the side~1 of tile~0 which belong to edges of tile~1(1). The other sectors are obtained
by appropriate rotations around the centre of tile~0{} in such a way that the union of the sectors
is the complement of tile~0{} in the tiling. Inside each sector, the tiles are numbered
as indicated in the right-hand side part of Figure~\ref{pentanum}. The tile~$\nu$ of 
sector~$\sigma$ is numbered $\nu(\sigma)$. Each sector is panned by the same tree. 
In each tile, the sides are increasingly and counter-clockwise numbered from~1 to~5. The
side~1 of the tile $\nu(\sigma)$ is the side it shares with its father in the tree rooted
in the tile $1(\sigma)$. In Section~\ref{rules}, we shall indicate another way to number the
sides of the tiles based on this numbering too. When we shall consider the cellular automaton
constructed for the proof of Theorem~\ref{letheo}, we shall speak of cell~$\nu(\sigma)$ instead
of tile~$\nu(\sigma)$. In this notation, tile~0 will be denoted by 0(0). 
Once the side~1 is fixed in a cell~$c$, its edges are numbered according to
what we mentioned. The neighbour of~$c$ which shares the side~$i$ with it is called
neighbour~$i$. As we also consider as neighbours of~$c$ the cells which share a vertex only
with~$c$, we call these neighbours \textbf{vertex-neighbours} and we number them as follows:
the vertices are increasingly and counter-clockwise numbered from~6 to~10, vertex~6 being shared
by sides~1 and~5. Accordingly, the vertex-neighbour of~$c$ sharing its vertex~$j$ will be
called neighbour~$j$ or vertex-neighbour~$j$.

\subsubsection{The fixed switch}
\label{fix}

    As the tracks are one-way and as an active fixed switch always sends the locomotive in
the same direction, there is no need of the other direction: there is no active fixed switch.
Now, passive fixed switches are still needed as just seen in the previous paragraph. 
Figure~\ref{stabfix} illustrates the passive fixed switch when there is no locomotive around. 
We can see that it consists of elements of the tracks which are simply assembled in the 
appropriate way in order to drive the locomotive to the bottom direction in the picture, 
whatever upper side the locomotive arrived at the switch. 
From our description of the working of the round-about, a passive fixed switch must be 
crossed by a double locomotive as well as a simple one. 

\vtop{
\ligne{\hfill
\includegraphics[scale=1]{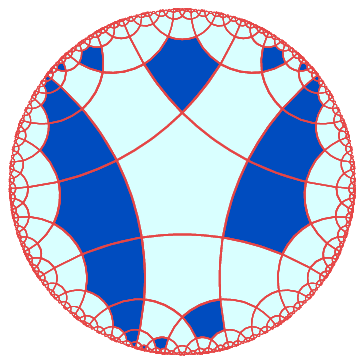}
\hfill}
\begin{fig}\label{stabfix}
The passive fixed switch in the pentagrid.
\end{fig}
}

   Later, in Section~\ref{fixedswitch}, we shall check that the structure illustrated by
Figure~\ref{stabfix} allows these crossings. Note that the cell~0(0) in Figure~\ref{stabfix}
has six black neighbours instead of five of them in an ordinary cell of the tracks in
a vertical track, see Figures~\ref{voievertd} and~\ref{voievertm}, further. Its black neighbours 
are neighbours~2, 4, 6, 7, 8 and~10. Note that neighbour~10 is a milestone belonging to the 
cell~1(5) which is
an element of the second track which arrives to the centre of the switch. 

\subsubsection{The doubler and the selector}
\label{doublesel}

    The doubler is illustrated by the left-hand side picture of Figure~\ref{stab_doublsel}. 
Note that the neighbourhood of the cell~$0(0)$
is different from an ordinary element of the tracks and it is also different from the cell~0(0)
of a fixed switch. Here, the black neighbours are neighbours~2, 4, 5, 7 and~8. It is slightly
simpler than the doubler in~\cite{mmpenta3st} although we have two states only. As will be seen
in Section~\ref{rules}, this is obtained by a particular numbering of the sides of a cell.
The working of the configuration is the same as in~\cite{mmpenta3st}, with this difference
that the locomotive is black.

The selector is illustrated by the right-hand side of Figure~\ref{stab_doublsel}. It is a very
symmetric picture, rather different from that for the selector in~\cite{mmpenta3st}. 
In fact, the symmetry of the picture and an appropriate choice of the sides~1 allow us
to make the cells~1(2) and~1(4) to react in a different way according to the number of
black cells in the locomotive. Whatever this number, when the first cell of the locomotive
is in cell~0(0), both cells~1(2) and~1(4) can see whether or not a second black cell follows
the first one in the cell~1(3). If a single black cell occurs, the cell~1(4) becomes black and
the cell~1(2) remains white. If a second black cell occurs, the cell~1(4) remains white
and the cell~1(2) turns to black.

\vtop{
\ligne{\hfill
\includegraphics[scale=1]{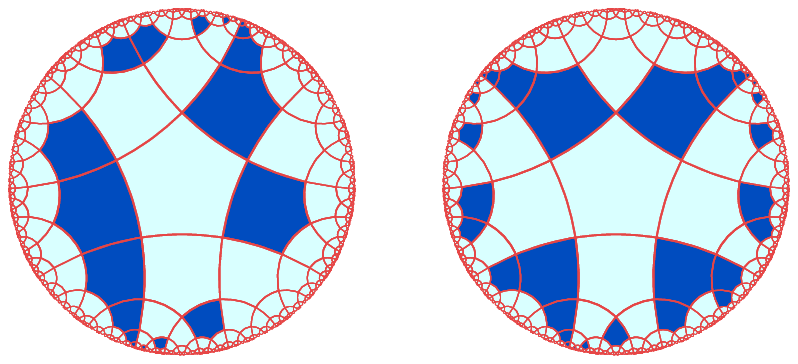}
\hfill}
\begin{fig}\label{stab_doublsel}
\leurre
To left: the doubler. To right: selector.
\end{fig}
}

    After this point, in both cases, the configuration is that of tracks so the locomotive
is sent on the appropriate path.

\subsection{The fork, the controller and the controller-sensor}
\label{forcontrol}

    In this Sub-section, we look at the decomposition of two active switches: the flip-flop 
and the active part of the memory switch. We follow the same implementation as 
in~\cite{mmpenta3st}, separating the working of the switch into two separate stages.
The fork is implemented as in~\cite{mmpenta3st}, but the controller is a bit different.
For the convenience of the reader, we reproduce the figures of~\cite{mmpenta3st}
which are again used by our automaton. In particular, we reproduce Figure~\ref{activswitches}.
Figure~\ref{stab_active} is different: the fork has the same configuration, but the
two possible stable configurations of the controller are different, distinguished by
the colour of the cell 1(1), especially when the
controller has to stop the locomotive.

\vtop{
\ligne{\hfill
\includegraphics[scale=0.35]{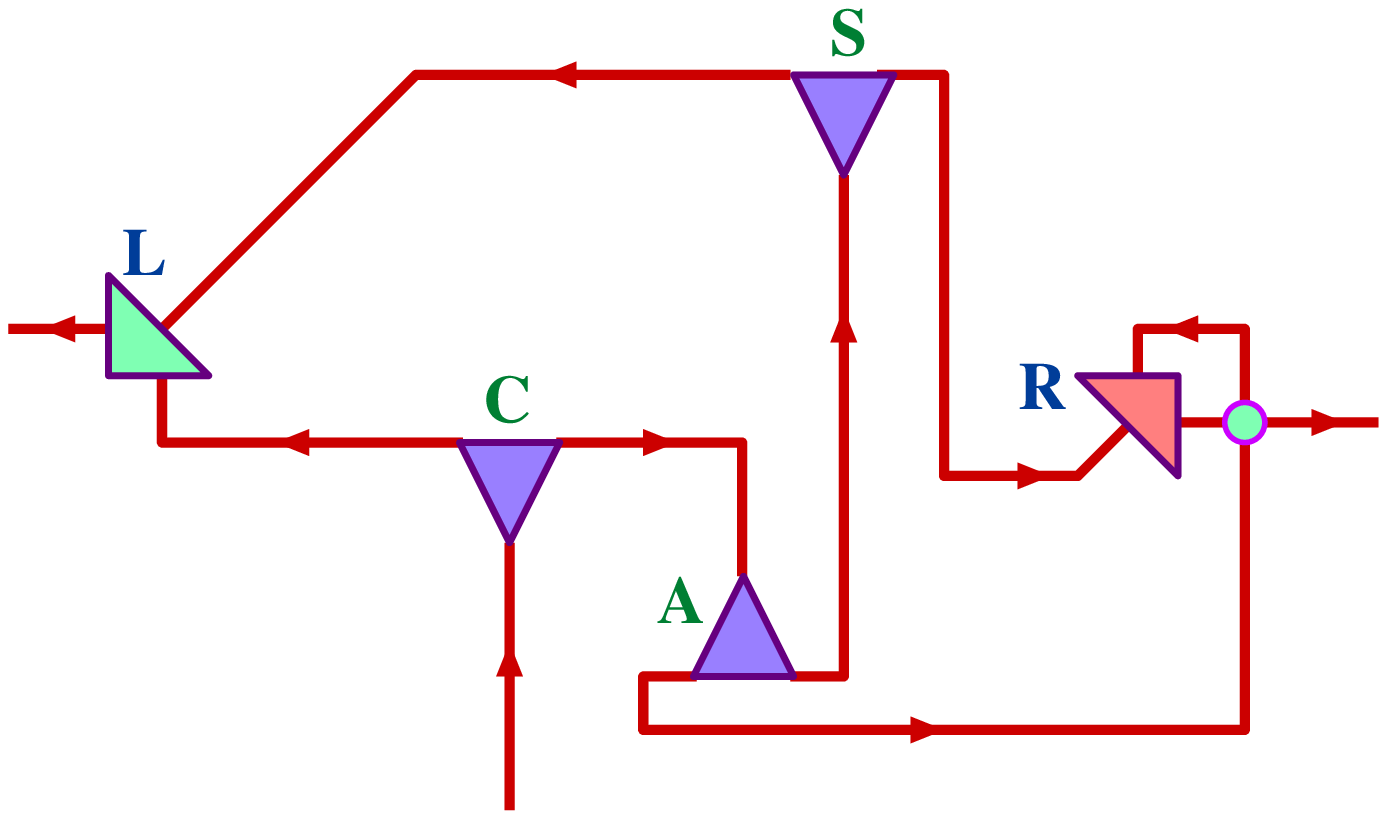}
\includegraphics[scale=0.35]{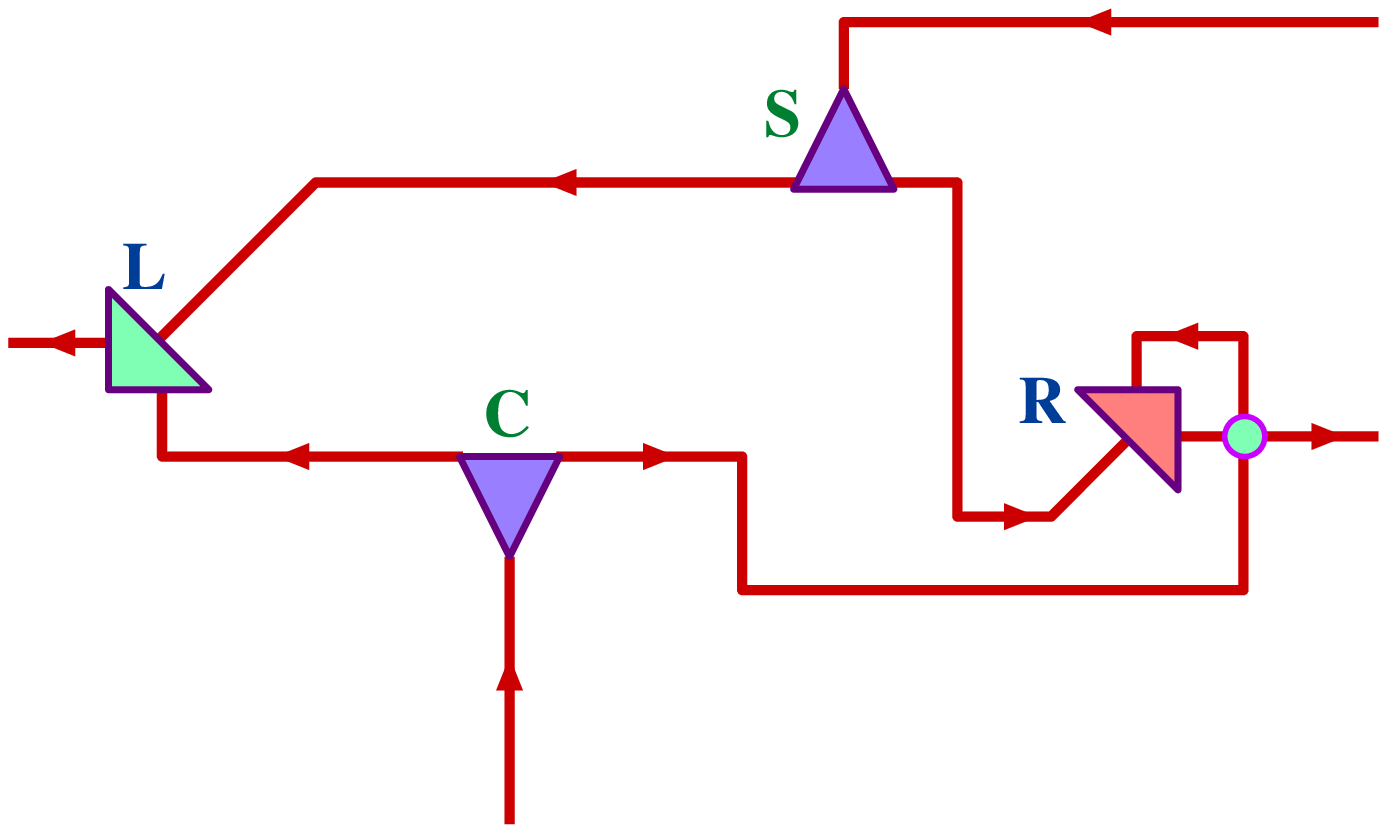}
\hfill}
\begin{fig}\label{activswitches}
\leurre
To left: the flip-flop. To right: the active memory switch.
\end{fig}
}

When the simple locomotive arrives at the fork~$C$,
it is duplicated into two simple locomotives, each one following its own path.
In the active memory switch, each locomotive goes to a controller. In the flip-flop,
one of these locomotives goes on its way to the controller, and the other goes to another
fork~$A$ which sends one locomotive to the other controller and the other, which is now a
third locomotive, is sent to a third fork~$S$.

The locomotives sent by~$S$ go both to a controller, one to the black controller, the other
to the white one. Now, when a locomotive sent by~$S$ arrives at a controller, it changes
the black one into a white one and the white one into a black one. Accordingly,
what should be performed by a flip-flop is indeed performed. It is enough to manage
things in such a way that the locomotives arriving to the controllers from~$S$ arrive
later than those sent by~$C$ and by~$A$. In the active memory switch, the locomotive
which arrives to~$S$ is sent by the passive memory switch. In this sense, the passive
memory switch is active while the active one is passive.

\vtop{
\ligne{\hfill
\includegraphics[scale=1]{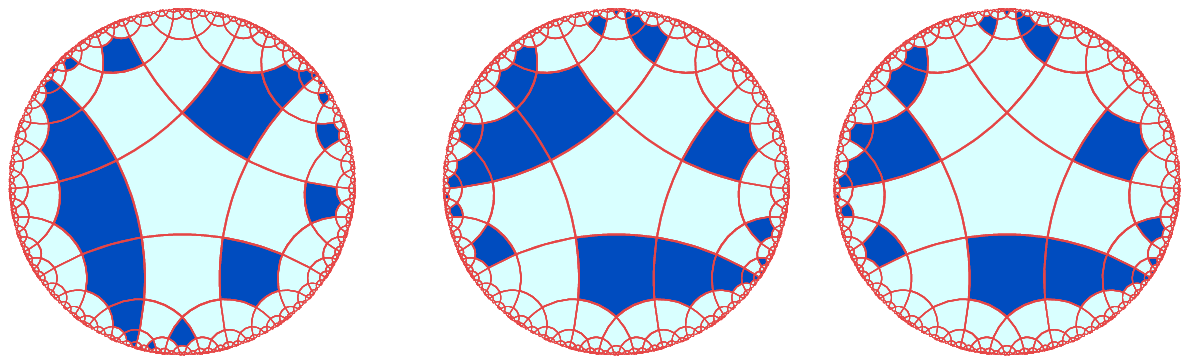}
\hfill}
\begin{fig}\label{stab_active}
\leurre
To left: the fork. To right, two configurations: the controller of the flip-flop and of the
active memory switch.
\end{fig}
}

In the fork, the locomotive, in its simple form, arrives through the cell~1(3). From the cell~0(0),
two simple locomotives arise at the same time at the cells 1(1) and~1(4).

In the controller, the locomotive arrives through the path defined by the cells
21(4), 8(4), 3(4) and~1(4), in this order. These cells constitute a branch of the tree 
spanning the sector. Along this path, the simple locomotive arrives at cell~0(0) 
through the cell~1(4). When the cell 1(1) is black, it goes on its way along the track 
through the cells 1(2), 3(2), 8(2) and 21(2). When the cell 1(1) is white, the locomotive 
is stopped at the cell 1(4) and does not reach cell~0(0).

   Figure~\ref{memopass} illustrates the construction of the passive memory switch with
the help of forks and controllers. That implementation is somehow different from the one
indicated in~\cite{mmpenta3st}. However, as in~\cite{mmpenta3st}, the controllers of 
Figure~\ref{memopass} are not the same as those of Figure~\ref{activswitches} and they are
different from those of~\cite{mmpenta3st}. The main reason is that in the case
of a passage of the locomotive through the non-selected track, in~\cite{mmpenta3st}, the 
controller let the locomotive go, it changes the selection and it sends a signal to
the other controller in order to order it to change its selection too. So that the controller
has to perform three tasks at once. The present idea is to slightly simplify the task
of the controller. As the locomotive should anyway not be stopped, we can place a fork on the
tracks passively arriving to the switch: this is the reason for placing the forks $S_1$ 
and~$S_2$ in Figure~\ref{memopass} where they are on the picture. Let us look at the locomotive
which arrives at the fork~$S_2$ on the figure. It corresponds to the non-selected track. The 
simple locomotive is duplicated: one goes to the fixed switch $F_1$ and goes further. The second
simple locomotive goes to the controller~$R$. As this controller signalizes a non-selected track,
the selection is changed and the controller let this locomotive go on along the track which leads
it to the other controller, $L$. There, at~$L$, the second locomotive changes the selection
from selected to non-selected. Note that the locomotive which goes on its way to~$L$ passes
through another fork, namely $S_3$ which sends a third locomotive to the active switch in order
to change its selection too.

\vtop{
\ligne{\hfill
\includegraphics[scale=0.55]{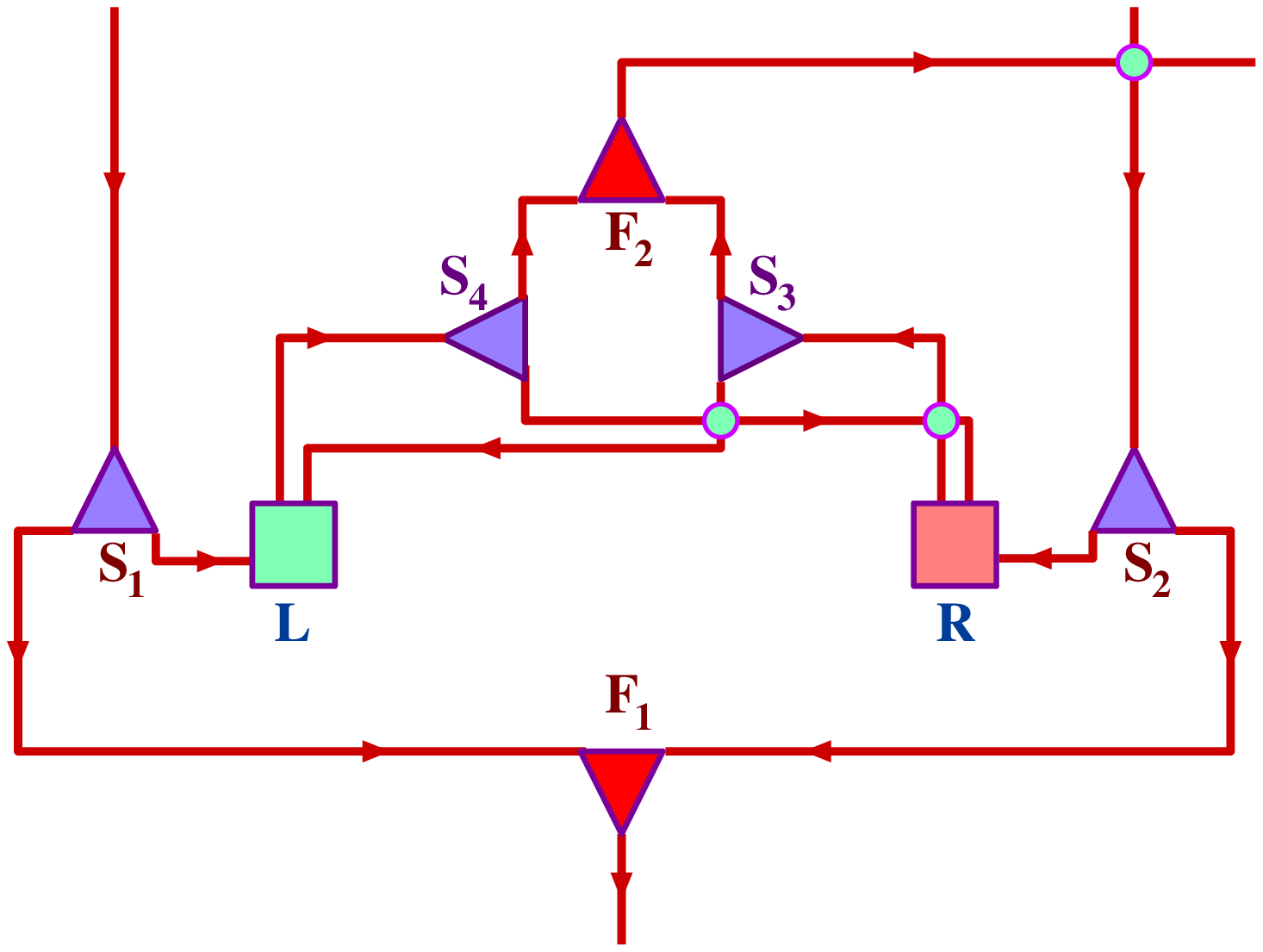}
\hfill}
\begin{fig}\label{memopass}
\leurre
Organization of the passive memory switch with forks and sensors. Note that the sensors
are not represented with the same symbol as the controllers in 
Figure~{\rm\ref{activswitches}}.
\end{fig}
}

   Consider the case when the simple locomotive arrives to~$S_1$, the fork which corresponds
on the figure to the selected track. Then, the locomotive is sent to~$L$ where it is stopped
as no change should be performed.

   Figure~\ref{stab_ctrlsgn} illustrates the controller of the passive memory switch. It is
presented when no locomotive is around and it presents the structure in its two basic 
configurations: when it is black and when it is blank. The difference is shown by the colour 
of the cell~1(1). The locomotive arrives at the cell~0(0) along the path 33(2), 12(2), 4(2)
and~1(2) in this order.
If the cell~1(1) is blank, which signalizes that the track is not selected, then
the cell~1(1) turns to black and the locomotive goes on its way through the cells 1(4), 4(4),
12(4) and 33(4). Accordingly, the constraints above described are satisfied. If the cell
1(1) is black when the locomotive arrives from 33(4), the locomotive is stopped at 1(2):
it does not reach even the cell~0(0) and the cell~1(1) remains white. In Section~\ref{rules}, we 
shall see that the working is slightly
more complex: this is induced by technical details raised by the rules themselves. Note that
when the locomotive arrived through the non-selected track, the second locomotive which arrives
to~$S_3$ or~$S_4$ in Figure~\ref{stab_ctrlsgn} arrives to the controller through the
path defined by 2(1), 2(6), 16(1) and 42(1). The signal stops at the cell~2(1) where it is
changed to blank. At the same time, the cell 1(1) is also changed to blank. It should be stressed
that when the cell~1(1) is black, and when the locomotive arrives through the selected track, 
no locomotive is sent to the other controller as there is nothing to change.

\vtop{
\ligne{\hfill
\includegraphics[scale=1]{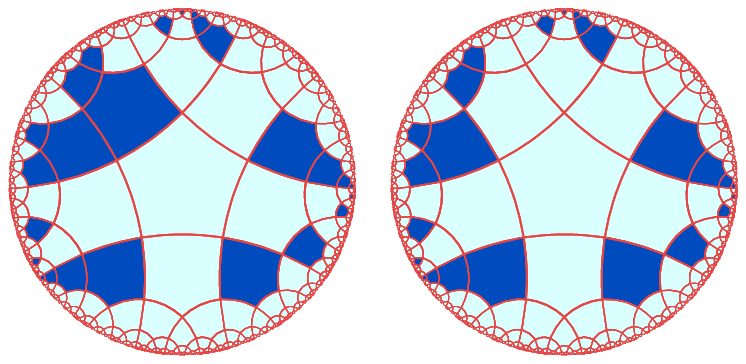}
\hfill}
\begin{fig}\label{stab_ctrlsgn}
\leurre
The controller-sensor of the passive memory switch.
\end{fig}
}

\section{Rules and figures}
\label{rules}

    The figures of Section~\ref{scenar} help us to establish the rules. The rules and the 
figures were established with the help of a computer program which checked the 
coherence of the rules. The program also wrote the PostScript files of the pictures from the 
computation of the application of the rules to the configurations of the various type 
of parts of the circuit. The computer program also established the traces of execution which
allow the reader to check the application of the rules.

    We have to revisit the format of the rules and also to explain what is allowed
from the relaxation from rotation invariance. We remind the reader that a rule has the form
\hbox{\footnotesize\tt$\underline{\hbox{X}}$$_o$X$_1$..X$_{12}$$\underline{\hbox{X}}$$_n$},
where \hbox{\footnotesize\tt X$_o$} is the state of the cell~$c$,
\hbox{\footnotesize\tt X$_i$} is the state of the neighbour~$i$ of~$c$ 
and \hbox{\footnotesize\tt X$_n$} is the new state of~$c$ once the rule
has been applied. As the rules no more observe the rotation invariance, we may freely choose
which is side~1 for each cell. We take this freedom from the format of the rule which
only requires to know which is neighbour~1. In order to restrict the number of rules,
it is decided that as a general rule, for a cell which is an element of the track, side~1
is the side shared by the cell and its next neighbour on the track. There can be exceptions
when the cell is in a switch or the neighbour of the central cell in a switch.
In particular, when a cell belongs to two tracks, side~1 is arbitrarily chosen among the two possible
cases. The milestones have their side~1 shared by an element of the track. This means that when 
a milestone~$\mu$ is a vertex-neighbour of an element of the track~$\eta$, the side~1 of $\mu$ 
is not a side of~$\eta$. 

    We have to keep in mind that there are two types of rules. Those
which keep the structure invariant when the locomotive is far from them, we call this
type of rules \textbf{conservative}, and those which control the motion of the locomotive.
These latter rules, which we call \textbf{motion rules}, are the rules applied to the
cells of  the tracks as well as their milestones and, sometimes to the cells of the structures
which may be affected by the passage of the locomotive.

   Table~\ref{consrules} provides us with rules which concern the white cells which are not on
 the track and who have a single or two black neighbours. It also concerns
black cells which have a single or two non white neighbours. A few other conservative
rules of the same kind also appear in further tables.

\def\aff #1 #2 #3 #4 {\ligne{\hfill\footnotesize\tt\hbox to 13pt{\hfill#1}
\hskip 5pt$\underline{\hbox{\tt#2}}$#3$\underline{\hbox{\tt#4}}$\hfill}\vskip-4pt
}
\def\laff #1 #2 #3 #4 {\hbox{{\footnotesize
$\underline{\hbox{\tt#1}}${\tt#2}$\underline{\hbox{\tt#3}}${}}}#4\hskip 4pt
}
\vtop{
\begin{tab}\label{consrules}
\leurre
Conservative rules.
\end{tab}
\ligne{\hfill
\vtop{\leftskip 0pt\parindent 0pt\hsize=78pt
\aff {  1} {W} {WWWWWWWWWW} {W}
\aff {  2} {B} {WWWWWWWWWW} {B}
\aff {  3} {W} {BWWWWWWWWW} {W}
\aff {  4} {W} {WWWWWBWWWW} {W}
}
\hfill
\vtop{\leftskip 0pt\parindent 0pt\hsize=78pt
\aff {  5} {W} {WWWWBWBWWW} {W}
\aff {  6} {W} {WWWWBBBWWW} {W}
\aff {  7} {W} {WWWWWWBWWW} {W}
}
\hfill
\vtop{\leftskip 0pt\parindent 0pt\hsize=78pt
\aff {  8} {B} {BWWWWWWWWW} {B}
\aff {  9} {W} {BWWWBWBWWW} {W}
\aff { 10} {W} {BBWWBBWWWW} {W}
}
\hfill}
}

\subsection{Defining tracks and their rules}
\label{trackrules}

   As explained in Section~\ref{scenar}, the construction of the tracks is very important.
The figures we give in this sub-section and in the following ones are \textit{local views}.
By that phrase, we mean that we can only see a very tiny part of the hyperbolic plane, that the
central cell is not the centre of the hyperbolic plane, such a centre does not exist,
but the cell on which we focus our attention. In these figures, we can see the central cell,
an element of the tracks in Figures~\ref{voievertd} and~\ref{voievertm}, as well as two other
cells of the track, on both sides of the central cell.

   We need to define verticals and horizontals in the pentagrid in order to simulate the
corresponding Euclidean notions, which will allow us to directly implement the basic
elements of the railway model, see~\cite{mmbook3,mmchapter} for references. We look at verticals 
in Sub-subsection~\ref{les_verticales} and we look at horizontals in 
Sub-subsection~\ref{les_horizontales}, defining in these
sub-subsections what we call verticals and horizontals in the pentagrid.

\subsubsection{The vertical tracks}
\label{les_verticales}

   We call \textbf{line of the pentagrid} a line of the hyperbolic plane which supports
a side of a tile of the pentagrid. Note that if a line~$\delta$ of the hyperbolic plane
supports a side~$s$ of some tile~$\tau$, it also supports the sides of a sequence
\hbox{$\{\tau_n\}_{n\in\mathbb{Z}}$} where $\tau=\tau_{k}$ for some \hbox{$k\in\mathbb{Z}$}
and such that each $\tau_n$ has a side on~$\delta$, that $\tau_n$ and $\tau_{n+1}$ share an
edge for all \hbox{$n\in\mathbb{Z}$} and that all $\tau_n$ are on the same side of~$\delta$.
Such a set of tiles is called a \textbf{tape}.
It is easy to see that a line of the pentagrid~$\delta$ defines two tapes: each one is in
its own side of~$\delta$. Say that a ray is a \textbf{ray of the pentagrid} if it is contained
in a line of the pentagrid and if it is issued from a vertex of a tile of the pentagrid.
We obtain a \textbf{half-tape} by replacing the line of the pentagrid~$\delta$
by a ray of the pentagrid in the above definition of a tape.
The tile of a half-tape defined by the ray of the pentagrid~$\rho$ which contains the vertex 
from which $\rho$ is issued is called the \textbf{head} of the half-tape. 

We say that a \textbf{vertical} is a finite sequence of consecutive tiles belonging to
a half-tape~$\eta$ and which also belong to a branch of the Fibonacci tree rooted at the head
of~$\eta$. We remind the reader that the Fibonacci tree is the tree which generates the
restriction of the pentagrid to a sector as illustrated in Figure~\ref{pentanum}.

\newdimen\largesept\largesept=21.5pt
\def\execneuf #1 #2 #3 #4 #5 #6 #7 #8 #9 {
\hbox{
\hbox to 25pt{\hfill#1\hskip 4pt}
\hbox to \largesept{\hfill#2\hskip 4pt}
\hbox to \largesept{\hfill#3\hskip 4pt}
\hbox to \largesept{\hfill#4\hskip 4pt}
\hbox to \largesept{\hfill#5\hskip 4pt}
\hbox to \largesept{\hfill#6\hskip 4pt}
\hbox to \largesept{\hfill#7\hskip 4pt}
\hbox to \largesept{\hfill#8\hskip 4pt}
\hbox to \largesept{\hfill#9\hskip 4pt}
}
}

\def\exechuit #1 #2 #3 #4 #5 #6 #7 #8 {
\hbox{
\hbox to 25pt{\hfill#1\hskip 4pt}
\hbox to \largesept{\hfill#2\hskip 4pt}
\hbox to \largesept{\hfill#3\hskip 4pt}
\hbox to \largesept{\hfill#4\hskip 4pt}
\hbox to \largesept{\hfill#5\hskip 4pt}
\hbox to \largesept{\hfill#6\hskip 4pt}
\hbox to \largesept{\hfill#7\hskip 4pt}
\hbox to \largesept{\hfill#8\hskip 4pt}
}
}

\def\execsept #1 #2 #3 #4 #5 #6 #7 {
\hbox{
\hbox to 25pt{\hfill#1\hskip 4pt}
\hbox to \largesept{\hfill#2\hskip 4pt}
\hbox to \largesept{\hfill#3\hskip 4pt}
\hbox to \largesept{\hfill#4\hskip 4pt}
\hbox to \largesept{\hfill#5\hskip 4pt}
\hbox to \largesept{\hfill#6\hskip 4pt}
\hbox to \largesept{\hfill#7\hskip 4pt}
}
}

\def\execsix #1 #2 #3 #4 #5 #6 {
\hbox{
\hbox to \largesept{\hfill#1\hskip 4pt}
\hbox to \largesept{\hfill#2\hskip 4pt}
\hbox to \largesept{\hfill#3\hskip 4pt}
\hbox to \largesept{\hfill#4\hskip 4pt}
\hbox to \largesept{\hfill#5\hskip 4pt}
\hbox to \largesept{\hfill#6\hskip 4pt}
}
}

\def\execcinq #1 #2 #3 #4 #5 {
\hbox{
\hbox to \largesept{\hfill#1\hskip 4pt}
\hbox to \largesept{\hfill#2\hskip 4pt}
\hbox to \largesept{\hfill#3\hskip 4pt}
\hbox to \largesept{\hfill#4\hskip 4pt}
\hbox to \largesept{\hfill#5\hskip 4pt}
}
}

\def\execquat #1 #2 #3 #4 {
\hbox{
\hbox to \largesept{\hfill#1\hskip 4pt}
\hbox to \largesept{\hfill#2\hskip 4pt}
\hbox to \largesept{\hfill#3\hskip 4pt}
\hbox to \largesept{\hfill#4\hskip 4pt}
}
}

\def\exectroi #1 #2 #3 {
\hbox{
\hbox to \largesept{\hfill#1\hskip 4pt}
\hbox to \largesept{\hfill#2\hskip 4pt}
\hbox to \largesept{\hfill#3\hskip 4pt}
}
}

\vtop{
\vspace{-15pt}
\ligne{\hskip 10pt
\includegraphics[scale=0.55]{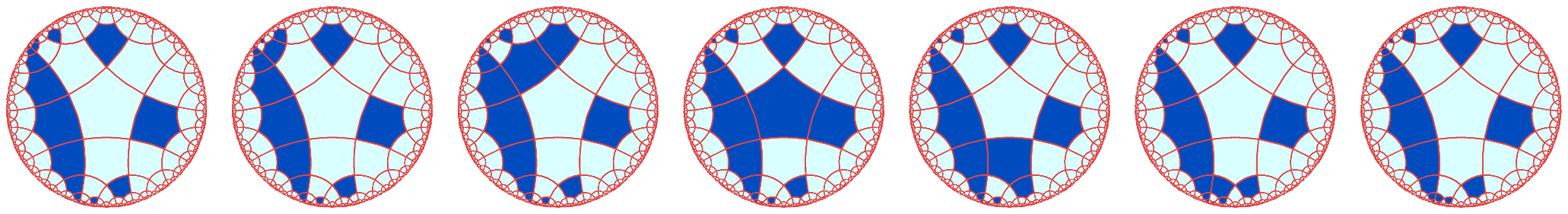} 
\hfill}
\vspace{-20pt}
\ligne{\hskip 10pt
\includegraphics[scale=0.55]{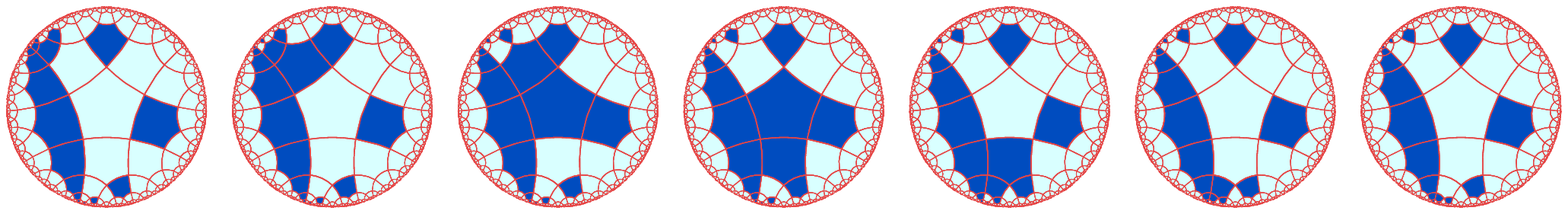} 
\hfill}
\begin{fig}\label{voievertd}
\leurre
Above: a single locomotive going down a vertical track.
Below: a double locomotive going down a vertical track.
\end{fig}
}

\vtop{
\vspace{-15pt}
\ligne{\hskip 10pt
\includegraphics[scale=0.55]{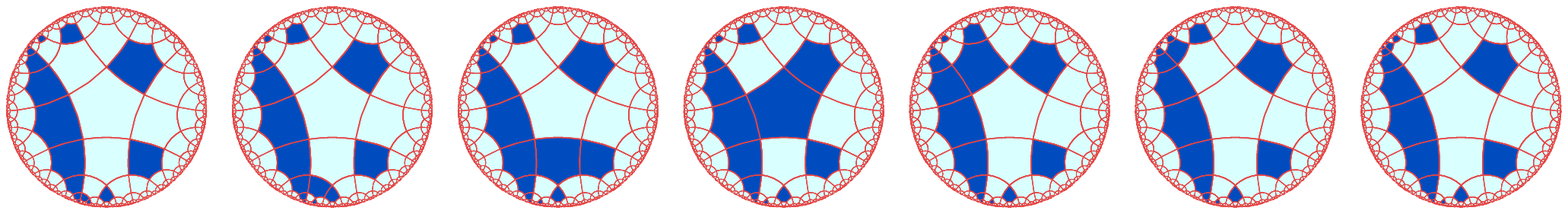} 
\hfill}
\vspace{-20pt}
\ligne{\hskip 10pt
\includegraphics[scale=0.55]{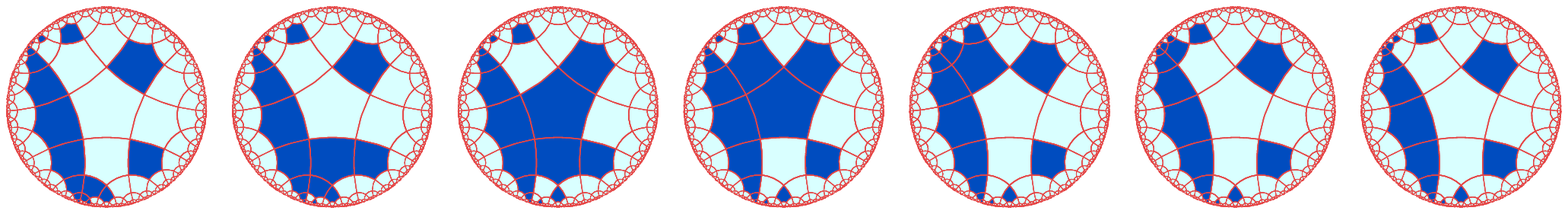} 
\hfill}
\begin{fig}\label{voievertm}
\leurre
Above: a single locomotive going up a vertical track.
Below: a double locomotive going up a vertical track.
\end{fig}
}
\vskip 5pt
   A \textbf{vertical track} as illustrated by Figures~\ref{voievertd} and~\ref{voievertm}
consists in two verticals which belong to half-tapes defined on both sides of the same
ray of the pentagrid~$\rho$. We say that $\rho$ is the \textbf{backbone} of the
vertical track. One vertical along~$\rho$ consists of blank cells while the other consists of
black cells: each black cell~$\beta$ is a milestone of the white cell~$\omega$ which shares a 
side on~$\rho$ with~$\beta$. If the side~1 of~$\omega$ is shared with the centre of the 
next element of the track, $\beta$ is the neighbour~5 of~$\omega$ while the second milestone
of~$\omega$ is its neighbour~2. Note that $\omega$ can see other milestones of its neighbouring
elements of the track. In a going down track, neighbour~6 is a milestone in the next element 
of the tracks, it is also a neighbour of the neighbour~5 of~$\omega$. Neighbours~9 and~10 
of~$\omega$ are milestones of the previous element of the track, neighbour~10 being also a 
neighbour of the neighbour~5  of~$\omega$. When the track is going up, the vertex-neighbours
are~7, a milestone of the next element of the track, and~8 with~9 which are milestones of the
previous element. 

    Note that if we take the reflection in its backbone of the vertical track illustrated by 
Figure~\ref{voievertd}, we obtain the image of the vertical track illustrated by
Figure~\ref{voievertm} by a reflection in a vertex of a pentagon lying on the backbone of that
vertical track. This shows us that Figures~\ref{voievertd} and~\ref{voievertm} illustrate
all possible vertical track: going down as in Figure~\ref{voievertd} or going up
as in Figure~\ref{voievertm} and again whichever the side where the track lies with respect
to its backbone. This is why the rules of Tables~\ref{verttracktab} and~\ref{mounttracktab} also 
apply to tracks which are displayed on another side compared 
to that of Figures~\ref{voievertd} and~\ref{voievertm}.

   Let us now turn to the rules displayed by Table~\ref{verttracktab} for a locomotive which goes
down and by Table~\ref{mounttracktab} for a locomotive which goes up. When the locomotive
is not there, the rule which is applied to the centre of an element of the track is 
rule~25{} in Table~\ref{verttracktab}: 
\laff {W} {WBWWBBWWBB} {W} {.}  
Note that, as already mentioned, the milestones are neighbours~2 and~5 and that
the milestones of neighbouring elements are seen by the centre as neighbours~6, 9 and~10
when the track goes down.
When a simple locomotive crosses an element~$\eta$ in a vertical track, rule 26 makes the 
locomotive enter~$\eta$. Indeed, the locomotive is seen by~$\eta$ as its neighbour~4. Rule~27
makes the locomotive go out: the cell turns from white to black. Rule~28 witnesses
that the locomotive went out: it is seen as the neighbour~1 of~$\eta$. Table~\ref{exvertd}
shows which rule are applied on the cells of the track around the central cell of 
Figure~\ref{voievertd}. We can see than the same sequence of rules: 25, 27, 28 and again 25
is applied along the cells of the tracks: 4(1), 1(1), 0(0), 1(3) and 3(3). These rules are
motion rules as they apply to the centres of the elements of the tracks. They allow
the locomotive to move on its way. But other rules are also needed. We can see there necessity
in Table~\ref{exvertd} which displays the rules applied to the cells~1(2) and~1(4) which are
the milestones of the cell~$0(0)$. When the locomotive is far from the element of the track,
rule~12 applies to 1(2) and rule~2 applies to~1(4). For that latter cell it is clear: all 
neighbours of~1(4) are white. For the cell~1(2), it can see two other milestones on the half-line
of black cells to which it belongs: its neighbour~2, a milestone of the cell of the track
before~0(0) and its neighbour~5, a milestone of the cell of the track after~0(0). This is
exactly what rule~12 says: with such a neighbouring, the cell remains blank. When the locomotive
approaches, the milestone can first see it as its neighbour~7: rule~15 applies. Then, the
milestone can see the locomotive as its neighbour~1: rule~16 applies. Then, the locomotive
is going further, it is seen as neighbour~6: rule~17 applies. After that time, the locomotive
is no more seen before a possible new passage on this very track, rule~12 applies again.
All these rules: 12, 15, 16 and~17 leave the milestone unchanged. For the cell~1(4),
other rules are applied with the same effect: the state is not changed. Rule~8 is applied when the
locomotive is in the cell~0(0) and not before: it is seen as neighbour~1 of the milestone. 
Then rule~11 is applied, indicating that the locomotive, seen as neighbour~7, has left the
cell~0(0). It is no more seen so that rule~2 again is applied.

   Table~\ref{verttracktab} also displays the rules when the locomotive is double. 
Table~\ref{exvertd} also indicates which rules are applied to the same cells as those seen 
previously. To some of the rules already examined, new rules are applied due to the fact
that now two contiguous black cells travel together. When the locomotive approaches the cell~0(0),
the same rule~26 is applied. But at the next time, instead of rule~27, rule~29 is applied: this
rule makes the second black cell enter the cell~0(0) as this second cell is seen as the neighbour~4
of~0(0). Then, rule~31 is applied which makes the second black cell leave~0(0) because the first
cell of the locomotive is seen as the neighbour~1 of~0(0). When the cell~0(0) is again white,
rule~28 is again applied: it can see the second cell of the locomotive in neighbour~1. After that
time, rule~25 is again applied. For the milestones, we have a similar situation. 
For~1(2), after rule~15, rule~33 is applied: it can see the first cell of the locomotive
in its neighbour~1 and the second one in its neighbour~7. Then rule~34 is applied which
can see the first cell of the locomotive in the neighbour~6 of the milestone and the second
cell of the locomotive in its neighbour~1. After that, rule~17 is applied as the first cell of the
locomotive is no more seen by the milestone. For the cell~1(4), similarly, rule~32 is applied
after rule~8 as two cells of the locomotive are seen: in the neighbour~7 of~1(4) and in its 
neighbour~1. Then rule~11 applies as a single cell of the locomotive can be seen. after
that, rule~2 is again applied.

\vskip 5pt
\vtop{
\begin{tab}\label{verttracktab}
\leurre
Rules forgoing down vertical tracks.
\end{tab}
\vspace{-12pt}
\ligne{\hfill a single locomotive:\hfill}
\vskip 2pt
\ligne{\hfill
\vtop{\leftskip 0pt\parindent 0pt\hsize=76pt
\aff { 11} {B} {WWWWWWBWWW} {B}
\aff { 12} {B} {WBWWBWWWWW} {B}
\aff { 13} {W} {BWWWWWBWWW} {W}
\aff { 14} {W} {BWWWWBWWWW} {W}
\aff { 15} {B} {WBWWBWBWWW} {B}
}\hfill
\vtop{\leftskip 0pt\parindent 0pt\hsize=76pt
\aff { 16} {B} {BBWWBWWWWW} {B}
\aff { 17} {B} {WBWWBBWWWW} {B}
\aff { 18} {B} {WWWWBBWWWW} {B}
\aff { 19} {B} {WWWWBWWWWW} {B}
\aff { 20} {W} {WBWWBWWWBB} {W}
}
\hfill
\vtop{\leftskip 0pt\parindent 0pt\hsize=76pt
\aff { 21} {W} {WBWBBWWWBB} {W}
\aff { 22} {B} {BWWWWBWWWW} {B}
\aff { 23} {W} {WWWWWBBWWW} {W}
\aff { 24} {W} {WBWWBBWWWW} {W}
}
\hfill
\vtop{\leftskip 0pt\parindent 0pt\hsize=76pt
\aff { 25} {W} {WBWWBBWWBB} {W}
\aff { 26} {W} {WBWBBBWWBB} {B}
\aff { 27} {B} {WBWWBBWWBB} {W}
\aff { 28} {W} {BBWWBBWWBB} {W}
}
\hfill}
\vskip 7pt
\ligne{\hfill a double locomotive:\hfill}
\vskip 2pt
\ligne{\hfill
\vtop{\leftskip 0pt\parindent 0pt\hsize=76pt
\aff { 29} {B} {WBWBBBWWBB} {B}
\aff { 30} {W} {BWWWBBBWWW} {W}
\aff { 31} {B} {BBWWBBWWBB} {W}
}\hfill
\vtop{\leftskip 0pt\parindent 0pt\hsize=76pt
\aff { 32} {B} {BWWWWWBWWW} {B}
\aff { 33} {B} {BBWWBWBWWW} {B}
\aff { 34} {B} {BBWWBBWWWW} {B}
}
\hfill}
}

\vskip 5pt
\def\rrr{\hbox{\footnotesize\tt R}}
\def\www{\hbox{\footnotesize\tt W}}

\vtop{
\begin{tab}\label{exvertd}
\leurre
Locomotive going down a vertical.
\end{tab}
\vskip-6pt
\ligne{\hfill\hbox to 180pt{\hfill simple locomotive:\hfill}
\hskip 5pt\hbox to 155pt{\hfill double locomotive:\hfill}\hfill}
\vskip 2pt
\ligne{\hfill
\vtop{\leftskip 0pt\parindent 0pt\hsize=180pt
\ligne{\hfill  \execsept     {}       0       1       2       3       4       5 \hfill}
\ligne{\hfill  \execsept  {4(1)}  {  26}  {  27}  {  28}  {  25}  {  25}  {  25}   \hfill}
\ligne{\hfill  \execsept  {1(1)}  {  25}  {  26}  {  27}  {  28}  {  25}  {  25}   \hfill}
\ligne{\hfill  \execsept  {0(0)}  {  25}  {  25}  {  26}  {  27}  {  28}  {  25}   \hfill}
\ligne{\hfill  \execsept  {1(3)}  {  25}  {  25}  {  25}  {  26}  {  27}  {  28}   \hfill}
\ligne{\hfill  \execsept  {3(3)}  {  25}  {  25}  {  25}  {  25}  {  26}  {  27}   \hfill}
\ligne{\hfill  \execsept  {1(2)}  {  12}  {  12}  {  15}  {  16}  {  17}  {  12}   \hfill}
\ligne{\hfill  \execsept  {1(4)}  {   2}  {   2}  {   2}  {   8}  {  11}  {   2}   \hfill}
}
\hskip 5pt
\vtop{\leftskip 0pt\parindent 0pt\hsize=155pt
\ligne{\hfill   \execsix       0       1       2       3       4       5 \hfill}
\ligne{\hfill   \execsix   {  29}  {  31}  {  28}  {  25}  {  25}  {  25}   \hfill}
\ligne{\hfill   \execsix   {  26}  {  29}  {  31}  {  28}  {  25}  {  25}   \hfill}
\ligne{\hfill   \execsix   {  25}  {  26}  {  29}  {  31}  {  28}  {  25}   \hfill}
\ligne{\hfill   \execsix   {  25}  {  25}  {  26}  {  29}  {  31}  {  28}   \hfill}
\ligne{\hfill   \execsix   {  25}  {  25}  {  25}  {  26}  {  29}  {  31}   \hfill}
\ligne{\hfill   \execsix   {  12}  {  15}  {  33}  {  34}  {  17}  {  12}   \hfill}
\ligne{\hfill   \execsix   {   2}  {   2}  {   8}  {  32}  {  11}  {   2}   \hfill}
}
\hfill}
}
\vskip 10pt

   In Table~\ref{mounttracktab}, we have the rules for going up. Table~\ref{exvertm}
shows the rules which where applied in a motion illustrated by Figure~\ref{voievertm},
where the locomotives are going up. This time, rules~35, 41, 42 and~48 are used. Indeed,
although side~1 is defined in the same way as in the case of Figure~\ref{voievertd},
the neighbourhood of the cell~0(0) is not exactly the same. This, time, as already mentioned,
the milestones of the cell are still its neighbours~2 and~5,
but the milestones of neighbouring elements are now seen in the neighbours~7, 8 and~9
instead of the neighbours~6, 9 and~10{} in the previous case. That explains the
differences we already noticed for a simple locomotive. Note that for the milestones,
the rules are the same but they are applied in the reverse order: this is explained by
the symmetry we already noticed.

\vtop{
\begin{tab}\label{mounttracktab}
\leurre
Rules for going up vertical tracks.
\end{tab}
\vspace{-12pt}
\ligne{\hfill a single locomotive:\hfill\hfill\hskip 80pt a double one:\hfill}
\vskip 2pt
\ligne{\hfill
\vtop{\leftskip 0pt\parindent 0pt\hsize=76pt
\aff { 35} {W} {WBWWBWBBBW} {W}
\aff { 36} {W} {WBWWWBWWWW} {W}
\aff { 37} {W} {WBWWBWWBBW} {W}
\aff { 38} {W} {WWWWWWWBBW} {W}
\aff { 39} {W} {WBWWWWWWWW} {W}
\aff { 40} {W} {WBWWWWBWWW} {W}
}\hfill
\vtop{\leftskip 0pt\parindent 0pt\hsize=76pt
\aff { 41} {W} {WBBWBWBBBW} {B}
\aff { 42} {B} {WBWWBWBBBW} {W}
\aff { 43} {W} {WBWWWBBWWW} {W}
\aff { 44} {W} {BBWWBWBWWW} {W}
\aff { 45} {W} {BBWWWBWWWW} {W}
\aff { 46} {B} {WBWWWWWWWW} {B}
}\hfill
\vtop{\leftskip 0pt\parindent 0pt\hsize=76pt
\aff { 47} {B} {WBWWWWBWWW} {B}
\aff { 48} {W} {BBWWBWBBBW} {W}
\aff { 49} {W} {WBWWBWBWWW} {W}
\aff { 50} {B} {WWWWWBWWWW} {B}
\aff { 51} {W} {WBBWBWWBBW} {W}
}\hfill\hskip 30pt
\vtop{\leftskip 0pt\parindent 0pt\hsize=76pt
\aff { 52} {B} {WBBWBWBBBW} {B}
\aff { 53} {B} {BBWWBWBBBW} {W}
\aff { 54} {W} {BBWWWBBWWW} {W}
}
\hfill}
}

\vskip 10pt
\vtop{
\begin{tab}\label{exvertm}
\leurre
Locomotive going up a vertical.
\end{tab}
\vskip-6pt
\ligne{\hfill\hbox to 180pt{\hfill simple locomotive:\hfill}
\hskip 5pt\hbox to 155pt{\hfill double locomotive:\hfill}\hfill}
\vskip 2pt
\ligne{\hfill
\vtop{\leftskip 0pt\parindent 0pt\hsize=180pt
\ligne{\hfill  \execsept     {}       0       1       2       3       4       5 \hfill}
\ligne{\hfill  \execsept  {3(3)}  {  41}  {  42}  {  48}  {  35}  {  35}  {  35}   \hfill}
\ligne{\hfill  \execsept  {1(3)}  {  35}  {  41}  {  42}  {  48}  {  35}  {  35}   \hfill}
\ligne{\hfill  \execsept  {0(0)}  {  35}  {  35}  {  41}  {  42}  {  48}  {  35}   \hfill}
\ligne{\hfill  \execsept  {1(1)}  {  35}  {  35}  {  35}  {  41}  {  42}  {  48}   \hfill}
\ligne{\hfill  \execsept  {4(1)}  {  35}  {  35}  {  35}  {  35}  {  41}  {  42}   \hfill}
\ligne{\hfill  \execsept  {1(2)}  {  12}  {  12}  {  17}  {  16}  {  15}  {  12}   \hfill}
\ligne{\hfill  \execsept  {1(5)}  {   2}  {   2}  {   2}  {   8}  {  50}  {   2}   \hfill}
}
\hskip 5pt
\vtop{\leftskip 0pt\parindent 0pt\hsize=155pt
\ligne{\hfill   \execsix       0       1       2       3       4       5 \hfill}
\ligne{\hfill   \execsix   {  52}  {  53}  {  48}  {  35}  {  35}  {  35}   \hfill}
\ligne{\hfill   \execsix   {  41}  {  52}  {  53}  {  48}  {  35}  {  35}   \hfill}
\ligne{\hfill   \execsix   {  35}  {  41}  {  52}  {  53}  {  48}  {  35}   \hfill}
\ligne{\hfill   \execsix   {  35}  {  35}  {  41}  {  52}  {  53}  {  48}   \hfill}
\ligne{\hfill   \execsix   {  35}  {  35}  {  35}  {  41}  {  52}  {  53}   \hfill}
\ligne{\hfill   \execsix   {  12}  {  17}  {  34}  {  33}  {  15}  {  12}   \hfill}
\ligne{\hfill   \execsix   {   2}  {   2}  {   8}  {  22}  {  50}  {   2}   \hfill}
}
\hfill}
}

\subsubsection{The horizontal tracks}
\label{les_horizontales}

   We now investigate the construction of the horizontal tracks. For this, we have to define what
we call \textbf{horizontals of the pentagrid}. These objects are not lines nor rays, they
are sets of tiles defined with the help of the Fibonacci tree. A \textbf{level} of the Fibonacci
tree is the set of nodes which are at the same distance from its root: this distance is the
number of nodes on the branch going from the root to the node, the root being not taken into 
account. A \textbf{horizontal of the pentagrid} is a finite sequence of tiles whose image
on in the Fibonacci tree belong to the same level of the tree, with this constraint, that two 
consecutive tiles in the sequence share a vertex. A \textbf{horizontal track} $\eta$ is defined
from a horizontal of the pentagrid $h$ which is called the \textbf{backbone} of~$\eta$.

   Figure~\ref{horizstab} shows us two possible configurations of a horizontal track depending
on whether the tile around which they turn corresponds to a black tile or to a white one.
Moreover, the figure is split into two parts: to left, the track goes in a clockwise way,
to right, it goes on a counter-clockwise one.

   Consider the case of a clockwise way around a black tile. The track consists of the 
following 
cells: 5(4), 4(3), 10(3), 3(3), 7(3), 2(3), 1(2), 2(2), 5(2), 4(1), 10(1), 3(1), 7(1),
18(1), 6(1), 15(1)  and 5(1). 
Their 'isolated' milestones are, in the same order: 13(4), 11(3), 28(3), 26(3), 8(3), 20(3), 18(3), 
5(3), 3(2), 6(2), 15(2), 13(2), 11(1), 28(1), 26(1), 8(1), 19(1), 49(1), 47(1), 16(1), 
41(1), 39(1) and 13(1). The other milestones, those which share a side with another one, are: 
2(1), 1(1), 0(0), 1(3) and 2(4).
In our setting, we consider that 2(1) and 2(4) are white nodes and that 0(0) is a black one.
As a consequence, 
1(3) is a black node. If we assume that these three nodes belong to the same generation,
we need to append the tiles 1(1) and 1(3) in order to avoid a cell of the track with
two consecutive neighbours numbered in 1..5. 
Indeed, in the same level of the tree, a black node is in between two white ones. 
The 'isolated' milestones are not always completely surrounded by blank cells.
For instance, 19(1) shares a side with 41(1).

Similarly, the cells which constitute the clockwise horizontal
track around a white cell are: 5(3), 4(2), 10(2), 3(2), 7(2), 2(2), 1(1), 2(1), 1(5), 2(5), 5(5),
4(4), 10(4), 3(4) and 7(4). The 'isolated' milestones are, in the same order: 
13(3), 11(2), 28(2), 
26(2), 8(2), 20(2), 18(2), 5(2), 3(1), 7(1), 5(1), 3(5), 6(5), 15(5), 13(5), 11(4), 28(4), 26(4),
8(4)
and 19(4). 
The other milestones are, sharing at least a side with another one: 18(3), 6(3), 2(3), 1(2), 0(0),
1(3), 1(4), 2(4), 5(4) and 4(3). Note that 1(3) is almost surrounded by black cells. We assume
that 2(4) is a black node, 0(0) and~2(3) are both white nodes, all of them the sons of a white
node, here 1(3). In our case, 1(2) and 1(4) cannot be the centres of elements of the tracks so that
we define these cells as milestones. This is a general feature: in between two consecutive
nodes $\nu_1$ and~$\nu_2$ on the same level which are milestones, the black node which shares a 
side with both $\nu_1$ and~$\nu_2$ must also be a milestone.

\vtop{
\ligne{\hfill
\includegraphics[scale=0.7]{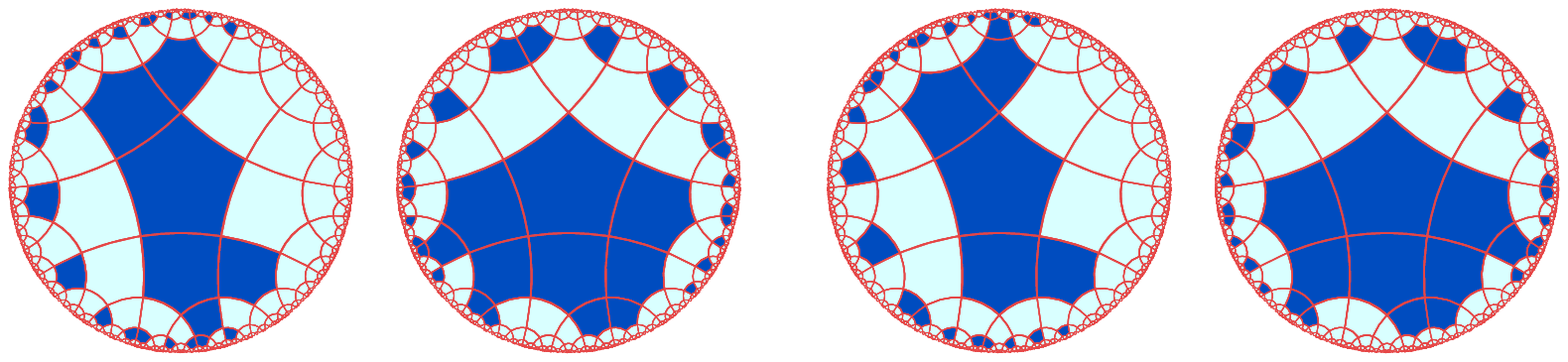} 
\hfill}
\begin{fig}\label{horizstab}
\leurre
To left: a clockwise horizontal track around a node: first, around a black node and then,
around a white one.
To right: a counter-clockwise horizontal track around a node: first, around a black node and then,
around a white one.
\end{fig}
}

For the counter-clockwise tracks, the centres of the elements of the track are the same
as in the clockwise case. The changes occur in the 'isolated' milestones. For the black node
track, 5(4), 11(3), 8(3), 5(3), 3(2), 6(2), 11(1), 8(1), 19(1), 16(1) and 13(1) are 
eplaced 
by 6(4), 12(3), 9(3), 6(3), 4(2), 7(2), 12(1), 9(1), 20(1), 17(1) and 14(1) respectively.
Similarly, for the the white node track, the milestones:
13(3), 11(2), 8(2), 5(2), 3(1), 3(5), 6(5), 11(4), 8(4) and 19(4) 
are replaced by
14(3), 12(2), 9(2), 6(2), 4(1), 4(5), 7(5), 12(4), 9(4) and 20(4) respectively.

   Tables~\ref{horizahstab} and~\ref{horizahdtab} display the rules for the motion
on a horizontal in a clockwise way around a black node.
The corresponding motion is illustrated by Figure~\ref{horizah}.
Also, Tables~\ref{exhorizahs} and~\ref{exhorizahd} show the application of the rules
of tables~\ref{horizahstab} and~\ref{horizahdtab} respectively, on cells of horizontal
tracks and on some of their milestones. The motion is illustrated by Figure~\ref{horizh}.

   We can see that the centres of the elements of the tracks are applied different rules from those
we have seen previously, although horizontal tracks consists of the same elements of the tracks.
Let us look at this more carefully. For instance, in Figure~\ref{horizah}, when it is visited
by a simple locomotive, the cell 3(1) is applied rules~58, 78, 83, 87 and 90
respectively. Rule 58, \laff {W} {WBWWBBWBBW} {W} {,} is a 
conservative rule: it leaves the cell blank and it is applied when the locomotive is not by 
the cell. Note that the milestones are in neighbours~2 and~5, but the cell can see the 
milestones of neighbouring cells in its vertex-neighbours~6, 8 and ~9, a very slight difference 
with rule~35. When the locomotive approaches 3(1), it is seen in its neighbour~3 and 
rule~78, \laff {W} {WBBWBBWBBW} {B} {,} is applied.
Then, rule~83, \laff {B} {WBWWBBWBBW} {W} {,} is applied which makes the locomotive leave 
the cell. Next, rule~87, \laff {W} {BBWWBBWBBW} {W} {,} is applied: it witnesses that the 
locomotive is now in neighbour~1. Then, rule~90, \laff {W} {WBWWBBBBBW} {W} {,} witnesses that the
locomotive is in the vertex-number~6 of~3(1). Note that for the cell~4(1), the neighbourhood
is a bit different. Besides the milestones of the element in neighbours~2 and~5, other
milestones, from neighbouring elements, can be seen in the vertex-neighbours~6 and~9 only.
Now, rule~58 again apply as the locomotive is seen in the vertex-neighbour of the cell. Later,
rule~88, \laff {W} {WBBWBBWWBW} {B} {,}
 makes the locomotive enter the cell as it is seen in its neighbour~3. Rule~61,
\laff {B} {WBWWBBWWBW} {W} {,} 
 makes it
leave the cell, rule~71, \laff {W} {BBWWBBWWBW} {W} {,}
 can see it in its neighbour~1 and rule~79, \laff {W} {WBWWBBBWBW} {W} {,}
 can see it in its 
vertex-neighbour~7.

\vtop{
\begin{tab}\label{horizahstab}
\leurre
Rules for a single locomotive, counter-clockwise moving on a horizontal track,
here, around the sons of a black node.
\end{tab}
\vspace{-12pt}
\ligne{\hfill
\vtop{\leftskip 0pt\parindent 0pt\hsize=76pt
\aff { 55} {B} {BWBWWBWWBW} {B}
\aff { 56} {B} {WWBBWWWWWW} {B}
\aff { 57} {B} {WBWWBWWBWW} {B}
\aff { 58} {W} {WBWWBBWBBW} {W}
\aff { 59} {W} {WBWWBBWWBW} {W}
\aff { 60} {W} {WBWWBBBWWW} {W}
\aff { 61} {B} {WBWWBBWWBW} {W}
\aff { 62} {W} {WWBWBWBBWW} {W}
\aff { 63} {W} {BWBWBWBBWW} {W}
\aff { 64} {B} {BWWWBWWWWW} {B}
\aff { 65} {W} {WBBWBWBBWW} {B}
}\hfill
\vtop{\leftskip 0pt\parindent 0pt\hsize=76pt
\aff { 66} {W} {WBWWBWBWBW} {W}
\aff { 67} {W} {BBWWWWBWWW} {W}
\aff { 68} {W} {WWWWWWBBWW} {W}
\aff { 69} {W} {BWWWBBWWWW} {W}
\aff { 70} {B} {WBWWWBWBWW} {B}
\aff { 71} {W} {BBWWBBWWBW} {W}
\aff { 72} {W} {WBBWBWBWBW} {B}
\aff { 73} {B} {WWBWBWBBWW} {W}
\aff { 74} {B} {BBWWWWWWWW} {B}
\aff { 75} {W} {BWWWWBBWWW} {W}
\aff { 76} {B} {WWBBWWWWWB} {B}
}\hfill
\vtop{\leftskip 0pt\parindent 0pt\hsize=76pt
\aff { 77} {B} {BBWWWWWBWW} {B}
\aff { 78} {W} {WBBWBBWBBW} {B}
\aff { 79} {W} {WBWWBBBWBW} {W}
\aff { 80} {B} {WBWWBWBWBW} {W}
\aff { 81} {B} {WWBBBWWWWW} {B}
\aff { 82} {B} {WBWWWWBBWW} {B}
\aff { 83} {B} {WBWWBBWBBW} {W}
\aff { 84} {W} {BBWWBWBWBW} {W}
\aff { 85} {B} {WWBBWBWWWW} {B}
\aff { 86} {B} {WBWWWWWBWW} {B}
}\hfill
\vtop{\leftskip 0pt\parindent 0pt\hsize=76pt
\aff { 87} {W} {BBWWBBWBBW} {W}
\aff { 88} {W} {WBBWBBWWBW} {B}
\aff { 89} {B} {BWBBWWWWWW} {B}
\aff { 90} {W} {WBWWBBBBBW} {W}
\aff { 91} {B} {WWBBWWBWWW} {B}
\aff { 92} {B} {BWBWWBBWBW} {B}
\aff { 93} {B} {WBBBWWWWWW} {B}
\aff { 94} {B} {BBBWWBWWBW} {B}
\aff { 95} {B} {WWBBWWWBWW} {B}
\aff { 96} {B} {BWBWWBWBBW} {B}
}
\hfill}
}

\vtop{
\begin{tab}\label{horizahdtab}
\leurre
Rules for a double locomotive, counter-clockwise moving on a horizontal track,
here, around the sons of a black node.
\end{tab}
\vspace{-12pt}
\ligne{\hfill
\vtop{\leftskip 0pt\parindent 0pt\hsize=76pt
\aff { 97} {B} {WBWWBBWBWW} {B}
\aff { 98} {B} {BBWWBBWWBW} {W}
\aff { 99} {B} {BWWWBBWWWW} {B}
\aff {100} {B} {WBBWBWBBWW} {B}
\aff {101} {B} {BBWWWWBWWW} {B}
\aff {102} {B} {BBWWWBWBWW} {B}
}
\hfill
\vtop{\leftskip 0pt\parindent 0pt\hsize=76pt
\aff {103} {W} {BBWWBBBWBW} {W}
\aff {104} {B} {WBBWBWBWBW} {B}
\aff {105} {B} {BWBWBWBBWW} {W}
\aff {106} {B} {WWBBBWWWWB} {B}
\aff {107} {B} {BBWWWWBBWW} {B}
\aff {108} {B} {WBBWBBWBBW} {B}
}
\hfill
\vtop{\leftskip 0pt\parindent 0pt\hsize=76pt
\aff {109} {B} {BBWWBWBWBW} {W}
\aff {110} {B} {WWBBBBWWWW} {B}
\aff {111} {B} {BBWWBBWBBW} {W}
\aff {112} {B} {BWBBWBWWWW} {B}
\aff {113} {W} {BBWWBBBBBW} {W}
\aff {114} {B} {WBBWBBWWBW} {B}
}
\hfill
\vtop{\leftskip 0pt\parindent 0pt\hsize=76pt
\aff {115} {B} {BWBBWWBWWW} {B}
\aff {116} {B} {WBBBWWBWWW} {B}
\aff {117} {B} {BBBWWBBWBW} {B}
\aff {118} {B} {WBBBWWWBWW} {B}
\aff {119} {B} {BBBWWBWBBW} {B}
}
\hfill
}
}
\vskip 10pt
Now, the cells~10(1) and 5(2) are applied the same rules as they are in the same situation,
playing the same role. The locomotive enters such a cell in a standard way but it leaves
the cell from an entry and not from the side which is in between two ones which are shared by
a black neighbour. This can be performed in one direction and also in the opposite one
as seen later, with Table~\ref{exhorizhs}. Rule~62,
\laff {W} {WWBWBWBBWW} {W} {,}
 is the conservative rule: it can see the 
milestones in the neighbours~3, 5 and vertex-neighbours~7 and~8. The locomotive enters through
side~2, rule~65, 
\laff {W} {WBBWBWBBWW} {B} {,}
and it leaves the cell by rule~73, 
\laff {B} {WWBWBWBBWW} {W} {.}
Rule~63, 
\laff {W} {BWBWBWBBWW} {W} {,}
can see the leaving locomotive in
the neighbour~1 of the cell. The milestones 12(1) and~13(1) are applied already used rules
in other contexts in the vertical motions. We leave this study to the reader. We also do the
same for the motions when the horizontal track goes through the sons of a white node:
see Tables~\ref{horizwahtab} and~\ref{horizwhtab} for the rules, 
Tables~\ref{exhorizwahs}, \ref{exhorizwahd}, \ref{exhorizwhs} and~\ref{exhorizwhd} for the 
application of the rules to cells of the tracks and a few milestones which are
taken from Figures~\ref{horizwah}, and \ref{horizwh}.
\vtop{
\begin{tab}\label{exhorizahs}
\leurre
Execution of the rules of Table~\textrm{\ref{horizahstab}}
for a simple locomotive counter-clockwise running around the sons of a black node.
\end{tab}
\vskip-6pt
\vskip 2pt
\ligne{\hfill
\vtop{\leftskip 0pt\parindent 0pt\hsize=180pt
\ligne{\hfill  \exechuit     {}       2       3       4       5       6       7       8 \hfill}
\ligne{\hfill  \exechuit  {3(1) } {  78}  {  83}  {  87}  {  90}  {  58}  {  58}  {  58}   \hfill}
\ligne{\hfill  \exechuit  {10(1)} {  62}  {  65}  {  73}  {  63}  {  62}  {  62}  {  62}   \hfill}
\ligne{\hfill  \exechuit  {4(1) } {  59}  {  58}  {  88}  {  61}  {  71}  {  79}  {  59}   \hfill}
\ligne{\hfill  \exechuit  {5(2) } {  62}  {  62}  {  62}  {  65}  {  73}  {  63}  {  62}   \hfill}
\ligne{\hfill  \exechuit  {2(2) } {  66}  {  66}  {  66}  {  35}  {  72}  {  80}  {  84}   \hfill}
\ligne{\hfill  \exechuit  {12(1)} {  19}  {  19}  {  19}  {  64}  {  18}  {  19}  {  19}   \hfill}
\ligne{\hfill  \exechuit  {13(2)} {  46}  {  46}  {  46}  {  47}  {  74}  {  46}  {  46}   \hfill}
}
\hfill}
}
\vskip 10pt

\vtop{
\begin{tab}\label{exhorizahd}
\leurre
Execution of the rules of Table~$\ref{horizahdtab}$
for a double locomotive counter-clockwise running around the sons of a black node.
\end{tab}
\vspace{-12pt}
\ligne{\hfill
\vtop{\leftskip 0pt\parindent 0pt\hsize=180pt
\ligne{\hfill  \exechuit     {}       2       3       4       5       6       7       8 \hfill}
\ligne{\hfill  \exechuit  {3(1) } { 108}  { 111}  { 113}  {  90}  {  58}  {  58}  {  58}   \hfill}
\ligne{\hfill  \exechuit  {10(1)} {  65}  { 100}  { 105}  {  63}  {  62}  {  62}  {  62}   \hfill}
\ligne{\hfill  \exechuit  {4(1) } {  58}  {  78}  { 114}  {  98}  { 103}  {  79}  {  59}   \hfill}
\ligne{\hfill  \exechuit  {5(2) } {  62}  {  62}  {  65}  { 100}  { 105}  {  63}  {  62}   \hfill}
\ligne{\hfill  \exechuit  {2(2) } {  66}  {  66}  {  35}  {  41}  { 104}  { 109}  {  84}   \hfill}
\ligne{\hfill  \exechuit  {12(1)} {  19}  {  19}  {  64}  {  99}  {  18}  {  19}  {  19}   \hfill}
\ligne{\hfill  \exechuit  {13(2)} {  46}  {  46}  {  47}  { 101}  {  74}  {  46}  {  46}   \hfill}
}
\hfill}
}
\vskip 10pt

\vtop{
\begin{tab}\label{exhorizhs}
\leurre
Execution of the rules of Table~$\ref{horizhtab}$
for a simple locomotive clockwise running around a black node: 
\end{tab}
\vskip-6pt
\ligne{\hfill\hbox to 180pt{\hfill simple locomotive:\hfill}
\hskip 5pt\hbox to 155pt{\hfill double locomotive:\hfill}\hfill}
\vskip 2pt
\ligne{\hfill
\vtop{\leftskip 0pt\parindent 0pt\hsize=180pt
\ligne{\hfill  \exechuit     {}       0       1       2       3       4       5       6 \hfill}
\ligne{\hfill  \exechuit  {4(3) } { 123}  { 129}  { 132}  { 121}  { 121}  { 121}  { 121}   \hfill}
\ligne{\hfill  \exechuit  {10(3)} { 124}  { 130}  { 133}  { 122}  { 122}  { 122}  { 122}   \hfill}
\ligne{\hfill  \exechuit  {3(3) } { 121}  { 128}  {  80}  {  84}  {  79}  {  66}  {  66}   \hfill}
\ligne{\hfill  \exechuit  {7(3) } { 122}  { 122}  { 124}  { 130}  { 133}  { 122}  { 122}   \hfill}
\ligne{\hfill  \exechuit  {2(3) } {  59}  {  59}  {  25}  { 135}  {  61}  {  71}  {  59}   \hfill}
\ligne{\hfill  \exechuit  {8(3) } {  46}  {  46}  {  74}  {  47}  {  46}  {  46}  {  46}   \hfill}
\ligne{\hfill  \exechuit  {20(3)} {  19}  {  19}  {  18}  {  64}  {  19}  {  19}  {  19}   \hfill}
}
\hfill}
}

\vtop{
\begin{tab}\label{exhorizhd}
\leurre
Execution of the rules of Table~$\ref{horizhtab}$
for a double locomotive clockwise running around a black node: 
\end{tab}
\vskip-6pt
\ligne{\hfill
\vtop{\leftskip 0pt\parindent 0pt\hsize=180pt
\ligne{\hfill  \exechuit     {}       0       1       2       3       4       5       6 \hfill}
\ligne{\hfill  \exechuit  {4(3) } { 141}  { 145}  { 132}  { 121}  { 121}  { 121}  { 121}   \hfill}
\ligne{\hfill  \exechuit  {10(3)} { 142}  { 146}  { 133}  { 122}  { 122}  { 122}  { 122}   \hfill}
\ligne{\hfill  \exechuit  {3(3) } { 137}  { 144}  { 109}  { 103}  {  79}  {  66}  {  66}   \hfill}
\ligne{\hfill  \exechuit  {7(3) } { 122}  { 124}  { 142}  { 146}  { 133}  { 122}  { 122}   \hfill}
\ligne{\hfill  \exechuit  {2(3) } {  59}  {  25}  {  26}  { 149}  {  98}  {  71}  {  59}   \hfill}
\ligne{\hfill  \exechuit  {8(3) } {  46}  {  74}  { 101}  {  47}  {  46}  {  46}  {  46}   \hfill}
\ligne{\hfill  \exechuit  {20(3)} {  19}  {  18}  {  99}  {  64}  {  19}  {  19}  {  19}   \hfill}
}
\hfill}
}

\vtop{
\vspace{-20pt}
\ligne{\hskip 10pt
\includegraphics[scale=0.55]{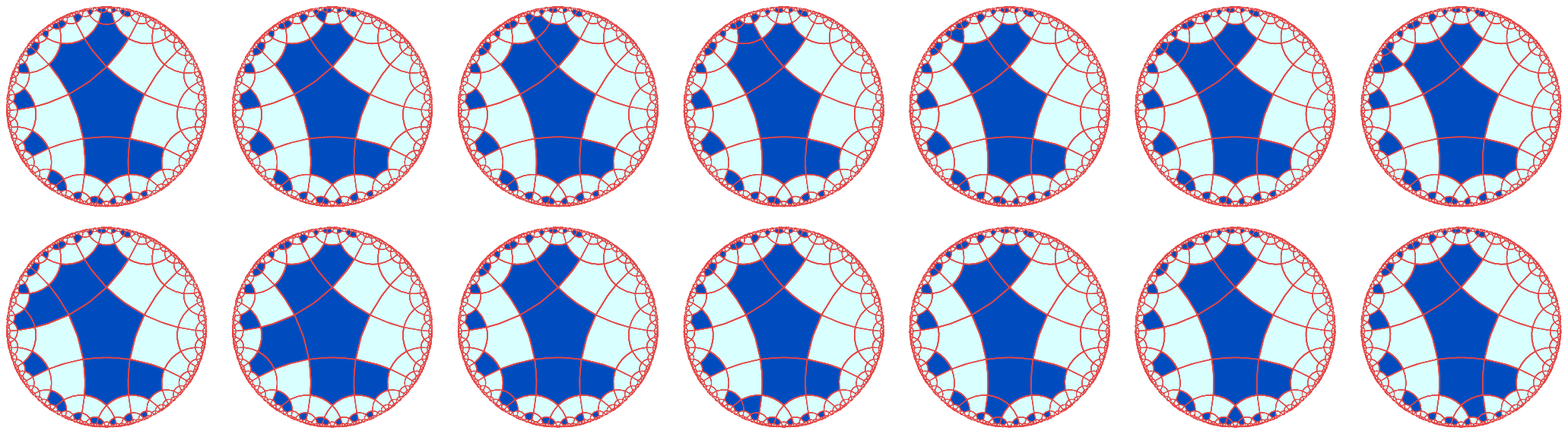} 
\hfill}
\vspace{-30pt}
\ligne{\hskip 10pt
\includegraphics[scale=0.55]{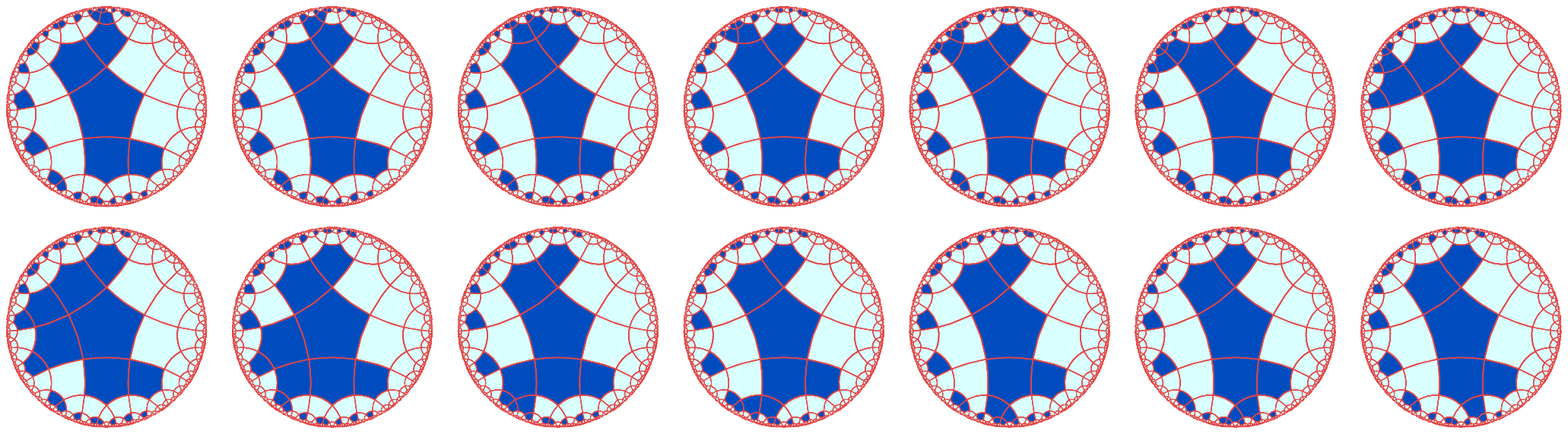} 
\hfill}
\begin{fig}\label{horizah}
\leurre
A locomotive \textbf{counter-clockwise} moving on a horizontal track, here around the sons
of a \textbf{black node}. Above: a simple locomotive; below: a double one.
\end{fig}
}

\vtop{
\vspace{-20pt}
\ligne{\hfill
\includegraphics[scale=0.55]{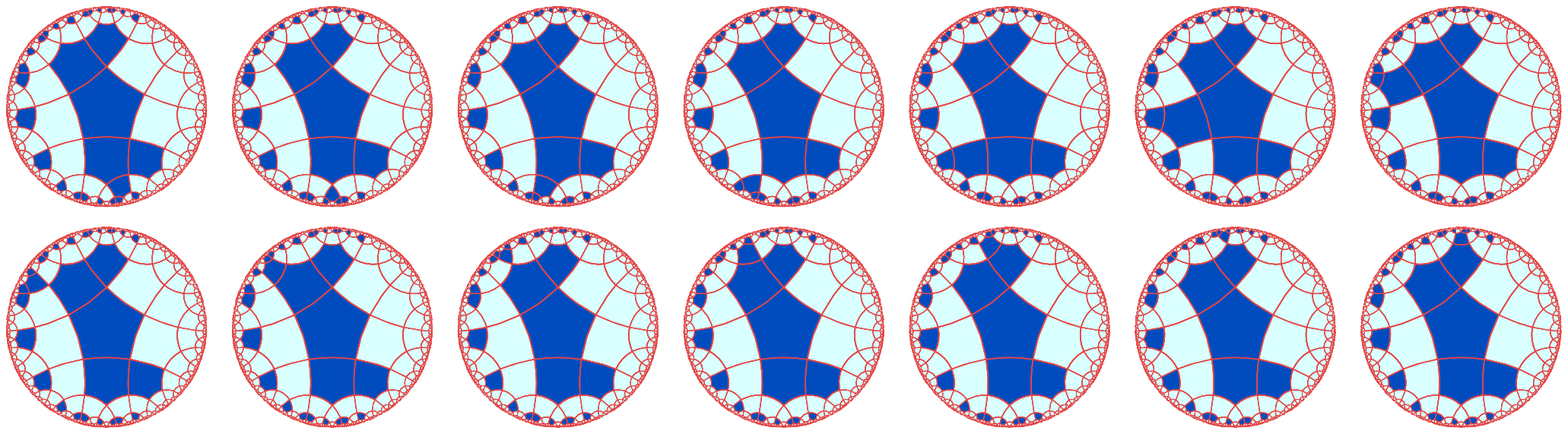} 
\hfill}
\vspace{-30pt}
\ligne{\hfill
\includegraphics[scale=0.55]{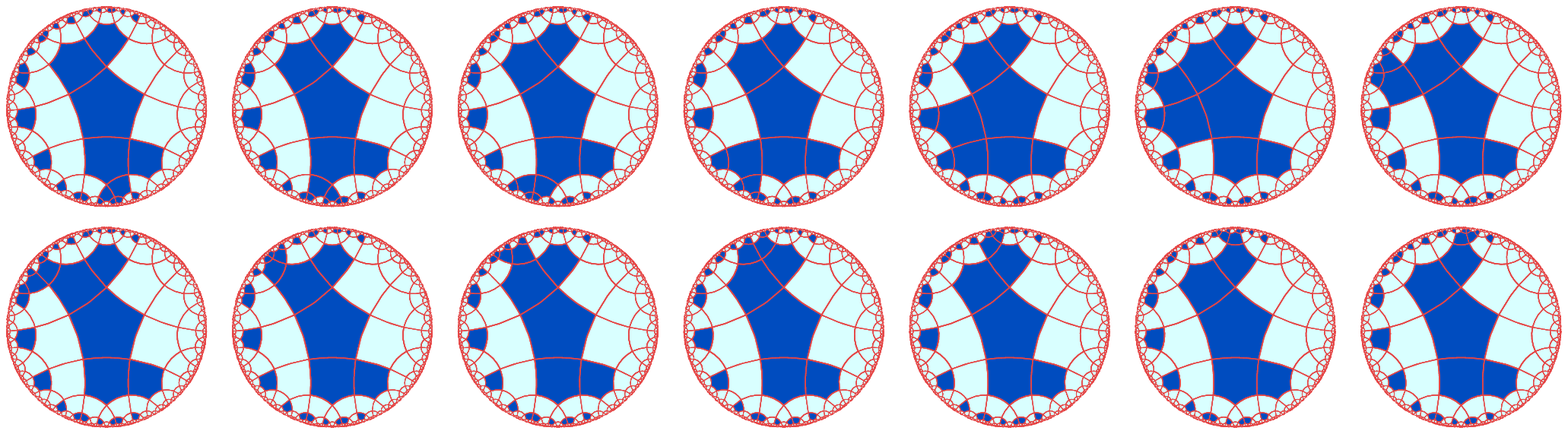} 
\hfill}
\begin{fig}\label{horizh}
\leurre
A locomotive \textbf{clockwise} moving on a horizontal track, here around the sons
of a \textbf{black node}. Above: a simple locomotive; below: a double one.
\end{fig}
}

In the abstract and in the introduction, we mentioned that we relaxed the condition on the
rotation invariance of the cellular automaton constructed for the proof of Theorem~\ref{letheo}.
In this part of our study, we can see six rules which are concerned with this point.
As an example: rule 21, \laff {W} {WBWBBWWWBB} {W} {,} 
has a rotated neighbourhood
which is also that of rule~65: \laff {W} {WBBWBWBBWW} {B} {.}  The state of
the cell is the same, but the new state is different. Rule~21 appeared in the motion
of a simple locomotive going down a vertical track, while rule~65 appears in the motion
of a simple locomotive counter-clockwise moving along a horizontal track. 
A higher number of rules are the rotated form of another one. For instance, 
rules~81, 93 and 136 are in this case: rule~93, \laff {B} {WBBBWWWWWW} {B} {,}
and rule~136, \laff {B} {BWWBBWWWWW} {B} {,} 
are both rotated forms of rule~81, \laff {B} {WWBBBWWWWW} {B} {,} as can
easily be seen. At the end of the paper we shall look at all rules which have the same 
neighbourhood, up to rotation, the same state but a different new state.

\vtop{
\begin{tab}\label{horizhtab}
\leurre
Rules for the clockwise motion of a locomotive around a black node.
\end{tab}
\vspace{-12pt}
\ligne{\hfill simple locomotive\hfill}
\vskip 4pt
\ligne{\hfill
\vtop{\leftskip 0pt\parindent 0pt\hsize=76pt
\aff {120} {W} {WBWWBWWWBW} {W}
\aff {121} {W} {WBWWBWBWBB} {W}
\aff {122} {W} {WBWBWBWWWB} {W}
\aff {123} {B} {WBWWBWBWBB} {W}
\aff {124} {W} {WBWBBBWWWB} {B}
}\hfill
\vtop{\leftskip 0pt\parindent 0pt\hsize=76pt
\aff {125} {W} {WWWWWBWWWB} {W}
\aff {126} {W} {WWWWBWWWWW} {W}
\aff {127} {B} {WWWBBWWBWW} {B}
\aff {128} {W} {WBWBBWBWBW} {B}
\aff {129} {W} {BBWWBWBWBB} {W}
}
\hfill
\vtop{\leftskip 0pt\parindent 0pt\hsize=76pt
\aff {130} {B} {WBWBWBWWWB} {W}
\aff {131} {B} {WBWBBWWWWW} {B}
\aff {132} {W} {WBWWBBBWBB} {W}
\aff {133} {W} {BBWBWBWWWB} {W}
\aff {134} {B} {WWWBBWBWWW} {B}
}
\hfill
\vtop{\leftskip 0pt\parindent 0pt\hsize=76pt
\aff {135} {W} {WBWBBBWWBW} {B}
\aff {136} {B} {BWWBBWWWWW} {B}
\aff {137} {W} {WBWBBWBWBB} {B}
\aff {138} {B} {WWWBBBWWWW} {B}
\aff {139} {B} {WWWBBWWWWW} {B}
}
\hfill}
\vskip 6pt
\ligne{\hfill double locomotive\hfill}
\vskip 4pt
\ligne{\hfill
\vtop{\leftskip 0pt\parindent 0pt\hsize=76pt
\aff {140} {B} {WWBBBWWBWW} {B}
\aff {141} {B} {BBWWBWBWBB} {W}
\aff {142} {B} {WBWBBBWWWB} {B}
}\hfill
\vtop{\leftskip 0pt\parindent 0pt\hsize=76pt
\aff {143} {B} {WBWBBWWBWW} {B}
\aff {144} {B} {WBWBBWBWBW} {B}
\aff {145} {W} {BBWWBBBWBB} {W}
}\hfill
\vtop{\leftskip 0pt\parindent 0pt\hsize=76pt
\aff {146} {B} {BBWBWBWWWB} {W}
\aff {147} {B} {WBWBBWBWWW} {B}
\aff {148} {B} {BWWBBWBWWW} {B}
}\hfill
\vtop{\leftskip 0pt\parindent 0pt\hsize=76pt
\aff {149} {B} {WBWBBBWWBW} {B}
\aff {150} {B} {BWWBBBWWWW} {B}
\aff {151} {B} {WBWBBWBWBB} {B}
}
\hfill}
}

\vtop{
\vspace{-20pt}
\ligne{\hskip 10pt
\includegraphics[scale=0.55]{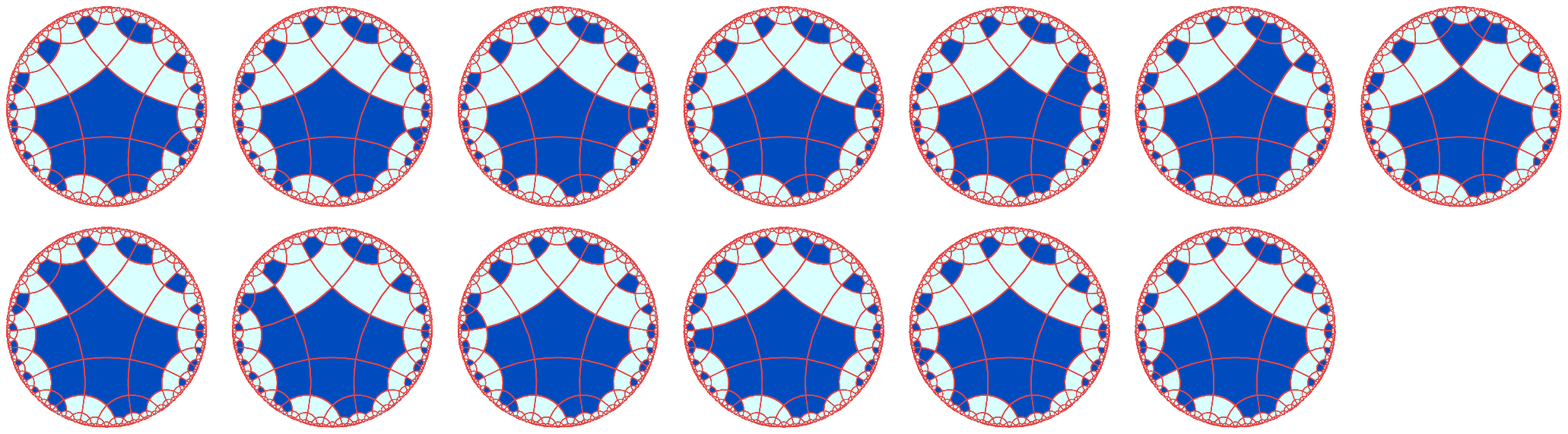} 
\hfill}
\vspace{-20pt}
\ligne{\hskip 10pt
\includegraphics[scale=0.55]{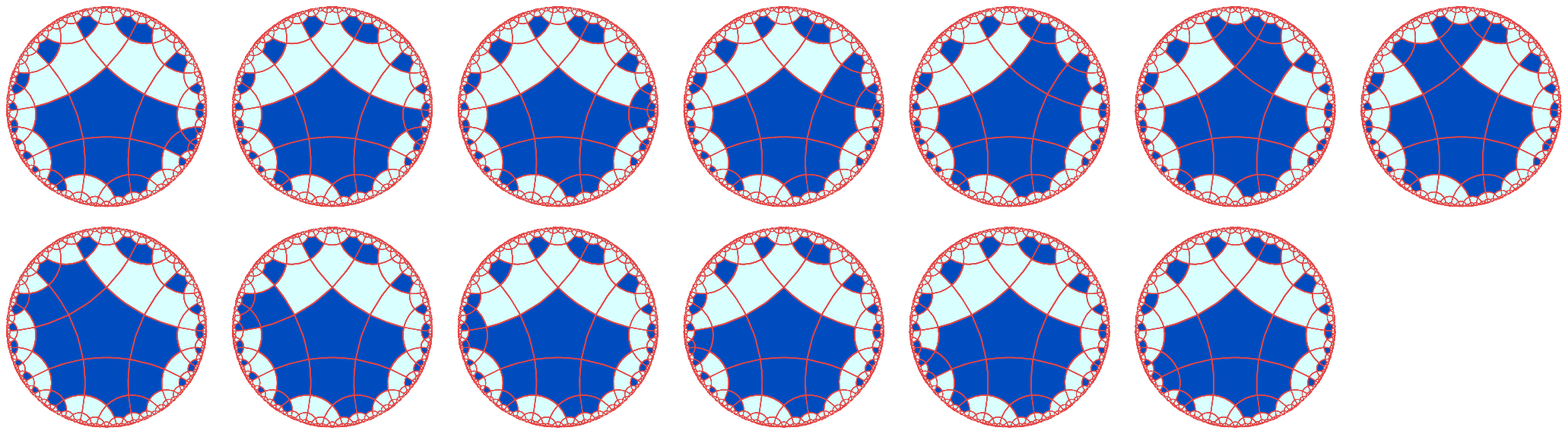} 
\hfill}
\begin{fig}\label{horizwah}
\leurre
A locomotive \textbf{counter-clockwise} moving on a horizontal track, here around the sons
of a \textbf{white node}. Above: a simple locomotive; below: a double one.
\end{fig}
}
\vskip 5pt

\vtop{
\vspace{-20pt}
\ligne{\hskip 10pt
\includegraphics[scale=0.55]{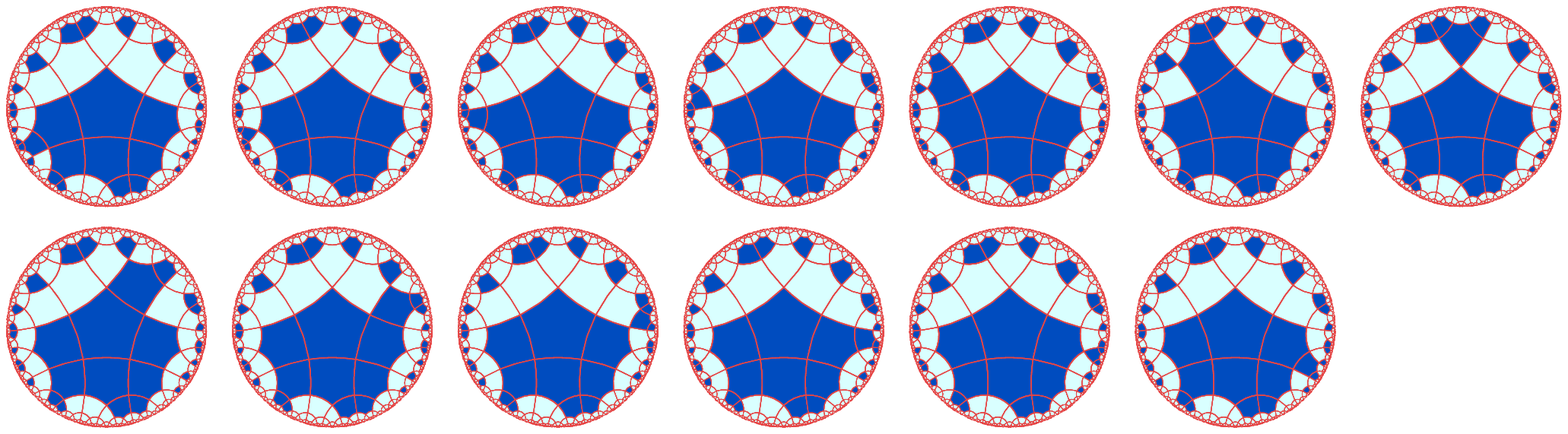} 
\hfill}
\vspace{-20pt}
\ligne{\hskip 10pt
\includegraphics[scale=0.55]{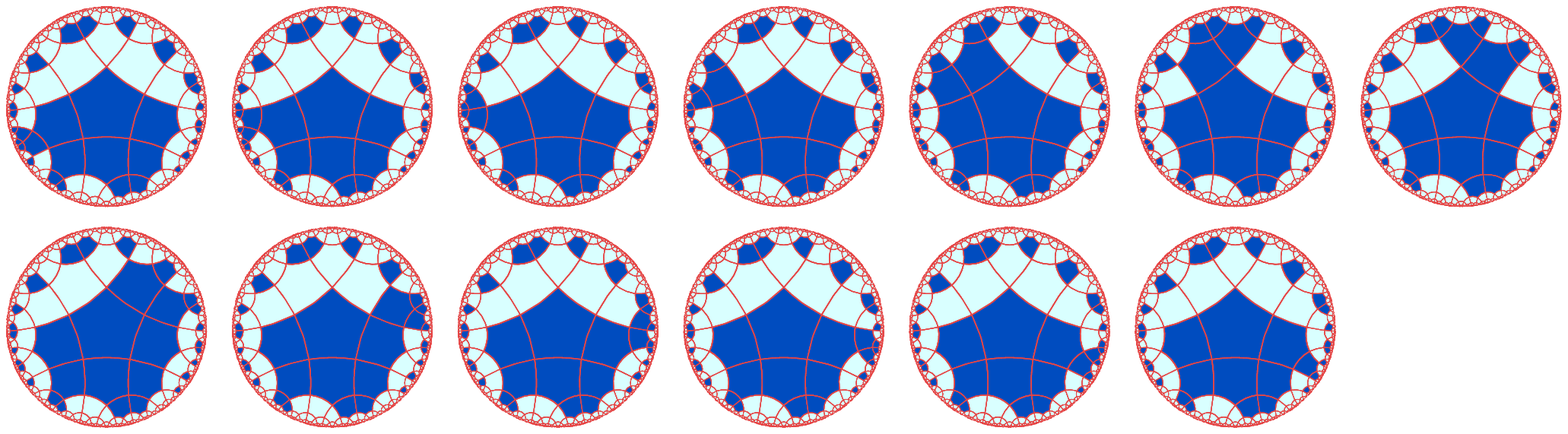} 
\hfill}
\begin{fig}\label{horizwh}
\leurre
A locomotive \textbf{clockwise} moving on a horizontal track, here around the sons
of a \textbf{white node}. Above: a simple locomotive; below: a double one.
\end{fig}
}
\vtop{
\begin{tab}\label{horizwahtab}
\leurre
Rules for the counter-clockwise motion of a locomotive around a white node.
\end{tab}
\vspace{-12pt}
\ligne{\hfill simple locomotive \hfill}
\vskip 4pt
\ligne{\hfill
\vtop{\leftskip 0pt\parindent 0pt\hsize=76pt
\aff {152} {B} {WBBBWWWBBW} {B}
\aff {153} {B} {WWWBBWWWWB} {B}
\aff {154} {B} {BBWBBBBWWB} {B}
\aff {155} {B} {WBWBBWWBWB} {B}
\aff {156} {W} {WBBWWWBBWW} {W}
\aff {157} {W} {BWWWWWBBWW} {W}
}
\hfill
\vtop{\leftskip 0pt\parindent 0pt\hsize=76pt
\aff {158} {B} {BBBWWBBBWW} {B}
\aff {159} {B} {WBBWWWWBWB} {B}
\aff {160} {B} {BBBWWWBBWW} {B}
\aff {161} {B} {WBBWBWWBWW} {B}
\aff {162} {B} {WBBWWBWBWW} {B}
\aff {163} {B} {WBBBWWWBBB} {B}
}
\hfill
\vtop{\leftskip 0pt\parindent 0pt\hsize=76pt
\aff {164} {B} {BBBWWWWBWW} {B}
\aff {165} {B} {WBBBBWWBBW} {B}
\aff {166} {B} {WBBWWWBBWW} {B}
\aff {167} {B} {WBBBWBWBBW} {B}
\aff {168} {B} {WBBWWWWBWW} {B}
\aff {169} {B} {BBBBWWWBBW} {B}
}
\hfill
\vtop{\leftskip 0pt\parindent 0pt\hsize=76pt
\aff {170} {B} {WWWBBBWWWB} {B}
\aff {171} {B} {WBBBWWBBBW} {B}
\aff {172} {B} {BWWBBWWWWB} {B}
\aff {173} {B} {WWWBBWBWWB} {B}
\aff {174} {B} {WBWBBWWWWB} {B}
\aff {175} {B} {WWWBBWWBWB} {B}
}
\hfill}
\vskip 4pt
\ligne{\hfill double locomotive \hfill}
\vskip 4pt
\ligne{\hfill
\vtop{\leftskip 0pt\parindent 0pt\hsize=76pt
\aff {176} {B} {WBBBWWWBWB} {B}
\aff {177} {B} {WBBWBWWBWB} {B}
\aff {178} {B} {WBBWBBWBWW} {B}
\aff {179} {B} {BBBWWBWBWW} {B}
}
\hfill
\vtop{\leftskip 0pt\parindent 0pt\hsize=76pt
\aff {180} {B} {WBBBBWWBBB} {B}
\aff {181} {B} {WBBBBBWBBW} {B}
\aff {182} {B} {BBBBWBWBBW} {B}
}
\hfill
\vtop{\leftskip 0pt\parindent 0pt\hsize=76pt
\aff {183} {B} {BBBBWWBBBW} {B}
\aff {184} {B} {BWWBBBWWWB} {B}
\aff {185} {B} {BWWBBWBWWB} {B}
}
\hfill
\vtop{\leftskip 0pt\parindent 0pt\hsize=76pt
\aff {186} {B} {WBWBBWBWWB} {B}
\aff {187} {B} {WWBBBWWBWB} {B}
\aff {188} {B} {WBWBBBWBWB} {B}
}
\hfill}
}

\vtop{
\begin{tab}\label{exhorizwahs}
\leurre
Execution of the rules~$\ref{horizwahtab}$ for a simple locomotive counter-clockwise running 
around the sons of a white node.
\end{tab}
\vskip -6pt
\ligne{\hfill
\vtop{\leftskip 0pt\parindent 0pt\hsize=210pt
\ligne{\hfill  \execsept     {}       6       7       8       9      10      11 \hfill}
\ligne{\hfill  \execsept  {2(2) } {  58}  {  78}  {  83}  {  87}  {  90}  {  58}   \hfill}
\ligne{\hfill  \execsept  {7(2) } {  62}  {  62}  {  65}  {  73}  {  63}  {  62}   \hfill}
\ligne{\hfill  \execsept  {3(2) } {  59}  {  59}  {  58}  {  88}  {  61}  {  71}   \hfill}
\ligne{\hfill  \execsept  {10(2)} {  62}  {  62}  {  62}  {  62}  {  65}  {  73}   \hfill}
\ligne{\hfill  \execsept  {4(2) } {  66}  {  66}  {  66}  {  66}  {  35}  {  72}   \hfill}
\ligne{\hfill  \execsept  {1(2) } { 153}  { 170}  { 172}  { 173}  { 174}  { 175}   \hfill}
\ligne{\hfill  \execsept  {0(0) } { 167}  { 169}  { 171}  { 152}  { 152}  { 152}   \hfill}
}
\hfill}
}

\vtop{
\begin{tab}\label{exhorizwahd}
\leurre
Execution of the rules~$\ref{horizwahtab}$ for a double locomotive counter-clockwise running 
around the sons of a white node.
\end{tab}
\vskip -6pt
\ligne{\hfill
\vtop{\leftskip 0pt\parindent 0pt\hsize=210pt
\ligne{\hfill  \execsept     {}       6       7       8       9      10      11 \hfill}
\ligne{\hfill  \execsept  {2(2) } {  78}  { 108}  { 111}  { 113}  {  90}  {  58}   \hfill}
\ligne{\hfill  \execsept  {7(2) } {  62}  {  65}  { 100}  { 105}  {  63}  {  62}   \hfill}
\ligne{\hfill  \execsept  {3(2) } {  59}  {  58}  {  78}  { 114}  {  98}  { 103}   \hfill}
\ligne{\hfill  \execsept  {10(2)} {  62}  {  62}  {  62}  {  65}  { 100}  { 105}   \hfill}
\ligne{\hfill  \execsept  {4(2) } {  66}  {  66}  {  66}  {  35}  {  41}  { 104}   \hfill}
\ligne{\hfill  \execsept  {1(2) } { 170}  { 184}  { 185}  { 186}  { 155}  { 187}   \hfill}
\ligne{\hfill  \execsept  {0(0) } { 182}  { 183}  { 171}  { 152}  { 152}  { 152}   \hfill}
}
\hfill}
}

\vtop{
\begin{tab}\label{horizwhtab}
\leurre
Rules for the clockwise motion of a locomotive around a white node. 
\end{tab}
\vspace{-12pt}
\vskip 4pt
\ligne{\hfill
\vtop{\leftskip 0pt\parindent 0pt\hsize=76pt
\aff {189} {W} {BBWWBBWWWB} {W}
}
\hfill
\vtop{\leftskip 0pt\parindent 0pt\hsize=76pt
\aff {190} {W} {WBWWBBWWWB} {W}
}
\hfill}
}

\vtop{
\begin{tab}\label{exhorizwhs}
\leurre
Execution of the rules~$\ref{horizwahtab}$ and~$\ref{horizwhtab}$ for a simple locomotive 
clockwise running around the sons of a white node.
\end{tab}
\vskip -6pt
\ligne{\hfill
\vtop{\leftskip 0pt\parindent 0pt\hsize=210pt
\ligne{\hfill  \exechuit     {}       0       1       2       3       4       5       6 \hfill}
\ligne{\hfill  \exechuit  {4(2) } { 123}  { 129}  { 132}  { 121}  { 121}  { 121}  { 121}   \hfill}
\ligne{\hfill  \exechuit  {10(2)} { 124}  { 130}  { 133}  { 122}  { 122}  { 122}  { 122}   \hfill}
\ligne{\hfill  \exechuit  {3(2) } { 121}  { 128}  {  80}  {  84}  {  79}  {  66}  {  66}   \hfill}
\ligne{\hfill  \exechuit  {7(2) } { 122}  { 122}  { 124}  { 130}  { 133}  { 122}  { 122}   \hfill}
\ligne{\hfill  \exechuit  {2(2) } {  59}  {  59}  {  25}  { 135}  {  61}  {  71}  {  59}   \hfill}
\ligne{\hfill  \exechuit  {1(2) } { 106}  { 175}  { 174}  { 173}  { 172}  { 170}  { 153}   \hfill}
\ligne{\hfill  \exechuit  {0(0) } { 152}  { 152}  { 152}  { 152}  { 171}  { 169}  { 167}   \hfill}
}
\hfill}
}

\vtop{
\begin{tab}\label{exhorizwhd}
\leurre
Execution of the rules~$\ref{horizwahtab}$ and~$\ref{horizwhtab}$ for a double locomotive 
clockwise running around the sons of a white node.
\end{tab}
\vskip -6pt
\ligne{\hfill
\vtop{\leftskip 0pt\parindent 0pt\hsize=210pt
\ligne{\hfill  \exechuit     {}       0       1       2       3       4       5       6 \hfill}
\ligne{\hfill  \exechuit  {4(2) } { 141}  { 145}  { 132}  { 121}  { 121}  { 121}  { 121}   \hfill}
\ligne{\hfill  \exechuit  {10(2)} { 142}  { 146}  { 133}  { 122}  { 122}  { 122}  { 122}   \hfill}
\ligne{\hfill  \exechuit  {3(2) } { 137}  { 144}  { 109}  { 103}  {  79}  {  66}  {  66}   \hfill}
\ligne{\hfill  \exechuit  {7(2) } { 122}  { 124}  { 142}  { 146}  { 133}  { 122}  { 122}   \hfill}
\ligne{\hfill  \exechuit  {2(2) } {  59}  {  25}  {  26}  { 149}  {  98}  {  71}  {  59}   \hfill}
\ligne{\hfill  \exechuit  {1(2) } { 187}  { 155}  { 186}  { 185}  { 184}  { 170}  { 153}   \hfill}
\ligne{\hfill  \exechuit  {0(0) } { 152}  { 152}  { 152}  { 171}  { 183}  { 182}  { 181}   \hfill}
}
\hfill}
}

\subsection{The fixed switch}
\label{fixedswitch}

   Figures~\ref{fixeg} and~\ref{fixed} illustrate how the locomotive passively crosses
a fixed switch. Both figures illustrate the case when the locomotive comes from the
left-hand side and when it comes from the right-hand side. Note that the locomotive may be
simple as well as double: indeed, the fixed switch plays a role in the round-about, this is
why it may be crossed by a double locomotive.

   Table~\ref{fixtab} displays the rules used for the passive fixed switch. More precisely,
the table gives the rules which were not yet used. In fact several rules from the vertical motions
are also used in this case.

\vtop{
\vspace{-10pt}
\begin{tab}\label{fixtab}
\leurre
Rules for the passive fixed switch:
\end{tab}
\vspace{-4pt}
\ligne{\hfill
\vtop{\leftskip 0pt\parindent 0pt\hsize=90pt 
\ligne{\hfill from left, simple one:\hfill}
\vskip 7pt
\aff {191} {W} {WBWWBBWBBB} {W}
\aff {192} {W} {WWWWWWWWBB} {W}
\aff {193} {W} {WBWBBBWBBB} {B}
\aff {194} {B} {WBWWBBWBBB} {W}
\aff {195} {W} {BBWWBBWBBB} {W}
\aff {196} {B} {WWWWBWBWWW} {B}
}
\hfill
\vtop{\leftskip 0pt\parindent 0pt\hsize=90pt 
\ligne{\hfill from left, double one:\hfill}
\vskip 2pt
\aff {197} {B} {WBWBBBWBBB} {B}
\aff {198} {B} {BBWWBBWBBB} {W}
\aff {199} {B} {BWWWBWBWWW} {B}
\vskip 5pt
\ligne{\hfill from right, simple one:\hfill}
\vskip 2pt
\aff {200} {W} {WBBWBBWBBB} {B}
\vskip 5pt
\ligne{\hfill from right, double one:\hfill}
\vskip 2pt
\aff {201} {B} {WBBWBBWBBB} {B}
}
\hfill}
}
\vskip 10pt

As illustrated by the figures, the switch is passive and it works both for a single and 
a double locomotive whatever the side from which they arrive to the switch.
Note that the exceptional cell is the centre of the switch, the cell~0(0) in the pictures
of Figures~\ref{fixeg} and~\ref{fixed}. Rule~191, \laff {W} {WBWWBBWBBB} {W} {,} is the conservative
rule which is applied when no locomotive crosses the switch. Its milestones are 
neighbours~2, and~5 and also the vertex-neighbours~6, 8, 9 and~10 which belong to milestones
of neighbouring elements of the tracks. Note that the cell~2(1) is a milestone shared by
two elements, each one belonging to one of the tracks which arrive to the switch.

\vtop{
\vspace{-15pt}
\ligne{\hskip 10pt
\includegraphics[scale=0.55]{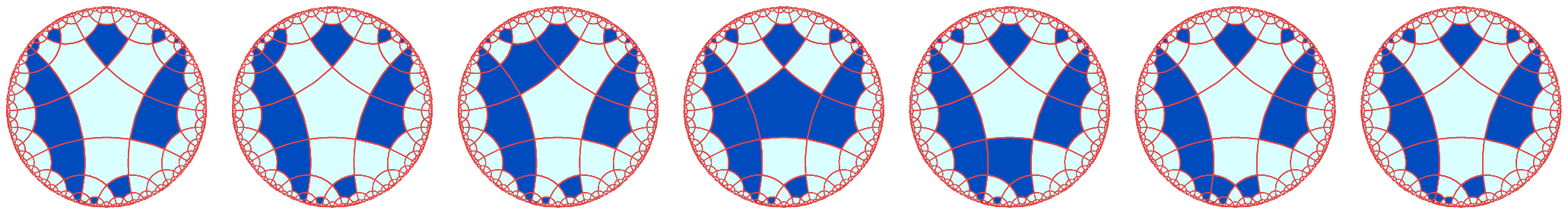} 
\hfill}
\vspace{-20pt}
\ligne{\hskip 10pt
\includegraphics[scale=0.55]{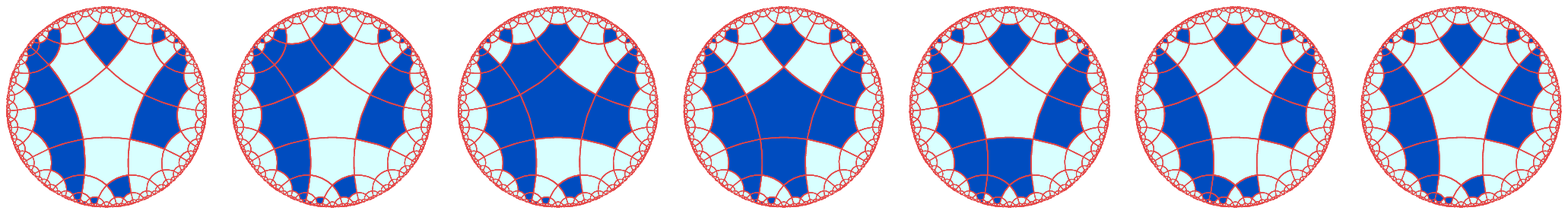} 
\hfill}
\vskip -5pt
\begin{fig}\label{fixeg}
\leurre
Locomotives passively crossing a fixed switch from the left-hand side. Above : a single
locomotive; below: a double one.
\end{fig}
}

\vtop{
\vspace{-15pt}
\ligne{\hskip 10pt
\includegraphics[scale=0.55]{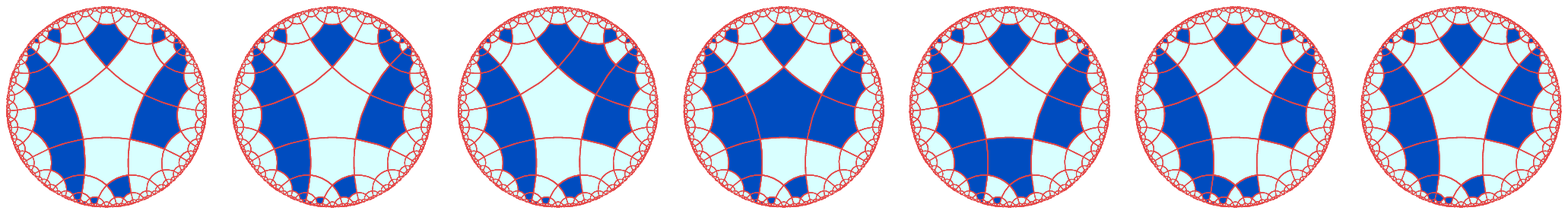} 
\hfill}
\vspace{-20pt}
\ligne{\hskip 10pt
\includegraphics[scale=0.55]{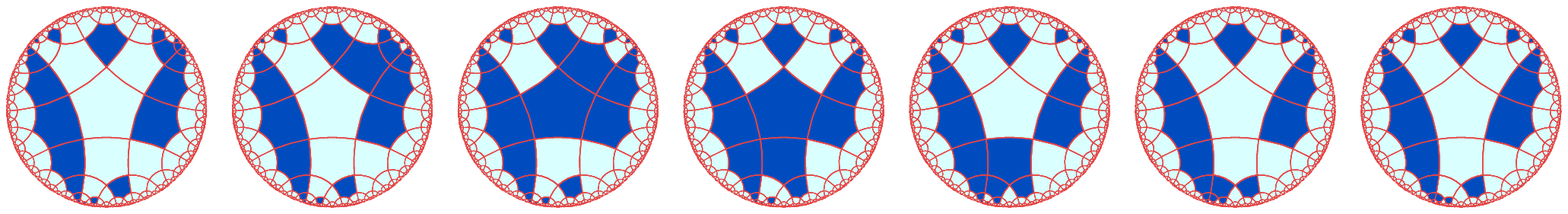} 
\hfill}
\vskip -5pt
\begin{fig}\label{fixed}
\leurre
Locomotives passively crossing a fixed switch from the right-hand side. Above : a single
locomotive; below: a double one.
\end{fig}
}
\vskip 10pt
Table~\ref{exfixs} gives traces of executions for a simple locomotive crossing the switch,
either from the left, left-hand side part of the table, or from the right, right-hand side part
of the table. Table~\ref{exfixd} does the same for the same cells when a double locomotive
crosses the switch.

\vtop{
\begin{tab}\label{exfixs}
\leurre
Execution of the rules on the cells~$1(1)$, $1(5)$, $0(0)$, $1(3)$, $1(2)$ and~$1(4)$
when a simple locomotive passively crosses the fixed switch.
\end{tab}
\vskip-6pt
\ligne{\hfill\hbox to 180pt{\hfill from the left:\hfill}
\hskip 5pt\hbox to 155pt{\hfill from the right:\hfill}\hfill}
\vskip 2pt
\ligne{\hfill
\vtop{\leftskip 0pt\parindent 0pt\hsize=180pt  
\ligne{\hfill  \execsept     {}       0       1       2       3       4       5 \hfill}
\ligne{\hfill  \execsept  {1(1)}  {  25}  {  26}  {  27}  {  28}  {  25}  {  25}   \hfill}
\ligne{\hfill  \execsept  {1(5)}  {  35}  {  35}  {  90}  {  48}  {  35}  {  35}   \hfill}
\ligne{\hfill  \execsept  {0(0)}  { 191}  { 191}  { 193}  { 194}  { 195}  { 191}   \hfill}
\ligne{\hfill  \execsept  {1(3)}  {  25}  {  25}  {  25}  {  26}  {  27}  {  28}   \hfill}
\ligne{\hfill  \execsept  {1(2)}  {  12}  {  12}  {  15}  {  16}  {  17}  {  12}   \hfill}
\ligne{\hfill  \execsept  {1(4)}  {  19}  {  19}  {  19}  {  64}  { 196}  {  19}   \hfill}
}
\hskip 5pt
\vtop{\leftskip 0pt\parindent 0pt\hsize=155pt  
\ligne{\hfill   \execsix       0       1       2       3       4       5 \hfill}
\ligne{\hfill   \execsix   {  25}  {  25}  { 132}  {  28}  {  25}  {  25}   \hfill}
\ligne{\hfill   \execsix   {  35}  {  41}  {  42}  {  48}  {  35}  {  35}   \hfill}
\ligne{\hfill   \execsix   { 191}  { 191}  { 200}  { 194}  { 195}  { 191}   \hfill}
\ligne{\hfill   \execsix   {  25}  {  25}  {  25}  {  26}  {  27}  {  28}   \hfill}
\ligne{\hfill   \execsix   {  12}  {  12}  {  12}  {  16}  {  17}  {  12}   \hfill}
\ligne{\hfill   \execsix   {  19}  {  19}  {  18}  {  64}  { 196}  {  19}   \hfill}
}
\hfill}
}
\vskip 10pt
\vtop{
\begin{tab}\label{exfixd}
\leurre
Execution of the rules on the cells~$1(1)$, $1(5)$, $0(0)$, $1(3)$, $1(2)$ and~$1(4)$
when a double locomotive passively crosses the fixed switch.
\end{tab}
\vskip-6pt
\ligne{\hfill\hbox to 180pt{\hfill from the left:\hfill}
\hskip 5pt\hbox to 155pt{\hfill from the right:\hfill}\hfill}
\vskip 2pt
\ligne{\hfill
\vtop{\leftskip 0pt\parindent 0pt\hsize=180pt  
\ligne{\hfill  \execsept     {}       0       1       2       3       4       5 \hfill}
\ligne{\hfill  \execsept  {1(1)}  {  26}  {  29}  {  31}  {  28}  {  25}  {  25}   \hfill}
\ligne{\hfill  \execsept  {1(5)}  {  35}  {  90}  { 113}  {  48}  {  35}  {  35}   \hfill}
\ligne{\hfill  \execsept  {0(0)}  { 191}  { 193}  { 197}  { 198}  { 195}  { 191}   \hfill}
\ligne{\hfill  \execsept  {1(3)}  {  25}  {  25}  {  26}  {  29}  {  31}  {  28}   \hfill}
\ligne{\hfill  \execsept  {1(2)}  {  12}  {  15}  {  33}  {  34}  {  17}  {  12}   \hfill}
\ligne{\hfill  \execsept  {1(4)}  {  19}  {  19}  {  64}  { 199}  { 196}  {  19}   \hfill}
}
\hskip 5pt
\vtop{\leftskip 0pt\parindent 0pt\hsize=155pt
\ligne{\hfill   \execsix       0       1       2       3       4       5 \hfill}
\ligne{\hfill   \execsix   {  25}  { 132}  { 145}  {  28}  {  25}  {  25}   \hfill}
\ligne{\hfill   \execsix   {  41}  {  52}  {  53}  {  48}  {  35}  {  35}   \hfill}
\ligne{\hfill   \execsix   { 191}  { 200}  { 201}  { 198}  { 195}  { 191}   \hfill}
\ligne{\hfill   \execsix   {  25}  {  25}  {  26}  {  29}  {  31}  {  28}   \hfill}
\ligne{\hfill   \execsix   {  12}  {  12}  {  16}  {  34}  {  17}  {  12}   \hfill}
\ligne{\hfill   \execsix   {  19}  {  18}  {  99}  { 199}  { 196}  {  19}   \hfill}
}
\hfill}
}
\vskip 10pt
These traces deal with the cells 1(1), 1(5), 0(0) and~1(3) for the tracks and
with the cells 1(2) 
and 1(4) for the particular milestones which are not in the situation of
milestones in a vertical track or around a set of cells. 

Note that the side from which the locomotive arrives to the cell~0(0) is indicated by
rule~193, \laff {W} {WBWBBBWBBB} {B} {,} for the left-hand one and rule~200,
\laff {W} {WBBWBBWBBB} {B} {,} for the right hand one. For rule~193, 
the locomotive can be seen in the neighbour~4 of 0(0) when it comes from the left-hand side.
For rule~200, 
it can be seen in neighbour~3 when it comes from the right-hand side. It can be checked that
many rules from Table~\ref{verttracktab} are used in these executions, this is especially the 
case for the conservative rules applied to all cells of the tables. The motion rules are also
those of the vertical tracks for cells~1(1) and~1(3).

\subsection{The round-about}

   In Section~\ref{scenar}, we have seen the idle configurations of the other pieces of the
round-about: the doubler, see Sub-subsection~\ref{subsdoubler} and the selector,
see Sub-subsection~\ref{subssel}.

\subsubsection{The doubler}
\label{subsdoubler}

   Figure~\ref{doubleur} illustrates the motion of the locomotive through the doubler.
The locomotive arrives in the form of a simple one. The creation of the second cell of
the locomotive happens in a simple way: when the locomotive enters the cell~1(1), at the same
moment, a black cell is created in the cell~0(0). That event constitutes the creation of the
double locomotive, which is clear in Figure~\ref{doubleur} as well as in Table~\ref{exdoubl}.
The rules are given in Table~\ref{doubltab}. However, rules of previous tables are also involved,
in particular rules from the motions on the vertical tracks.

The cells of the track involved in the doubler are, in the order they are traversed: 
2(1), 1(1), 0(0), 1(3), 3(3) and 8(3). The milestones are in the cells: 4(5), 15(1), 1(5),
7(1), 3(1), 2(2), 1(2), 1(4), 2(3), 4(3), 7(3) and 9(3).
Here, the conservative rule for the central cell is rule~202: \laff {W} {WBBWBBWWWB} {W} {.}
We can see the milestones in neighbours~2, 3 and~5 as well as in vertex-neighbours~6 and~10.
The creation of the double locomotive is performed by rules~207, \laff {W} {WBBWBBBBWW} {B} {,}
and rule~206, \laff {W} {WBBWBBWWBB} {B} {,} simultaneously applied 
to the cell~1(1) and 
to~0(0), respectively. Note that in rule~206, the occurrence of a black cell in the
vertex-neighbour~9 triggers the creation of the front cell of the locomotive.

\vtop{
\vspace{-10pt}
\ligne{\hskip 10pt
\includegraphics[scale=0.55]{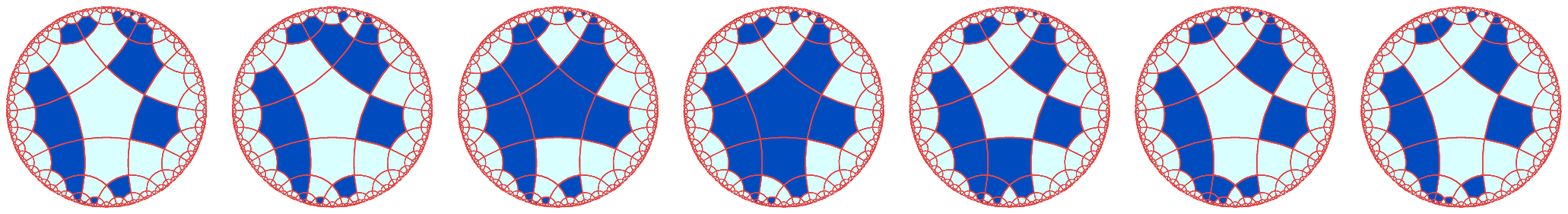} 
\hfill}
\vspace{-5pt}
\begin{fig}\label{doubleur}
\leurre
The structure which doubles a locomotive: a single locomotive enters the structure;
a double one leaves it.
\end{fig}
}

\vtop{
\begin{tab}\label{doubltab}
\leurre
Rules for the doubler.
\end{tab}
\vspace{-6pt}
\ligne{\hfill
\vtop{\leftskip 0pt\parindent 0pt\hsize=76pt 
\aff {202} {W} {WBBWBBWWWB} {W}
\aff {203} {W} {WWBWBBBBWW} {W}
\aff {204} {B} {WWWBWWBWWB} {B}
\aff {205} {W} {BBWWWWWWWW} {W}
}
\hfill
\vtop{\leftskip 0pt\parindent 0pt\hsize=76pt 
\aff {206} {W} {WBBWBBWWBB} {B}
\aff {207} {W} {WBBWBBBBWW} {B}
\aff {208} {B} {WBBBBBWWWB} {B}
\aff {209} {B} {BWBWBBBBWW} {W}
}
\hfill
\vtop{\leftskip 0pt\parindent 0pt\hsize=76pt 
\aff {210} {B} {BWWBWBBWWW} {B}
\aff {211} {B} {BBBWBBWWWB} {W}
\aff {212} {W} {BWBWBBBBWW} {W}
\aff {213} {B} {BWWWWBBWWW} {B}
}
\hfill
\vtop{\leftskip 0pt\parindent 0pt\hsize=76pt 
\aff {214} {B} {BWWBWWBWWW} {B}
\aff {215} {W} {BBBWBBWWWB} {W}
\aff {216} {B} {WWWWWBBWWW} {B}
\aff {217} {B} {WWWBWWBWWW} {B}
}
\hfill
}
}

\vskip 10pt
\vtop{
\begin{tab}\label{exdoubl}
\leurre
Execution of a run through the doubler.
\end{tab}
\vskip-6pt
\ligne{\hfill
\vtop{\leftskip 0pt\parindent 0pt\hsize=180pt
\ligne{\hfill  \execsept     {}       0       1       2       3       4       5 \hfill}
\ligne{\hfill  \execsept  {2(1)}  {  78}  {  83}  { 113}  {  90}  {  58}  {  58}   \hfill}
\ligne{\hfill  \execsept  {1(1)}  { 203}  { 207}  { 209}  { 212}  { 203}  { 203}   \hfill}
\ligne{\hfill  \execsept  {0(0)}  { 202}  { 206}  { 208}  { 211}  { 215}  { 202}   \hfill}
\ligne{\hfill  \execsept  {3(1)}  {  25}  {  25}  {  26}  {  29}  {  31}  {  28}   \hfill}
\ligne{\hfill  \execsept  {2(2)}  {  19}  {  19}  {  99}  {  18}  {  19}  {  19}   \hfill}
\ligne{\hfill  \execsept  {1(2)}  {  12}  {  12}  {  33}  {  34}  {  17}  {  12}   \hfill}
\ligne{\hfill  \execsept  {2(3)}  {  12}  {  12}  {  15}  {  33}  {  34}  {  17}   \hfill}
\ligne{\hfill  \execsept  {1(5)}  { 204}  { 134}  { 210}  { 214}  { 217}  { 217}   \hfill}
\ligne{\hfill  \execsept  {1(4)}  {  50}  {  50}  {  22}  { 213}  { 216}  {  50}   \hfill}
}
\hfill}
}

\subsubsection{The fork}

   Although the fork is not connected with the round-about, we place it here as it appears
at that point in the table of the rules. However, it has some link with the doubler as, like
for that structure, two locomotives are issued from the fork. Now, in contrast with what happens
in the doubler where both locomotives travel together, always in contact with one another,
in the fork each created simple locomotive goes its own way: they travel in different directions.
Figure~\ref{fourche} illustrates this structure. Table~\ref{forktab} gives the rules 
for the fork and Table~\ref{exfork} displays the rules which are applied in the visit
of the fork by a locomotive.

   The cells of the tracks involved in the structure are: 8(3), 3(3), 1(3), 0(0)
and then 1(1), 4(1), 12(1) and 33(1) in one direction and 1(4), 2(5), 7(5), 20(5) and 54(5)
in the other direction. The milestones are: 20(3), 23(3), 7(3), 10(3), 2(3), 2(4)
and then 1(2), 1(5), 2(2), 3(1), 5(2), 11(1), 13(2) and 32(1)  in the first direction and
4(4), 6(5), 3(5), 19(5), 8(5) and 53(5) for the second one. 

\vtop{
\begin{tab}\label{forktab}
\leurre
Rules for the fork.
\end{tab}
\vspace{-12pt}
\ligne{\hfill
\vtop{\leftskip 0pt\parindent 0pt\hsize=76pt 
\aff {218} {W} {WWBWBWBWWW} {W}
\aff {219} {B} {WWBWWWWWWW} {B}
\aff {220} {W} {BWWBWBBBWW} {W}
}
\hfill
\vtop{\leftskip 0pt\parindent 0pt\hsize=76pt 
\aff {221} {W} {WBWWWWWBBW} {W}
\aff {222} {W} {WBBWBWBWWW} {B}
\aff {223} {B} {BWBWWWWWWW} {B}
}
\hfill
\vtop{\leftskip 0pt\parindent 0pt\hsize=76pt
\aff {224} {B} {WBBWWWWWWW} {B}
\aff {225} {W} {BBWBBWBBBW} {W}
\aff {226} {B} {WWBWBWBWWW} {W}
}
\hfill
\vtop{\leftskip 0pt\parindent 0pt\hsize=76pt 
\aff {227} {B} {WWBWWBBWWW} {B}
\aff {228} {W} {WBWWBWBBBB} {W}
\aff {229} {W} {BWBWBWBWWW} {W}
}
\hfill
}
}

\vskip 10pt
\vtop{
\begin{tab}\label{exfork}
\leurre
Execution of the rules 
of the fork. 
\end{tab}
\vspace{-9pt}
\ligne{\hfill
\vtop{\leftskip 0pt\parindent 0pt
\ligne{\hfill  \execsept     {}       0       1       2       3       4       5 \hfill}
\ligne{\hfill  \execsept  {1(3)}  {  35}  {  41}  {  42}  {  48}  {  90}  {  35}   \hfill}
\ligne{\hfill  \execsept  {0(0)}  {  35}  {  35}  {  41}  {  42}  { 225}  { 228}   \hfill}
\ligne{\hfill  \execsept  {1(1)}  {  35}  {  35}  {  35}  {  41}  {  42}  {  48}   \hfill}
\ligne{\hfill  \execsept  {1(4)}  { 218}  { 218}  {  62}  { 222}  { 226}  { 229}   \hfill}
\ligne{\hfill  \execsept  {2(5)}  {  66}  {  66}  {  66}  {  35}  {  72}  {  80}   \hfill}
\ligne{\hfill  \execsept  {1(5)}  { 219}  { 219}  { 219}  { 223}  { 227}  { 224}   \hfill}
}
\hfill}
}

\vtop{
\vspace{-10pt}
\ligne{\hskip 10pt
\includegraphics[scale=0.55]{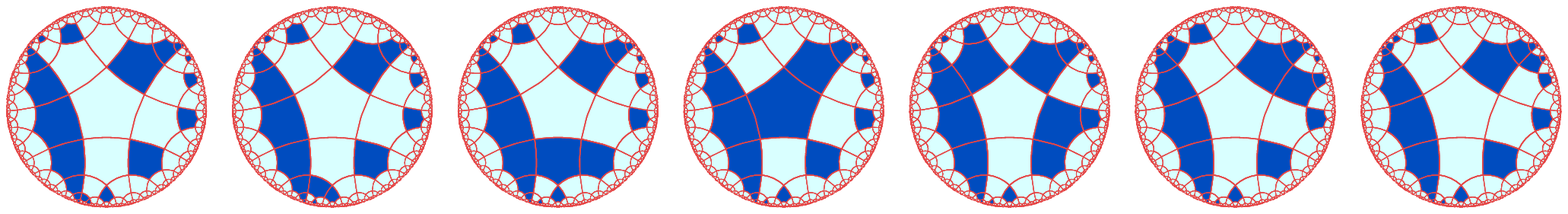} 
\hfill}
\vspace{-5pt}
\begin{fig}\label{fourche}
\leurre
The fork: a single locomotive enters the fork. Two single one leave it in different directions.
\end{fig}
}
\vskip 10pt
The neighbourhood of the cell~0(0) is exactly that of a cell of going up vertical tracks.
It is given by the conservative rule~35:  \laff {W} {WBWWBWBBBW} {W} {.} The milestones
are in neighbours~2 and~5 and the cell can see the milestones of the neighbouring elements
in the vertex-neighbours~ 7, 8 and~9. When the simple locomotive is in the cell~0(0),
rules~41, \laff {W} {WBBWBWBBBW} {B} {,} and rule~222, \laff {W} {WBBWBWBWWW} {B} {,} 
apply to cell~1(1) and~1(4) respectively, simultaneously creating a locomotive at the 
corresponding place. In the cell~1(1), the locomotive can be seen in its neighbour~3 while
in the cell~1(4), it can be seen in its neighbour~2. Note that the cell~1(4) behaves
like  the cells 10(1) and~5(2) we have looked at in Sub-section~\ref{les_horizontales}.
The cell~1(4) allows the locomotive to perform a turn at right-angle.

\subsubsection{The selector}
\label{subssel}

    Figure~\ref{select} illustrates the working of the selector of the 
round-about. The case of a single locomotive is illustrated by the first two rows 
of the figure while the case of a double locomotive is illustrated by the
last row. Tables~\ref{selectstab} and~\ref{selectdtab} give the rules
for the locomotive and for the cells of the selector, Table~\ref{selectstab} for a simple
locomotive, Table~\ref{selectdtab} for a double one.

\vtop{
\vspace{-30pt}
\ligne{\hskip 10pt
\includegraphics[scale=0.55]{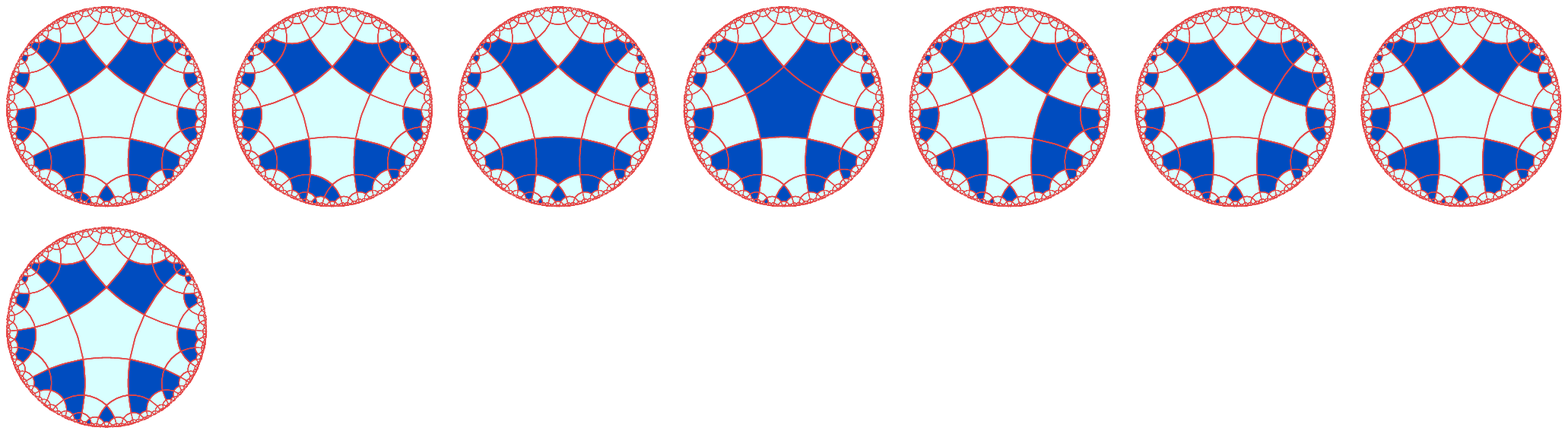} 
\hfill}
\vspace{-20pt}
\ligne{\hskip 10pt
\includegraphics[scale=0.55]{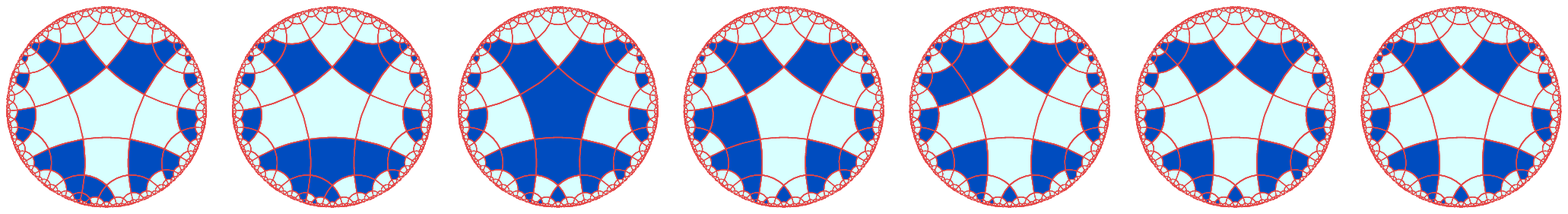} 
\hfill}
\begin{fig}\label{select}
\leurre
The selector of the round about. Above: a single locomotive enters the selector. It leaves
it to go on its way on the right track. Below: a double locomotive enters the selector. 
It leaves it on the way going to the next selector.
\end{fig}
}

\vtop{
\begin{tab}\label{selectstab}
\leurre
Rules for the selector: the case of a simple locomotive.  
\end{tab}
\vspace{-12pt}
\ligne{\hfill
\vtop{\leftskip 0pt\parindent 0pt\hsize=76pt 
\aff {230} {W} {WWWBBWBBWW} {W}
\aff {231} {B} {WWWWBWWBWW} {B}
\aff {232} {W} {WBWBWBWBBW} {W}
\aff {233} {W} {WWWWBWWWBB} {W}
\aff {234} {W} {WBWWBWWBBB} {W}
\aff {235} {B} {WBWBWWWWWW} {B}
\aff {236} {W} {WWBWBWBWBB} {W}
\aff {237} {B} {WWBWBWWWWB} {B}
\aff {238} {B} {WBWWWWWWWB} {B}
}
\hfill
\vtop{\leftskip 0pt\parindent 0pt\hsize=76pt 
\aff {239} {W} {WBWWWWBBBW} {W}
\aff {240} {W} {WBBWBWWBBB} {B}
\aff {241} {B} {WBWBWWWWWB} {B}
\aff {242} {W} {WBWBBWBBWW} {B}
\aff {243} {W} {WBWBWBWBBB} {W}
\aff {244} {B} {WBWWBWWBBB} {W}
\aff {245} {W} {WWBWBWBBBB} {W}
\aff {246} {B} {WBBWBWWWWB} {B}
\aff {247} {B} {WWWBBWBBWW} {W}
}
\hfill
\vtop{\leftskip 0pt\parindent 0pt\hsize=76pt 
\aff {248} {W} {WBWBBBWBBW} {W}
\aff {249} {W} {BBWWBWWBBB} {W}
\aff {250} {B} {WBWBWBWWWW} {B}
\aff {251} {W} {WBBWBWBWBB} {B}
\aff {252} {B} {WWBWBWBWWB} {B}
\aff {253} {B} {WBWWBWWWWB} {B}
\aff {254} {W} {WWBBBWBBWW} {W}
\aff {255} {B} {WWBWBWBWBB} {W}
\aff {256} {B} {BWBWBWWWWB} {B}
}
\hfill
\vtop{\leftskip 0pt\parindent 0pt\hsize=76pt 
\aff {257} {W} {BBWWBBBWWW} {W}
\aff {258} {B} {WBBWWWBWWW} {B}
\aff {259} {B} {WBWWWBWWWB} {B}
\aff {260} {W} {WWWBBWBBBW} {W}
\aff {261} {W} {BWBWBWBWBB} {W}
\aff {262} {B} {WBWWWBWWWW} {B}
\aff {263} {B} {BBWWWWWWWB} {B}
\aff {264} {B} {WBWWWWBWWB} {B}
}
\hfill
}
}

\vtop{
\begin{tab}\label{selectdtab}
\leurre
Rules for the selector: the case of double locomotive.  
\end{tab}
\vspace{-12pt}
\ligne{\hfill
\vtop{\leftskip 0pt\parindent 0pt\hsize=76pt 
\aff {265} {B} {WBBWBWWBBB} {B}
\aff {266} {B} {WBWBBWBBWW} {W}
\aff {267} {B} {BBWWBWWBBB} {W}
\aff {268} {B} {WBWBBBWWWW} {B}
}
\hfill
\vtop{\leftskip 0pt\parindent 0pt\hsize=76pt 
\aff {269} {W} {WBBWBWBBBB} {W}
\aff {270} {B} {WBBWBWBWWB} {B}
\aff {271} {W} {BWWBBWBBWW} {W}
\aff {272} {B} {WWWWBWBBWW} {B}
}
\hfill
\vtop{\leftskip 0pt\parindent 0pt\hsize=76pt 
\aff {273} {B} {WBWBWBWBBW} {W}
\aff {274} {B} {BBWBWWWWWW} {B}
\aff {275} {W} {WWWBBBBBWW} {W}
}
\hfill
\vtop{\leftskip 0pt\parindent 0pt\hsize=76pt 
\aff {276} {B} {BWWWBWWBWW} {B}
\aff {277} {W} {BBWBWBWBBW} {W}
\aff {278} {B} {WWWWBBWBWW} {B}
}
\hfill
}
}
\vskip 10pt
   Note the very symmetric structure of the exit configurations when no locomotive is around.
As shown by the figures, the cells of tracks involved in this structure are 8(3), 3(3), 1(3) 
and 0(0) for the arrival and then 1(2), 2(2), 5(2) and 13(2) for the path leading to the
next selector and 1(4), 2(5), 7(5) and 20(5) for the track leading to the continuation of
the required path. The milestones are 20(3), 23(3), 7(3), 10(3), 2(3) and 2(4) for the entry path,
1(1), 6(2), 4(1), 14(2), 12(1) and 35(2) for the path leading to the next selector and
1(5), 6(5), 3(5), 19(5), 8(5) and 53(5) for the track to the required path. We remain with
the milestones of cells~1(2) and~1(4) as these cells play the selection role: besides the already
mentioned 1(1) and 2(3), we have 5(3), 3(2) and 10(2) for~1(2); besides the already mentioned
1(5) and 2(4), we have 5(4), 7(4), 18(4), 4(4) and 10(4) for~1(4).

\vtop{
\begin{tab}\label{exsels}
\leurre
Execution of the rules 
of the selector when it is crossed by a simple locomotive. 
\end{tab}
\vspace{-9pt}
\ligne{\hfill
\vtop{\leftskip 0pt\parindent 0pt
\ligne{\hfill  \execsept     {}       0       1       2       3       4       5 \hfill}
\ligne{\hfill  \execsept  {1(3)}  { 234}  { 240}  { 244}  { 249}  { 191}  { 234}   \hfill}
\ligne{\hfill  \execsept  {0(0)}  { 230}  { 230}  { 242}  { 247}  { 254}  { 260}   \hfill}
\ligne{\hfill  \execsept  {1(2)}  { 232}  { 232}  { 243}  { 248}  { 232}  { 232}   \hfill}
\ligne{\hfill  \execsept  {2(2)}  {  59}  {  59}  {  59}  {  25}  {  59}  {  59}   \hfill}
\ligne{\hfill  \execsept  {1(4)}  { 236}  { 236}  { 245}  { 251}  { 255}  { 261}   \hfill}
\ligne{\hfill  \execsept  {2(5)}  {  66}  {  66}  {  66}  {  35}  {  72}  {  80}   \hfill}
\ligne{\hfill  \execsept  {2(3)}  { 235}  { 241}  { 131}  { 250}  { 235}  { 235}   \hfill}
\ligne{\hfill  \execsept  {2(4)}  { 237}  { 237}  { 246}  { 252}  { 256}  { 237}   \hfill}
}
\hfill}
}
\vskip 10pt
   The conservative rule for~0(0) is rule~230, \laff {W} {WWWBBWBBWW} {W} {,} and we can see
that the milestones of the cell are neighbours~4, 5, 7 and~8. Rule~232, 
\laff {W} {WBWBWBWBBW} {W} {,} and rule~236, \laff {W} {WWBWBWBWBB} {W} {,} are the conservative
rules for~1(2) and~1(4) respectively. Note that they are rotated forms of one another.
When the locomotive is a simple one, 1(2) remains white while the locomotive passes by.
This can checked on rule~243, \laff {W} {WBWBWBWBBB} {W} {,} and on rule~248,
\laff {W} {WBWBBBWBBW} {W} {,} the locomotive being seen in the neighbour~10 of~1(2) for the first
rule and in its neighbour~5 for the second one. For the same passage, 1(4) very
differently reacts: rule~245, \laff {W} {WWBWBWBBBB} {W} {,} can see the locomotive through
its vertex-neighbour~8 while rule~251, \laff {W} {WBBWBWBWBB} {B} {,} now seeing the locomotive
in the neighbour~2 of the cell makes it become black. This can be checked on Table~\ref{exsels}.

\vtop{
\begin{tab}\label{exseld}
\leurre
Execution of the rules 
of the selector when it is crossed by a double locomotive. 
\end{tab}
\vspace{-9pt}
\ligne{\hfill
\vtop{\leftskip 0pt\parindent 0pt
\ligne{\hfill  \execsept     {}       0       1       2       3       4       5 \hfill}
\ligne{\hfill  \execsept  {1(3)}  { 240}  { 265}  { 267}  { 228}  { 234}  { 234}   \hfill}
\ligne{\hfill  \execsept  {0(0)}  { 230}  { 242}  { 266}  { 271}  { 275}  { 230}   \hfill}
\ligne{\hfill  \execsept  {1(2)}  { 232}  { 243}  { 193}  { 273}  { 277}  { 232}   \hfill}
\ligne{\hfill  \execsept  {2(2)}  {  59}  {  59}  {  25}  { 135}  {  61}  {  71}   \hfill}
\ligne{\hfill  \execsept  {1(4)}  { 236}  { 245}  { 269}  { 236}  { 236}  { 236}   \hfill}
\ligne{\hfill  \execsept  {2(5)}  {  66}  {  66}  {  35}  {  66}  {  66}  {  66}   \hfill}
\ligne{\hfill  \execsept  {2(3)}  { 241}  { 174}  { 268}  { 274}  { 235}  { 235}   \hfill}
\ligne{\hfill  \execsept  {2(4)}  { 237}  { 246}  { 270}  { 237}  { 237}  { 237}   \hfill}
}
\hfill}
}
\vskip 10pt
When the locomotive is double, 1(4) remains white. It is witnessed by rule~245 again,
the locomotive is in the neighbour~8 of the cell, and then
by rule~269, \laff {W} {WBBWBWBBBB} {W} {,} applied when the locomotive is in the neighbours~2
and~8 of the cell. But this time, 1(2) behaves differently: rule~243 again applies as the cell
can only see the front of the locomotive through a vertex neighbour: it cannot know whether it
is simple or double. At the next time, rule~193, \laff {W} {WBWBBBWBBB} {B} {,}
applies as the locomotive can be seen in the neighbours~5 and~10 respectively, making
the cell becoming black.

\subsection{Flip-flop and active memory switch}
\label{flflp}

    In Section~\ref{scenar} we have seen that we can implement both a flip-flop and
the active memory switch by using the same basic devices provided that they are
suitably arranged. Figure~\ref{control} shows how the controller is working, depending
on its cell~1(1) whose state we call the \textbf{colour} of the controller. 
Figure~\ref{controls} shows how the colour of controller is changed by the arrival 
of the appropriate signal: we remind the reader that the signal has the form of a 
simple locomotive. Table~\ref{controltab} gives the rules corresponding to 
the illustration given by Figures~\ref{control} 
and~\ref{controls}. Also, Table~\ref{excontrol} gives the traces of executions
of the rules applied to significant cells when the locomotive arrives to the controller while
Table~\ref{excontrols} gives similar traces when the signal arrives to the controller in
order to change its colour.

\vtop{
\vspace{-15pt}
\ligne{\hskip 10pt
\includegraphics[scale=0.55]{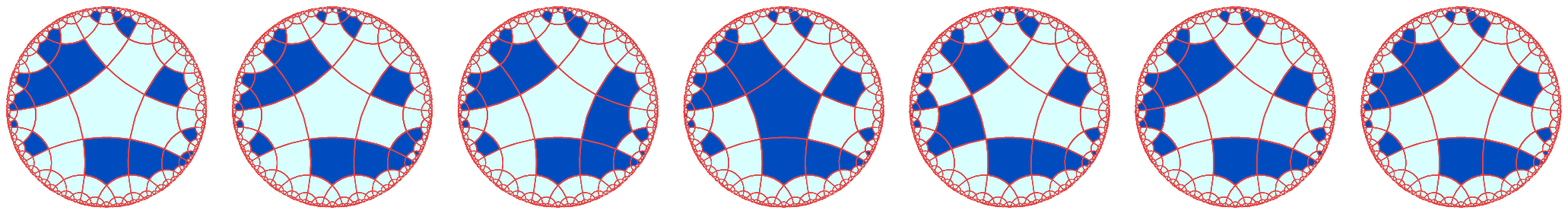} 
\hfill}
\vspace{-20pt}
\ligne{\hskip 75pt
\includegraphics[scale=0.55]{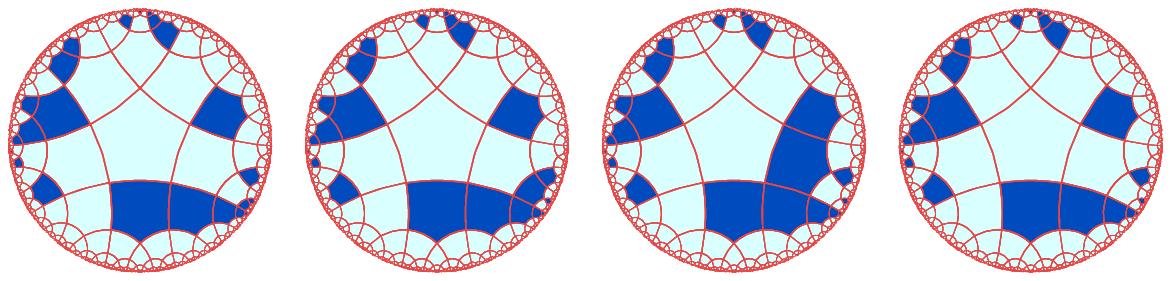} 
\hfill}
\vspace{-5pt}
\begin{fig}\label{control}
\leurre
The controller of the flip-flop and of the active memory switch. Above, the
controller is black: the locomotive passes without problem. Below, the controller is
white: it stops the locomotive which vanishes.
\end{fig}
}

   Note that the tracks which crosses the controller passes through the cells~8(4), 3(4), 1(4),
0(0), 1(2), 3(2) and~8(2) in this order. The signal for changing the colour of the cell~1(1) 
passes through the cells 16(1), 6(1) and 2(1). The milestones of the path which
crosses the controller are 54(4), 57(4), 20(4), 23(4), 7(4), 10(4), 2(4), 2(5), 1(3), 1(1),
4(2), 2(2), 9(2), 7(2), 20(2) and 22(2). The milestones for the path followed by the signal
are the cells 41(1), 44(1), 15(1), 18(1) and~2(5). There are three additional milestones
which are specific to the controller: the cells 4(1), 10(1) and 6(2). 

   Figure~\ref{control} shows us that when the cell~1(1) is black, the locomotive passes through
the tracks, because 1(1) constitutes a milestone of the track. When 1(1) is white, the
expected milestone being missing, the locomotive is stopped, see
rule~315, \laff {W} {WBBWWBWBBW} {W} {.} The rule shows that the cell~0(0) remains white despite
the fact that the locomotive is seen in the neighbour~3 of the cell. Indeed, the conservative
rule for~0(0) when the cell 1(1) is white is rule~279, \laff {W} {WBWWWBWBBW} {W} {.} That rule 
can be compared with the one which is used when the cell~1(1) is black, namely rule~58,
\laff {W} {WBWWBBWBBW} {W} {.}

Table~\ref{excontrol} gives the traces of execution for the rules which apply to
the cells 1(4), 0(0) and 1(2) for the tracks, the cells 2(2) and 1(1) for milestones and 
the cell 2(1) for a witness of transformations induced by the passage of the locomotive.
   
\vtop{
\begin{tab}\label{controltab}
\leurre
Rules for the controllers of the flip-flop and the active memory switch.
\end{tab}
\vspace{-9pt}
\ligne{\hfill managing the control \hfill}
\vskip 3pt
\ligne{\hfill \hbox to 200pt{\hfill removing lock \hfill}
\hfill
\hbox to 80pt{\hfill setting lock\hfill}
}
\vskip 3pt
\ligne{\hfill
\vtop{\leftskip 0pt\parindent 0pt\hsize=76pt 
\aff {279} {W} {WBWWWBWBBW} {W}
\aff {280} {W} {WWWBBWWWBW} {W}
\aff {281} {W} {WWBWWWWWBB} {W}
\aff {282} {W} {WWWWBBWWWW} {W}
\aff {283} {W} {WWBWWWWWWW} {W}
\aff {284} {B} {WBWBWWWBBW} {B}
\aff {285} {W} {BBWWBWWWWW} {W}
}
\hfill
\vtop{\leftskip 0pt\parindent 0pt\hsize=76pt 
\aff {286} {W} {WBWWWWWWWB} {W}
\aff {287} {W} {WWBBWWWWBB} {B}
\aff {288} {W} {WWBWWWWBWW} {W}
\aff {289} {W} {WBWWWBWBBB} {W}
\aff {290} {W} {WBWBBWWWBW} {B}
\aff {291} {B} {WWBWWWWWBB} {W}
\aff {292} {W} {WBBWWWWWWW} {W}
}
\hfill
\vtop{\leftskip 0pt\parindent 0pt\hsize=76pt 
\aff {293} {W} {WBWWBWWWWB} {W}
\aff {294} {B} {WWWBBWWWBW} {B}
\aff {295} {W} {BWBWWWWWBB} {W}
\aff {296} {B} {BBWWWBWWWW} {B}
\aff {297} {W} {WWBWWWBWWW} {W}
\aff {298} {B} {WBWWBWWBBW} {B}
\aff {299} {W} {WBWWWBWWWB} {W}
}
\hfill\hskip 30pt
\vtop{\leftskip 0pt\parindent 0pt\hsize=76pt 
\aff {300} {W} {BWBBWWWWBB} {B}
\aff {301} {W} {WWBWWWBBWW} {W}
\aff {302} {B} {WBWBBWWWBW} {W}
\aff {303} {B} {BWBWWWWWBB} {W}
\aff {304} {W} {WBBWWWBWWW} {W}
}
\hfill}
\vskip 7pt
\ligne{\hfill managing the motion of the locomotive \hfill}
\vskip 3pt
\ligne{\hfill\hbox to 200pt{\hfill selected track \hfill}
\hfill\hbox to 80pt{\hfill non-selected track\hskip 10pt}
}
\vskip 3pt
\ligne{\hfill
\vtop{\leftskip 0pt\parindent 0pt\hsize=76pt 
\aff {305} {W} {WBWWWBBWWB} {W}
\aff {306} {B} {BWWBBWWWBW} {B}
\aff {307} {W} {BWBWWWBWBB} {W}
\aff {308} {W} {BBWWWBWWWB} {W}
}
\hfill
\vtop{\leftskip 0pt\parindent 0pt\hsize=76pt 
\aff {309} {W} {BBWWBWWBBW} {W}
\aff {310} {B} {WWWBWBWWBW} {B}
\aff {311} {B} {WBWWWBWWBB} {W}
}
\hfill
\vtop{\leftskip 0pt\parindent 0pt\hsize=76pt 
\aff {312} {W} {BBWBBWWBWW} {B}
\aff {313} {B} {WBWBBBWBWW} {B}
\aff {314} {B} {BBWWBWWBWW} {B}
}
\hfill\hskip 30pt
\vtop{\leftskip 0pt\parindent 0pt\hsize=76pt 
\aff {315} {W} {WBBWWBWBBW} {W}
\aff {316} {W} {WBWWWWBWWB} {W}
}
\hfill}
}

\vskip 10pt
\def\exectabff #1 #2 #3 #4 {%
\hbox to 25pt{\hfill#1\hskip 5pt}
\hbox to 25pt{\hfill#2\hskip 5pt}
\hbox to 25pt{\hfill#3\hskip 5pt}
\hbox to 25pt{\hfill#4\hskip 5pt}
}

\vtop{
\begin{tab}\label{excontrol}
\leurre
Execution of the rules 
of the controller when a simple locomotive crosses the structure.
\end{tab}
\vspace{-6pt}
\ligne{\hfill\hbox to 150pt{\hskip 10pt 2(1) is black:\hfill}
\hskip 20pt\hbox to 75pt{\hskip 10pt 2(1) is white:\hfill}
\hfill}
\vskip 2pt
\ligne{\hfill
\vtop{\leftskip 0pt\parindent 0pt\hsize=150pt
\ligne{\hfill   \execsix     {}       0       1       2       3       4 \hfill}
\ligne{\hfill   \execsix  {1(2)}  {  25}  {  25}  {  25}  {  26}  { 311}   \hfill}
\ligne{\hfill   \execsix  {0(0)}  {  58}  {  58}  {  78}  {  83}  { 309}   \hfill}
\ligne{\hfill   \execsix  {1(4)}  {  35}  {  41}  {  42}  {  48}  {  35}   \hfill}
\ligne{\hfill   \execsix  {2(2)}  { 143}  { 143}  { 143}  { 266}  { 312}   \hfill}
\ligne{\hfill   \execsix  {1(1)}  { 294}  { 294}  { 294}  { 306}  { 310}   \hfill}
\ligne{\hfill   \execsix  {2(1)}  { 295}  { 295}  { 295}  { 307}  { 295}   \hfill}
}
\hskip 20pt
\vtop{\leftskip 0pt\parindent 0pt\hsize=75pt
\ligne{\hfill  \exectroi       0       1       2 \hfill}
\ligne{\hfill  \exectroi   {  59}  {  59}  {  59}   \hfill}
\ligne{\hfill  \exectroi   { 279}  { 279}  { 315}   \hfill}
\ligne{\hfill  \exectroi   {  35}  {  41}  {  42}   \hfill}
\ligne{\hfill  \exectroi   { 127}  { 127}  { 127}   \hfill}
\ligne{\hfill  \exectroi   { 280}  { 280}  { 280}   \hfill}
\ligne{\hfill  \exectroi   { 281}  { 281}  { 281}   \hfill}
}
\hfill}
}
\vskip 10pt
When the colour of the controller is black, the conservative rule for~1(1)
is rule~294,
\laff {B} {WWWBBWWWBW} {B} {.} When the locomotive crosses the controller, cell~1(1)
can see it first through in its neighbour~1, rule~306,
\laff {B} {BWWBBWWWBW} {B} {,} and then in its vertex-neighbour~6, see rule~310,
\laff {B} {WWWBWBWWBW} {B} {.} That rule and Figure~\ref{control} show us that the
cell~2(2) became white after seeing the locomotive in~0(0), which is the neighbour~7 
of~2(2). Indeed, in this situation, rule~266, \laff {B} {WBWBBWBBWW} {W} {,} applies 
to~2(2). When 2(2) is white, applying rule~312, \laff {W} {BBWBBWWBWW} {B} {,} allows us 
to restore the black state in the cell. This problem comes from the fact that rule~266 
was needed in the selector for the cell~0(0) when a double locomotive crosses that cell.
Note that after rule~310, rule~294 again applies to the cell~1(1).

Figure~\ref{controls} illustrates the action of the signal on the cell~1(1): when the cell is
black, the signal turns it to white and conversely.
   Table~\ref{excontrols} displays which rules are used to which cells when the signal arrives
in order to change the colour of the cell~1(1). The conservative rules for the cell~1(1) 
are rule~294, we have already seen, when the cell is black, it is rule~280, 
\laff {W} {WWWBBWWWBW} {W} {,} when the cell is white. We can see that the neighbourhood is
the same for rules~294 and~280: the milestones are in the neighbours~4, 5 and~9 of the cell,
and the difference lies in the state of the cell itself. The change from black to blank
is performed by rule~302, \laff {B} {WBWBBWWWBW} {W} {,} which can see the signal in its
neighbour~2. The reverse change is obtained by rule~290, \laff {W} {WBWBBWWWBW} {B} {,}
again as the signal is seen in the neighbour~2 of the cell.

\vtop{
\vspace{-15pt}
\ligne{\hfill\hskip 50pt
\includegraphics[scale=0.55]{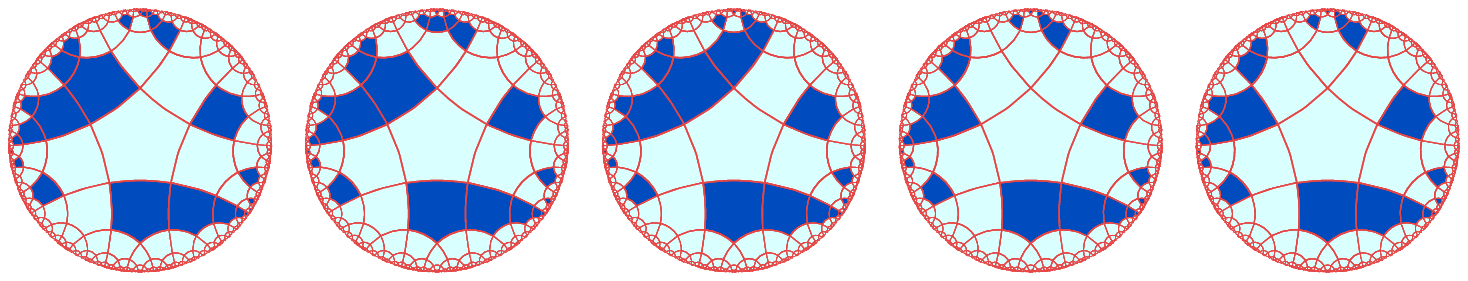} 
\hfill}
\vspace{-20pt}
\ligne{\hfill\hskip 50pt
\includegraphics[scale=0.55]{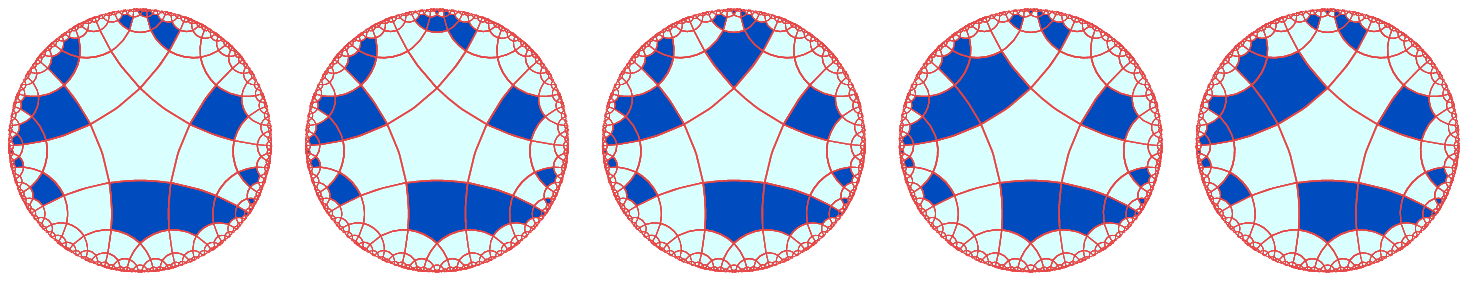} 
\hfill}
\begin{fig}\label{controls}
\leurre
The controller of the flip-flop and of the active memory switch. The signal arrives to
change the selection. Above, the controller is changed to white. Below, it is changed to black.
\end{fig}
}

\def\exectabffs #1 #2 #3 #4 {%
\hbox to 25pt{\hfill#1\hskip 5pt}
\hbox to 25pt{\hfill#2\hskip 5pt}
\hbox to 25pt{\hfill#3\hskip 5pt}
\hbox to 25pt{\hfill#4\hskip 5pt}
}

\vtop{
\begin{tab}\label{excontrols}
\leurre
Execution of the rules 
of the controller when a signal comes for changing the colour of the controller.
\end{tab}
\vspace{-6pt}
\ligne{\hfill\hbox to 130pt{\hskip 10pt from black to blank:\hfill}
\hskip 20pt\hbox to 95pt{\hskip 10pt from blank to black:\hfill}\hfill}
\vskip 2pt
\ligne{\hfill
\vtop{\leftskip 0pt\parindent 0pt\hsize=130pt
\ligne{\hfill  \execcinq     {}       0       1       2       3 \hfill}
\ligne{\hfill  \execcinq  {6(1)}  {  41}  {  42}  {  48}  {  35}   \hfill}
\ligne{\hfill  \execcinq  {2(1)}  { 295}  { 300}  { 303}  { 281}   \hfill}
\ligne{\hfill  \execcinq  {0(0)}  {  58}  {  58}  { 191}  { 279}   \hfill}
\ligne{\hfill  \execcinq  {2(2)}  { 143}  { 143}  { 143}  { 127}   \hfill}
\ligne{\hfill  \execcinq  {1(1)}  { 294}  { 294}  { 302}  { 280}   \hfill}
\ligne{\hfill  \execcinq  {1(5)}  { 299}  { 299}  { 190}  { 286}   \hfill}
}
\hskip 20pt
\vtop{\leftskip 0pt\parindent 0pt\hsize=95pt
\ligne{\hfill  \execquat       0       1       2       3 \hfill}
\ligne{\hfill  \execquat   {  41}  {  42}  {  48}  {  35}   \hfill}
\ligne{\hfill  \execquat   { 281}  { 287}  { 291}  { 295}   \hfill}
\ligne{\hfill  \execquat   { 279}  { 279}  { 289}  {  58}   \hfill}
\ligne{\hfill  \execquat   { 127}  { 127}  { 127}  { 143}   \hfill}
\ligne{\hfill  \execquat   { 280}  { 280}  { 290}  { 294}   \hfill}
\ligne{\hfill  \execquat   { 286}  { 286}  { 293}  { 299}   \hfill}
}
\hfill}
}
\vskip 10pt
   Other neighbours of the cell~1(1) can see those changes. As an example which is not
given in Table~\ref{excontrol}, we take the cell~1(5) which is always blank. The conservative 
rule is rule~299 \laff {W} {WBWWWBWWWB} {W} {,} when the cell 1(1) is black, and it is rule~286
\laff {W} {WBWWWWWWWB} {W} {,} when 1(1) is white. When the locomotive crosses the controller
in the black colour, rule~305,
\laff {W} {WBWWWBBWWB} {W} {,}
can see the locomotive in its neighbour~7, then rule~308,
\laff {W} {BBWWWBWWWB} {W} {,} can see it in its neighbour~1 and then, rule~299 is again
applied.
When 1(1) is white, rule~316, \laff {W} {WBWWWWBWWB} {W} {,} is applied when the locomotive
appears: it can be seen in its neighbour~7 and then the locomotive disappeared, so that
rule~286 again applies. Table~\ref{excontrols} shows the rules applied when the signal
arrives, changing the colour of the cell~1(1).

\subsection{The passive memory switch}

    As we said in Section~\ref{scenar}, the controller of the passive memory switch
works in a different way than that of the active switches.  For this reason, we call it 
the \textbf{controller-sensor} as it does not stop the locomotive when it runs through 
the non-selected track. As in the case of the active switches, this controller is 
characterized by the state of a cell, again the cell~1(1), see Figure~\ref{controlsign}.
Here, the cell 1(1) is called the \textbf{sensor} as its role is to detect a passage,
not to stop it. As in the case of the controller, we say that the state of 1(1) is the
\textbf{colour} of the sensor.
The tracks goes along the cells 33(2), 12(2), 4(2), 1(2), 0(0) and 1(4), 4(4), 12(4) and
33(4) in this order. This order is the opposite of what he have seen for the passage of a locomotive
through the controller of an active switch. This raises no problem thanks to the possibility
to use round-abouts if needed: the round-abouts do not use the controllers nor the 
controller-sensors. The difference of direction induces a sharp difference in the milestones.
Here the milestones of the track of the locomotive are: 34(3), 86(2), 13(3), 31(2), 5(3), 10(2),
2(3), 2(2), 2(5), 3(4), 5(5), 11(4), 13(5), 32(4), 34(5) and 87(4). 
Now, there is a common point between the controller-sensor and the controller of the active
switches: it is the path of the signal for changing the colour of the sensor. The path and the
milestones are the same. However a difference again: here, in the case of the controller-sensor,
the colour is changed only if the sensor is black. When it is white, it is changed to black 
by the very passage of the locomotive. Here, there are also additional milestones:
they consists of the cells 7(2), 6(2), 4(1), 10(1) and 2(4).
Those actions are illustrated by Figure~\ref{controlsign}. Table~\ref{passmemotab} gives
all the new rules induced by those constraints.
 
\vtop{
\vspace{-15pt}
\ligne{\hfill\hskip 60pt
\includegraphics[scale=0.55]{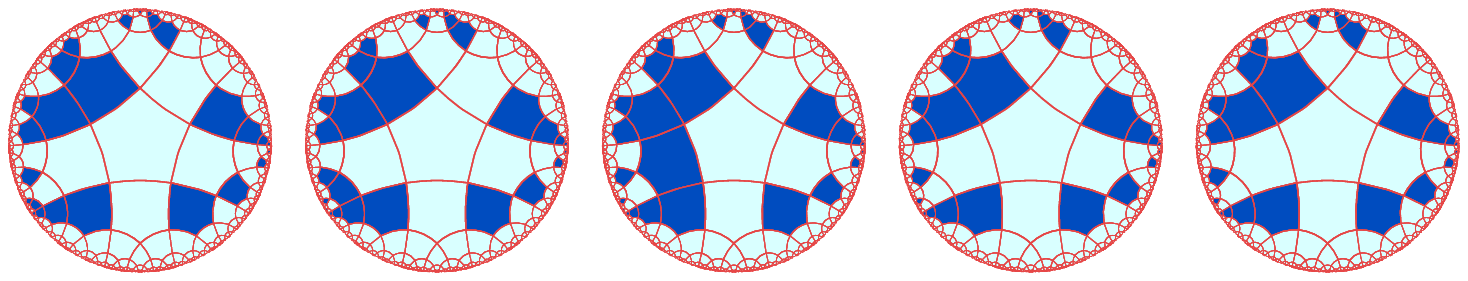} 
\hfill}
\vspace{-20pt}
\ligne{\hfill\hskip 10pt
\includegraphics[scale=0.55]{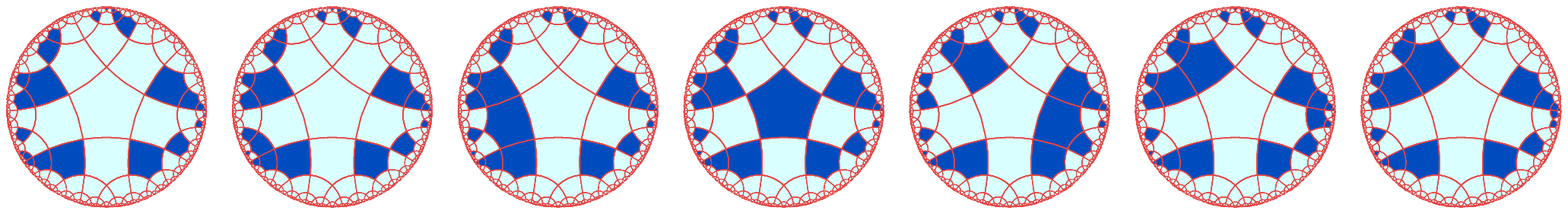} 
\hfill}
\vspace{-20pt}
\ligne{\hfill\hskip 60pt
\includegraphics[scale=0.55]{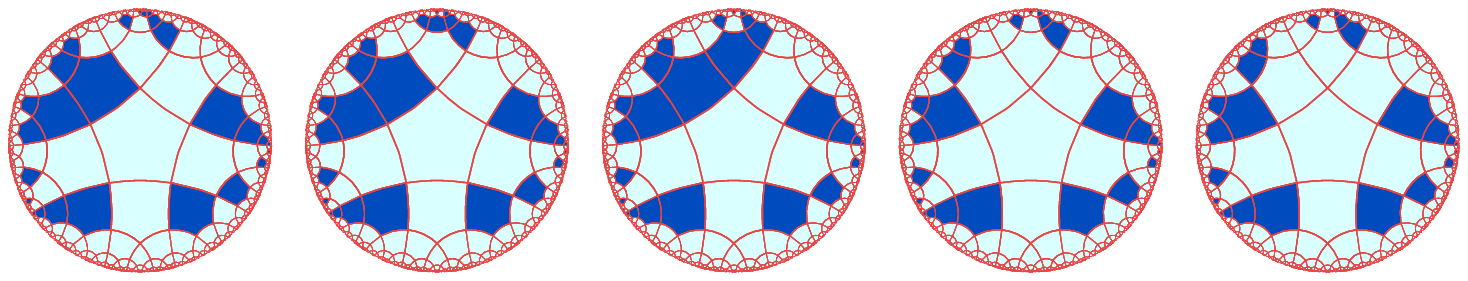} 
\hfill}
\begin{fig}\label{controlsign}
\leurre
The controller-sensor of the passive memory switch. Above: a single locomotive passively
crossed the switch through its selected track. Middle row: the locomotive crossed the passive
switch through the non-selected track. Below: the change of signal induced by a passive crossing
through the non-selected track.
\end{fig}
}
\vskip 10pt
   Consider the case when the sensor is white.
   The conservative rule for~0(0) is rule~317, \laff {W} {WWWWWBWBBB} {W} {.} Its milestones
are in its vertex-neighbours~6, 8, 9 and~10.

\vtop{
\begin{tab}\label{passmemotab}
\leurre
Rules for the passive memory switch.
\end{tab}
\vspace{-9pt}
\ligne{\hfill rules for the control-sensor\hfill}
\vskip 3pt
\ligne{\hfill non-selected track\hfill\hbox to 96pt{selected track\hfill}}
\vskip 3pt
\ligne{\hfill
\vtop{\leftskip 0pt\parindent 0pt\hsize=76pt 
\aff {317} {W} {WWWWWBWBBB} {W}
\aff {318} {W} {WBWWBWWWWW} {W}
\aff {319} {B} {BWBWWWWBWW} {B}
\aff {320} {B} {BBBWWWWWWW} {B}
\aff {321} {W} {BWBWBWWBWW} {W}
\aff {322} {W} {WBWBBWWBBB} {B}
\aff {323} {B} {WWBWWWWBWW} {B}
\aff {324} {B} {BWBWWWBWWW} {B}
\aff {325} {W} {WWBWWBWBBB} {B}
\aff {326} {W} {WWWBBBWWBW} {W}
\aff {327} {B} {BWWBBWWBWW} {B}
\aff {328} {B} {WWWWWBWBBB} {W}
\aff {329} {W} {BWWBBWWWBW} {B}
}
\hfill
\vtop{\leftskip 0pt\parindent 0pt\hsize=76pt 
\aff {330} {W} {WWBWWWBWBB} {W}
\aff {331} {B} {WWBWWWBWWW} {B}
\aff {332} {W} {BWBBBWWBWW} {B}
\aff {333} {W} {BBWWWWWWWB} {W}
\aff {334} {B} {WWWBWWWWBW} {B}
\aff {335} {W} {BBBWWWWBBB} {W}
\aff {336} {W} {WWWWBWBBBB} {W}
\aff {337} {W} {WBWBBWWBWW} {B}
\aff {338} {B} {WWWWWWWBWW} {B}
\aff {339} {B} {BWBWBWWBWW} {W}
\aff {340} {W} {BWBWWWWBBB} {W}
\aff {341} {W} {BBBWBWWBWW} {W}
\aff {342} {W} {WBBWWWWBBW} {W}
}
\hfill
\vtop{\leftskip 0pt\parindent 0pt\hsize=76pt 
\aff {343} {W} {WBWBBWBBBB} {B}
\aff {344} {W} {WBBWWBWBBB} {W}
\aff {345} {B} {WWWBBBWWBW} {B}
\aff {346} {B} {WBWWBWBBBB} {W}
\aff {347} {B} {BBWBBWWBWW} {B}
\vskip 3pt
\ligne{\hfill change of selection\hfill}
\vskip 3pt
\aff {348} {W} {WBWWWBBBBB} {W}
\aff {349} {B} {BWBBWWWWBB} {W}
\aff {350} {B} {BBBWBWBBBW} {W}
\aff {351} {B} {BBWWBBBWWW} {B}
\aff {352} {W} {BWWWWBWBBB} {W}
}
\hfill
}
}
\vskip 10pt

Table~\ref{exctrlsn} displays the rules used for the cells of the controller-sensor when the 
locomotive passes through the non-selected track. When the locomotive arrives at the cell~0(0),
it is seen from the neighbour~3 of the cell see rule~325, \laff {W} {WWBWWBWBBB} {B} {~:}
here, the side~1 of cell~0(0) is shared with 1(5). Then, rule~328,
\laff {B} {WWWWWBWBBB} {W} {,} makes the locomotive leave the cell. Rule~25, 
\laff {W} {WBWWBBWWBB} {W} {,}
witnesses
that the locomotive left~0(0) as it can be seen in the neighbour~5 of~0(0). The rule
also witnesses that the sensor turned to black: the cell~1(1) is the neighbour~2 of~0(0) which
is now black. Rule~25 is a rule of 
the vertical track is also applied here because at that very moment, the cell~2(2) became white
and the cell~2(2) is the vertex-number~8 of the cell which, usually, is black. After that,
rule~289,
\laff {W} {WBWWWBWBBB} {W} {,} is applied. It is the conservative rule of~0(0) when the sensor is 
black: remember that the cell~1(1) is the neighbour~2 of~0(0). The rule also witnesses
that the cell~2(2) returned to black.

\vtop{
\begin{tab}\label{exctrlsn}
\leurre
Execution of the rules for the controller of the passive memory switch when the sensor is white.
\end{tab}
\vskip -6pt
\ligne{\hfill
\vtop{\leftskip 0pt\parindent 0pt\hsize=186pt
\ligne{\hfill  \execsept     {}       0       1       2       3       4       5 \hfill}
\ligne{\hfill  \execsept  {4(2)}  {  26}  {  27}  {  28}  {  25}  {  25}  {  25}   \hfill}
\ligne{\hfill  \execsept  {2(2)}  { 234}  { 322}  { 244}  { 249}  { 336}  { 228}   \hfill}
\ligne{\hfill  \execsept  {0(0)}  { 317}  { 317}  { 325}  { 328}  {  25}  { 289}   \hfill}
\ligne{\hfill  \execsept  {4(1)}  { 321}  { 321}  { 321}  { 332}  { 339}  { 341}   \hfill}
\ligne{\hfill  \execsept  {4(4)}  {  35}  {  35}  {  35}  {  35}  {  41}  {  42}   \hfill}
\ligne{\hfill  \execsept  {2(2)}  { 127}  { 127}  { 327}  { 247}  { 337}  { 143}   \hfill}
\ligne{\hfill  \execsept  {1(1)}  { 280}  { 280}  { 326}  { 329}  { 334}  { 294}   \hfill}
}
\hfill}
}
\vskip 10pt
We leave to the reader the examination of the application of the rules given by
Table~\ref{passmemotab} to the cells indicated in Table~\ref{exctrlsn}. 
The first two rows of Figure~\ref{controlsign} are of help for such a task.

Table~\ref{exctrlbands} gives traces of execution for
the rules applied to the cells when the locomotive crosses a black sensor
and for those concerned by the change of the sensor from black to white.
In the latter case, the comparison of the right-hand side part of Table~\ref{exctrlbands}
with the left-hand side part of Table~\ref{excontrols} show us that the same rules apply.
This explains why the last row of Figure~\ref{controlsign} is very similar to that of
the first line of Figure~\ref{controls} for the part of these figures involved by
the cells concerned by the signal: the track where the signal goes, their milestones
and the cell~1(1) with its neighbours. Similarly, the behaviour of the cell~2(2) is
the same in the first line of Figure~\ref{control} and in the first one of 
Figure~\ref{controlsign}.

An important difference should also be noticed between the first line of Figure~\ref{controlsign}
and the first line of Figure~\ref{control}. It is the fact that the sensor is white in the first
case while it is black in the second one. Also, in the first case the sensor changes its colour
while its colour is unchanged in the second one. This explains why a few rules in the
Tables~\ref{exctrlsn} and~\ref{exctrlbands} are taken from Table~\ref{passmemotab}.

\vtop{
\begin{tab}\label{exctrlbands}
\leurre
Executions of the rules for the control of the passive memory switch when the sensor is blue.
\end{tab}
\vskip-6pt
\ligne{\hfill\hbox to 170pt{\hfill the locomotive passes\hfill}\hskip 15pt
\hbox to 120pt{\hfill change from blue to red\hfill}\hfill}
\ligne{\hfill
\vtop{\leftskip 0pt\parindent 0pt\hsize=176pt
\ligne{\hfill  \execcinq     {}       0       1       2       3 \hfill}
\ligne{\hfill  \execcinq  {4(2)}  {  26}  {  27}  {  28}  {  25}   \hfill}
\ligne{\hfill  \execcinq  {1(2)}  { 228}  { 343}  { 346}  { 228}   \hfill}
\ligne{\hfill  \execcinq  {0(0)}  { 289}  { 289}  { 344}  { 289}   \hfill}
\ligne{\hfill  \execcinq  {2(2)}  { 143}  { 143}  { 347}  { 143}   \hfill}
\ligne{\hfill  \execcinq  {1(1)}  { 294}  { 294}  { 345}  { 294}   \hfill}
}
\hskip 15pt
\vtop{\leftskip 0pt\parindent 0pt\hsize=125pt 
\ligne{\hfill  \execcinq     {}       0       1       2       3 \hfill}
\ligne{\hfill  \execcinq  {6(1)}  {  41}  {  42}  {  48}  {  35}   \hfill}
\ligne{\hfill  \execcinq  {2(1)}  { 295}  { 300}  { 303}  { 281}   \hfill}
\ligne{\hfill  \execcinq  {2(2)}  { 143}  { 143}  { 143}  { 127}   \hfill}
\ligne{\hfill  \execcinq  {1(1)}  { 294}  { 294}  { 302}  { 280}   \hfill}
}
\hfill}
}
\vskip 10pt
As examples, consider the cells~0(0), 1(1) and~2(2) when the locomotive passes while the
sensor is black.
As already mentioned, the conservative rule for~0(0) when the cell~1(1) is black is
rule~289. When the locomotive is seen by~0(0), the locomotive is in the neighbour~3 of~0(0),
so that rule~344, 
\laff {W} {WBBWWBWBBB} {W} {,}
is applied. At the next time, as the locomotive vanished, rule~289 is again applied.
In the case of~1(1) when it is black, the conservative rule is rule~294,
\laff {B} {WWWBBWWWBW} {B} {,} a rule from Table~\ref{controltab}. When the cell can see the
locomotive which appears as the neighbour~6 of~1(1), rule~345,
\laff {B} {WWWBBBWWBW} {B} {,} is applied. Again,as the locomotive vanishes at the next time,
rule~294 is again applied. At last, the conservative rule for~2(2) is rule~143,
\laff {B} {WBWBBWWBWW} {B} {,} a rule for the horizontal tracks. Here too, the presence of the
locomotive is noticed for one time only: the locomotive is in the neighbour~1 of~2(2):
rule 347,
\laff {B} {BBWBBWWBWW} {B} {,} is applied. At the next time, rule~143 is applied: the locomotive
vanished.
\vskip 15pt
The tables of this section completes the proof of Theorem~\ref{letheo}.\hfill\boxempty

\subsection{About rotation invariance}

We already mentioned in Section~\ref{rules} that the automaton we constructed to prove 
Theorem~\ref{letheo} is not rotation invariant. We used this property in order to suitably
number the sides of each cell. Among the rules displayed in the various tables of 
Section~\ref{rules}, many of them are rotated forms of other rules. As an example, 
rules~3, 39, 126 and 286, \textit{i.e.}
\laff {W} {BWWWWWWWWW} {W} {,}
\laff {W} {WBWWWWWWWW} {W} {,}
\laff {W} {WWWWBWWWWW} {W} {} and
\laff {W} {WWBWWWWWWW} {W} {,} respectively, are rotated form of each other. They apply to white
cells with a single black neighbour, neighbour~1, 2, 5 and 3 respectively. We shall say
that in the corresponding neighbourhoods there is a shift by 1, 4 and~2 for rules~39, 126 
and~286 respectively with respect to that of rule~3. 

\def\lignure #1 #2 #3 #4 #5 #6 #7 {%
\ligne{\vtop{\leftskip 0pt\parindent 0pt\hsize=74pt
             #1}\hskip 3pt 
       \hbox to 25pt{\hfill #2}\hskip 3pt
       \hbox to 25pt{\hfill\tt #3}\hskip 7pt
       \vtop{\leftskip 0pt\parindent 0pt\hsize=74pt 
             #4}\hskip 3pt 
       \hbox to 25pt{\hfill #5}\hskip 3pt
       \hbox to 25pt{\hfill\tt #6}\hskip 5pt
       \hbox to 7pt{\hfill #7}
\hfill}
}

\vtop{
\begin{tab}\label{diffs1}
\leurre
Rules which contradict rotation invariance, part {\rm I}.
\end{tab}
\vspace{-9pt}
\ligne{\hfill
\vtop{\leftskip 0pt\parindent 0pt\hsize=300pt
\lignure {\aff { 21} {W} {WBWBBWWWBB} {W} } {21(3)} {Vdd} 
         {\aff { 65} {W} {WBBWBWBBWW} {B} } {$^{\ast1}$} {Has} 3
\lignure {\ligne{\hfill}} {7(4)} {Hwcd} {\ligne{\hfill}} {$^{\ast2}$} {Had} {}
\lignure {\ligne{\hfill}} {21(3)} {fxld} {\ligne{\hfill}} {$^{\ast3}$} {Hwas} {}
\lignure {\ligne{\hfill}} {21(3)} {fxrd} {\ligne{\hfill}} {$^{\ast4}$} {Hwad} {}
\lignure {\ligne{\hfill}} {21(3)} {D} {\ligne{\hfill}} {} {} {}
\lignure {\ligne{\hfill}} {8(2)} {Cb} {\ligne{\hfill}} {} {} {}
\vskip 3pt
\lignure {\aff { 30} {W} {BWWWBBBWWW} {W} } {$^{\ast5}$} {Vdd}
         {\aff {287} {W} {WWBBWWWWBB} {B} } {2(1)} {Cws} 3
\lignure {\ligne{\hfill}} {$^{\ast6}$} {Hcd}
         {\ligne{\hfill}} {} {} {}
\lignure {\ligne{\hfill}} {$^{\ast7}$} {Had}
         {\ligne{\hfill}} {} {} {}
\lignure {\ligne{\hfill}} {$^{\ast8}$} {fxld}
         {\ligne{\hfill}} {} {} {}
\lignure {\ligne{\hfill}} {$^{\ast9}$} {fxrd}
         {\ligne{\hfill}} {} {} {}
\lignure {\ligne{\hfill}} {$^{\ast9}$} {D}
         {\ligne{\hfill}} {} {} {}
\lignure {\ligne{\hfill}} {$^{\ast10}$} {Sls}
         {\ligne{\hfill}} {} {} {}
\lignure {\ligne{\hfill}} {$^{\ast10}$} {Sld}
         {\ligne{\hfill}} {} {} {}
\lignure {\ligne{\hfill}} {$^{\ast11}$} {Cbs}
         {\ligne{\hfill}} {} {} {}
\lignure {\ligne{\hfill}} {$^{\ast11}$} {snbs}
         {\ligne{\hfill}} {} {} {}
\lignure {\aff { 51} {W} {WBBWBWWBBW} {W} } {33(1)} {Vud}
         {\aff {124} {W} {WBWBBBWWWB} {B} } {$^{\ast12}$} {Hcs} 2
\lignure {\ligne{\hfill}} {5(4)} {Had}
         {\ligne{\hfill}} {$^{\ast13}$} {Hcd} {}
\lignure {\ligne{\hfill}} {} {}
         {\ligne{\hfill}} {$^{\ast14}$} {Hwcs} {}
\lignure {\ligne{\hfill}} {} {}
         {\ligne{\hfill}} {$^{\ast15}$} {Hwcd} {}
\lignure {\ligne{\hfill}} {} {}
         {\aff {300} {W} {BWBBWWWWBB} {B} } {2(1)} {Cbs} 1
\lignure {\ligne{\hfill}} {} {}
         {\ligne{\hfill}} {2(1)} {Snbs} {}
\lignure {\aff { 80} {B} {WBWWBWBWBW} {W} } {$^{\ast16}$} {Has}
         {\aff {252} {B} {WWBWBWBWWB} {B} } {2(4)} {Sls} 3
\lignure {\ligne{\hfill}} {$^{\ast17}$} {Hcs}
         {\ligne{\hfill}} {} {} {}
\lignure {\ligne{\hfill}} {$^{\ast18}$} {Hwas}
         {\ligne{\hfill}} {} {} {}
\lignure {\ligne{\hfill}} {$^{\ast19}$} {Hwcs}
         {\ligne{\hfill}} {} {} {}
\lignure {\ligne{\hfill}} {2(5)} {Sls}
         {\ligne{\hfill}} {} {} {}
\lignure {\ligne{\hfill}} {2(5)} {fk}
         {\ligne{\hfill}} {} {} {}
\lignure {\aff {112} {B} {BWBBWBWWWW} {B} } {$^{\ast20}$} {Had}
         {\aff {339} {B} {BWBWBWWBWW} {W} } {1(4)} {Snw} 2
\lignure {\ligne{\hfill}} {1(1)} {Hcd}
         {\ligne{\hfill}} {} {} {}
}
\hfill}
\vspace{2pt}
{\leftskip 0pt\parindent 0pt\footnotesize\tt
\noindent
$^{\ast1}$ 18(1), 10(1), 5(2), 7(3), 10(3) \hskip 20pt
$^{\ast2}$ 10(1), 5(2), 7(3), 10(3) \\
$^{\ast3}$ 10(4), 5(5), 2(1), 7(2), 10(2)\hskip 20pt
$^{\ast4}$ 5(5), 2(1), 7(2), 10(2) \\
$^{\ast5}$ 11(1), 3(1), 1(5), 2(4), 10(3), 23(3) \\ 
$^{\ast6}$ 9(3), 6(3), 4(2), 7(2), 12(1), 9(1), 20(1), 17(1)\\
$^{\ast7}$ 9(2), 6(2), 4(1), 4(5), 7(5), 12(4), 9(4)\\
$^{\ast8}$ 11(1), 3(1), 2(4), 10(3), 23(3)
$^{\ast9}$ 2(4), 10(3), 23(3)
$^{\ast10}$ 6(4)
$^{\ast11}$ 3(1) \\
$^{\ast12}$ 10(3), 7(3), 5(2), 10(1), 18(1)
$^{\ast13}$ 7(3), 5(2), 10(1), 18(1), 15(1) \\
$^{\ast14}$ 10(2), 7(2), 2(1), 5(5), 10(4)
$^{\ast15}$ 7(2), 2(1), 5(5), 10(4)
$^{\ast16}$ 7(1), 2(2)\\
$^{\ast17}$ 3(3), 4(1)
$^{\ast18}$ 2(5), 1(1)
$^{\ast19}$ 3(2), 4(4)
$^{\ast20}$ 1(1), 1(3)
\par}
}
\vskip 10pt
   Now, there are 14 pairs of rules in which the neighbourhoods are rotated, the state of the cell
is the same but its new state is different. Tables~\ref{diffs1} and~\ref{diffs2} give the rules 
together with the cells to which they apply and in which part of the circuit. 
The part of the circuit is identified by a few letters as follows. For the tracks, {\tt V}
means vertical and {\tt H} means horizontal; {\tt d} as second letter means down while {\tt u}
means up; {\tt d}, {\tt s} as a last letter indicates a double, simple locomotive, 
respectively; {\tt c}, {\tt a} means a clockwise, counter-clockwise running, respectively;
{\tt w} means that the horizontal goes around a white node; when {\tt w} is missing,
this means that the horizontal goes around a black node.
The fixed switch is represented by {\tt fx}, {\tt l}, {\tt r} indicates from the left-, right-hand 
side respectively. The fork is represented by {\tt fk}, the doubler by {\tt D}. The controller
is denoted by~{\tt C} while the controller-sensor is defined by {\tt Sn}, {\tt b}, {\tt w}
indicate the colour of the controller or the sensor and {\tt s} indicates the run involving the 
signal. At last, {\tt Sl} denotes the selector, {\tt s} and {\tt d} indicating whether it is crossed
by a simple or a double locomotive.

\vtop{
\begin{tab}\label{diffs2}
\leurre
Rules which contradict rotation invariance, part {\rm II}.
\end{tab}
\vspace{-9pt}
\ligne{\hfill
\vtop{\leftskip 0pt\parindent 0pt\hsize=300pt
\lignure {\aff {147} {B} {WBWBBWBWWW} {B} } {1(3)} {Hcd}
         {\aff {339} {B} {BWBWBWWBWW} {W} } {1(4)} {Snw} 1
\lignure {\aff {202} {W} {WBBWBBWWWB} {W} } {0(0)} {D}
         {\aff {242} {W} {WBWBBWBBWW} {B} } {0(0)} {Sls} 2
\lignure {\ligne{\hfill}} {} {}
         {\ligne{\hfill}} {0(0)} {seld} {}
\lignure {\aff {209} {B} {BWBWBBBBWW} {W} } {1(1)} {D}
         {\aff {265} {B} {WBBWBWWBBB} {B} } {1(3)} {Sld} 2
\lignure {\aff {212} {W} {BWBWBBBBWW} {W} } {1(1)} {D}
         {\aff {240} {W} {WBBWBWWBBB} {B} } {1(3)} {Sls} 2
\lignure {\ligne{\hfill}} {} {}
         {\ligne{\hfill}} {1(3)} {seld} {}
\lignure {\aff {226} {B} {WWBWBWBWWW} {W} } {1(4)} {fk}
         {\aff {250} {B} {WBWBWBWWWW} {B} } {2(3)} {Sls} 4
\lignure {\aff {246} {B} {WBBWBWWWWB} {B} } {2(4)} {Sls}
         {\aff {339} {B} {BWBWBWWBWW} {W} } {1(4)} {Snw} 3
\lignure {\ligne{\hfill}} {2(4)} {Sld}
         {\ligne{\hfill}} {} {} {}
\lignure {\aff {251} {W} {WBBWBWBWBB} {B} } {1(4)} {Sls}
         {\aff {277} {W} {BBWBWBWBBW} {W} } {1(2)} {Sld} 4
\lignure {\aff {256} {B} {BWBWBWWWWB} {B} } {2(4)} {Sls}
         {\aff {302} {B} {WBWBBWWWBW} {W} } {1(1)} {Cbs} 4
\lignure {\ligne{\hfill}} {} {}
         {\ligne{\hfill}} {1(1)} {Snbs} {}
}
\hfill}
}

   The tables indicate that the relaxation of the rotation invariance plays an important role.
However, the number of rules concerned is a bit less than ten percents of the rules. Also
the high number of rotated forms of rules is an interesting feature. At last, besides
the first rules numbered from~1 to~10, several rules are used in many different places.
As an example, rule~190, \laff {W} {WBWWBBWWWB} {W} {,} is used in many places:
5(3) in horizontal tracks which clockwise go around a white node, both for the simple
and the double locomotive. The rule is also applied to the cell 1(5) in the controller
and in the controller-sensor when the black colour is changed to the white one. We have already
seen that these motions of the signal are very similar in both those cases.

Three rules have all their rotated forms present in the tables: rule~16, rule~23 and rule~56.
As an interesting example, we have chosen rule~16, 
\laff {B} {BBWWBWWWWW} {B} {,} whose other forms are rule~81,
\laff {B} {WWBBBWWWWW} {B} {,} rule~93,
\laff {B} {WBBBWWWWWW} {B} {,} rule~136,
\laff {B} {BWWBBWWWWW} {B} {} and rule~320,
\laff {B} {BBBWWWWWWW} {B} {.}
With respect to rule~16, the shifts applied to rule~81, 93, 136 and~320 are 3, 2, 4 and~1,
respectively. 
Rule~16 is present very often, in many situations, the horizontal motions excepted. Rule~81
is applied to 1(1) and 1(3) each time once in the horizontal motions with the simple locomotive
around a black node. Rule~93 is applied once to~1(1) in the same motions. Rule~136 is applied
once to 1(3) in the clockwise motion only. Rule~320 is applied to 5(3) and 13(3), once in the
sensor, in the passage of the locomotive, whatever the colour of the sensor.
  
   It should be noted that the rules of Table~\ref{diffs1} apply to many cells, in particular
the first five pairs of the table. This is due to the fact that the rules mainly concern 
elements of the tracks. By contrast, in each pair of Table~\ref{diffs2}, each rule concerns 
at most two cells.


\section*{Conclusion}

Several questions are raised by this result. How to reduce the number of states in this
situation? Is it possible to obtain two states with a Moore neighbourhood and still observing
rotation invariance?  Still hard work ahead.

\end{document}